\documentstyle[preprint,eqsecnum,prd,aps]{revtex}
\begin{document}
\draft
%%%%%End of Preamble
%%%%Start of Text%%%%%%%%%%%%%%%%%%%%%%%%%%%%%%%%%%%%%%%%%%%%%%%%%%%%%%%
\preprint{
\vbox{
\halign{&##\hfil\cr
	& ANL-HEP-PR-94-24 \cr
	& FERMILAB-PUB-94-073-T \cr
	& NUHEP-TH-94-5 \cr
	& July 1994  \cr
	& Revised January 1997 \cr}}
}
\title{
Rigorous QCD Analysis \\
of Inclusive Annihilation and Production \\
of Heavy Quarkonium\\
}
\author{Geoffrey T. Bodwin}
\address{
High Energy Physics Division, Argonne National Laboratory,
Argonne, IL 60439
}
\author{Eric Braaten\footnotemark[1]}
\footnotetext[1]{On leave from Dept. of Physics and Astronomy,
	Northwestern University, Evanston, IL 60208.}
\address{
Theory Group, Fermilab, Batavia, IL 60510
}
\author{G. Peter Lepage}
\address{
Newman Laboratory of Nuclear Studies, Cornell University,
Ithaca, NY 14853
}
\maketitle
\begin{abstract}

A rigorous QCD analysis of the inclusive annihilation decay rates of
heavy quarkonium states is presented.
The effective-field-theory framework of nonrelativistic QCD
is used to separate the short-distance scale of annihilation,
which is set by the heavy quark mass $M$, from the longer-distance scales
associated with quarkonium structure.  The annihilation decay
rates are expressed in terms of nonperturbative matrix elements
of 4-fermion operators in nonrelativistic QCD, with coefficients that
can be computed using perturbation theory in the
coupling constant $\alpha_s(M)$. The matrix elements are organized into a
hierarchy according to their scaling with $v$,
the typical velocity of the heavy quark.
An analogous factorization formalism is developed for the production
cross sections of heavy quarkonium in processes involving momentum transfers
of order $M$ or larger.
The factorization formulas are applied to the annihilation decay rates
and production cross sections of S-wave 
states, up to corrections of relative order $v^3$,
and of P-wave states, up to corrections of relative order $v^2$.

\end{abstract}

\narrowtext
\section{Introduction}

Calculations of the decay rates of heavy quarkonium states into
light hadrons and into photons and
lepton pairs are among the earliest applications of
perturbative  quantum chromodynamics (QCD) \cite{aprg,barbb,itep,rem}.
In these early analyses, it was assumed that the
decay rate of the meson factored into a short-distance part that is
related to the annihilation rate of the heavy quark and antiquark,
and a long-distance factor containing all the nonperturbative effects
of QCD.  The short-distance factor
was calculated in terms of the running
coupling constant $\alpha_s(M)$ of QCD, evaluated at the scale of
the heavy-quark mass $M$, while the long-distance factor was
expressed in terms of the meson's nonrelativistic
wavefunction, or its derivatives, evaluated at the origin.
In the case of S-waves \cite{barba,ml}
and in the case of P-wave decays into photons \cite{barbe},
the factorization assumption was supported by explicit calculations at
next-to-leading order in $\alpha_s$. However, no general argument was
advanced for its validity in higher orders of perturbation theory. In
the case of P-wave decays into light hadrons, the factorization is
spoiled by logarithmic infrared divergences that appear in the
$Q \overline{Q}$~annihilation rates at order $\alpha_s^3$ \cite{barbe,barbc}.
Logarithmic infrared divergences also appear in relativistic corrections
to the annihilation decays of S-wave states \cite{barbc}.  These
divergences cast a shadow over applications of perturbative QCD to the
calculation of annihilation rates of heavy quarkonium states.

In this paper, we present a rigorous QCD analysis of the annihilation decays
of heavy quarkonium.  We derive a  general factorization formula
for the annihilation rates of S-wave, P-wave, and
higher orbital-angular-momentum states, which
includes not only perturbative corrections to all orders in $\alpha_s$,
but relativistic corrections as well.
Factorization occurs in the annihilation decay rates
because the heavy quark and antiquark can annihilate only when they are
within a distance of order $1/M$, where $M$ is the heavy-quark mass.
Since, in the meson rest frame, the heavy quark and antiquark are
nonrelativistic, with typical velocities~$v \ll 1$, this distance is
much smaller than the size of the meson, which is of order $1/(Mv)$.
Factorization involves separating the relativistic physics of
annihilation (which involves momenta $p \sim M$) from the
nonrelativistic physics of quarkonium structure (which involves $p
\sim Mv$). A particularly elegant approach for separating relativistic
from nonrelativistic scales is
to recast the analysis in terms of nonrelativistic quantum chromodynamics
(NRQCD) \cite{caswell-lepage}, an effective field theory designed
precisely for this purpose. NRQCD consists of a nonrelativistic
Schr\"odinger field theory for the heavy quark and antiquark that is
coupled to the usual relativistic field theory for light quarks and
gluons.  The theory is made precisely equivalent to full QCD
through the addition of local interactions that systematically
incorporate relativistic corrections through any given order in the
heavy-quark velocity~$v$.  It is an effective field theory, with a
finite ultraviolet cutoff of order~$M$ that excludes relativistic
states --- states that are poorly described by nonrelativistic dynamics.
A heavy quark in the meson can fluctuate into a relativistic state, but
these fluctuations are necessarily short-lived. This means that the
effects of the excluded relativistic states  can be mimicked by local
interactions and can, therefore, be incorporated into NRQCD through
renormalizations of its infinitely many coupling constants.
Thus, nonrelativistic
physics is correctly described by the nonperturbative dynamics of NRQCD,
while all relativistic effects are absorbed into
coupling constants that can be computed as perturbation series
in $\alpha_s(M)$.

The main advantage offered by NRQCD over ordinary QCD in this context is
that it is easier to separate contributions of different orders in $v$
in NRQCD. Thus, we are able not only to organize calculations to all
orders in $\alpha_s$, but also to elaborate systematically the
relativistic corrections to the conventional formulas. Furthermore, we
provide nonperturbative definitions of the long-distance factors in
terms of matrix elements of NRQCD, making it possible to evaluate them
in numerical lattice calculations. Analyzing S-wave decays within this
framework, we recover, up to corrections of relative order $v^2$, the
standard factorization formulas, which contain a single nonperturbative
parameter.  At relative order $v^2$, the decay rates satisfy a more
general factorization formula, which contains two additional independent
nonperturbative matrix elements. Our results for P-wave decays into
light hadrons are even more striking, as we have discussed in Ref.
\cite{bbl}. Up to corrections of relative order $v^2$, the factorization
formula for these decay rates is the sum of two terms.  In addition to
the conventional term, which takes into account the annihilation of the
$Q \overline{Q}$ pair from a color-singlet P-wave state, there is a
second term that involves annihilation from a color-octet S-wave state.
The infrared divergences encountered in previous calculations are
absorbed into the matrix element of the color-octet term. 

Our presentation is organized as follows. In Section~\ref{sec:nrqcd}, we
first review NRQCD in general, emphasizing the velocity-scaling rules,
which are used in separating contributions of different orders in $v$.
We then discuss the space-time structure of the annihilation of heavy
quarks
and antiquarks and explain how the effects of annihilation
can be taken into account in NRQCD by adding local 4-fermion operators
to the effective lagrangian.
In Section~\ref{sec:matrix-el}, we analyze the matrix elements of the
4-fermion operators.  We discuss their scaling with $v$,
the constraints on them that follow from heavy-quark spin symmetry,
their relations to Coulomb-gauge wavefunctions,
and their dependences on the factorization scale.
In Section~\ref{sec:annih}, we apply our formalism to the
annihilation decays of S-wave quarkonium 
states, up to corrections of relative order $v^3$,
and to P-wave decays,
up to corrections of relative order $v^2$.
In Section~\ref{sec:pert-fac}, we sketch the derivation of our results
in a more conventional perturbative approach to factorization.
In Section~\ref{sec:prod}, we develop an analogous
factorization formalism for calculating
the production cross sections of heavy quarkonium. In the concluding section,
we compare our formalism with previous approaches to the annihilation
and production of heavy quarkonium, and we summarize the current status
of calculations of annihilation and production rates.

\vfill \eject

\section{NRQCD}
\label{sec:nrqcd}

We begin this section with a brief discussion of the various momentum
scales involved in heavy quarkonia.  Nonrelativistic QCD (NRQCD)
\cite{caswell-lepage} is our major tool for resolving the different
momentum scales involved in their annihilation decays. We review this
effective field theory and its application to heavy-quarkonium physics.
Then we discuss the space-time structure of the $Q \overline{Q}$ annihilation
process and develop a general factorization formula for the
annihilation decay rates of heavy quarkonia in terms of matrix elements
of NRQCD.

\subsection{Energy Scales in Heavy Quarkonium}
\label{sec:energy-scales}

In a meson containing a heavy quark and antiquark, there are
several different momentum scales that play important roles in the dynamics.
The most important scales are the mass $M$ of the heavy quark,
its typical 3-momentum $Mv$ (in the meson rest frame),
and its typical kinetic energy $Mv^2$.
The heavy-quark mass $M$ sets the overall scale of the
rest energy of the bound state
and also provides the short-distance scale for annihilation processes.
The size of the bound state is the inverse of the momentum $Mv$,
while $Mv^2$ is the scale of the energy splittings between
radial excitations and between orbital-angular-momentum excitations.
Spin splittings within a given radial and orbital-angular-momentum
excitation are
of order $Mv^4$, but this scale plays no significant role in the dynamics.

The typical velocity $v$ of the heavy quark decreases as the mass $M$
increases.  If $M$ is large enough, $v$ is proportional to the running
coupling constant $\alpha_s(M)$, and it therefore decreases asymptotically
like $1/\log(M)$.  Thus, if $M$ is sufficiently large,
the heavy quark and antiquark are nonrelativistic,
with typical velocities $v \ll 1$. We assume in this paper that
the mass $M$ is heavy enough that the momentum scales $M$, $Mv$ and $Mv^2$
are well-separated: $ (Mv^2)^2 \ll (Mv)^2 \ll M^2$.
Quark potential model calculations indicate that the average
value of $v^2$ is about 0.3 for charmonium and about 0.1 for
bottomonium \cite{qr}, and these estimates
are confirmed by lattice QCD simulations.  Thus, the
assumption $(Mv^2)^2 \ll (Mv)^2 \ll M^2$ is very good for bottomonium,
and reasonably good even for charmonium.  For lighter quarkonium states,
such as the $s \bar s$ system, our analysis does not apply.

Another momentum  scale that plays a role in the physics of heavy
quarkonium is $\Lambda_{QCD}$, the scale associated with nonperturbative
effects involving gluons and light quarks.  It determines, for example,
the long-range behavior of the potential between the heavy quark and
antiquark, which is approximately linear, with a coefficient of $(450 \;
{\rm MeV})^2$ \cite{qr}.  We can use this coefficient as an estimate for
the nonperturbative scale: $\Lambda_{QCD} \approx 450 \; {\rm MeV}$.
For both charmonium and bottomonium, the first radial excitation
and the first orbital-angular-momentum excitation are both about 500 MeV
above the ground state.  Taking this value as an estimate for the scale
$Mv^2$, we see that $\Lambda_{QCD}$ and $Mv^2$ are comparable for both
charmonium and bottomonium.

Our analysis of heavy quarkonium annihilation is based on
separating the effects at the momentum scale $M$ from those at the lower
momentum scales $Mv$, $Mv^2$ and $\Lambda_{QCD}$.
The effects at the scale $M$ are taken into account through the
coupling constants of 4-fermion operators in the lagrangian
for NRQCD.  We assume that
$\alpha_s(M) \ll 1$, so that these coupling constants can be calculated
using perturbation theory in $\alpha_s(M)$.  The assumption that
$\alpha_s(M) \ll 1$ is well-satisfied for bottomonium,
for which $\alpha_s(M) \approx 0.18$,  and reasonably well-satisfied
for charmonium, for which $\alpha_s(M) \approx 0.24$.

The effects of the lower momentum scales $Mv$, $Mv^2$, and
$\Lambda_{QCD}$ are factored into matrix elements that can be calculated
using nonperturbative methods, such as lattice-QCD simulations. These
matrix elements are organized into a hierarchy in terms of their
dependence on $v$. Our final expression for the annihilation rate
therefore takes the form of a double expansion in $\alpha_s(M)$ and $v$.
These expansion parameters are not independent for quarkonium. The
typical velocity $v$ of the heavy quark is determined by a
nonperturbative balance between its kinetic energy $Mv^2/2$ and the
potential energy, which, for sufficiently large $M$, is dominated by a
color-Coulomb term proportional to $\alpha_s(1/r)/r$. Setting $r \sim
1/(Mv)$ in the potential and equating it with the kinetic energy, we
obtain the identification 
\begin{equation}
v  \;\sim\; \alpha_s(Mv).
\label{valpha}
\end{equation}
This equation can be solved self-consistently to obtain an approximate
value for the typical velocity $v$. The identification (\ref{valpha})
has a simple, but important, implication for calculations of
annihilation rates.  Since the running coupling constant in QCD
decreases with the momentum scale, $v$ is greater than or of order
$\alpha_s(M)$.  Thus relativistic corrections of order $(v^2)^n$ can be
expected to be more important than perturbative corrections of order
$\alpha_s^{2n}(M)$. In particular, there is little to be gained  by
calculating perturbative corrections at next-to-next-to-leading order in
$\alpha_s(M)$, unless relativistic corrections through relative order
$v^2$ are included as well. 

\subsection{The NRQCD Lagrangian}
\label{sec:lagrangian}

The most important energy scales for the structure and spectrum
of a heavy quarkonium system
are $Mv$ and $Mv^2$, where $M$ is the mass of the heavy quark $Q$ and
$v\ll 1$ is its average velocity in the meson rest frame.
Momenta of order~$M$ play only a minor role
in the complex binding dynamics of the system. We can take advantage of
this fact in our analysis of heavy-quark mesons by modifying QCD in two
steps.

We start with full QCD, in which the heavy quarks are described by
4-component Dirac spinor fields.  In the first step,
we introduce an ultraviolet momentum
cutoff~$\Lambda$ that is of order $M$. This cutoff explicitly excludes
relativistic heavy quarks from the theory, as well as gluons and light
quarks with momenta of order $M$. It is appropriate to our analysis of
heavy quarkonium, since the important nonperturbative physics involves
momenta of order $Mv$ or less. Of course, the relativistic states we are
discarding do have some effect on the low-energy physics of the theory.
However, any interaction involving relativistic intermediate states is
approximately local, since the intermediate states are necessarily
highly virtual and so cannot propagate over long distances. Thus,
generalizing standard renormalization
procedures, we systematically compensate for the removal of relativistic
states by adding new local interactions to the lagrangian. To leading
order in~$1/\Lambda$ or, equivalently, $1/M$,
these new interactions are identical in form to interactions already
present in the theory, and so the net effect is simply to shift bare
masses and charges. Beyond leading order in~$1/M$, one must extend the
lagrangian to include nonrenormalizable interactions that correct the
low-energy dynamics order-by-order in $1/M$. In this cutoff formulation
of QCD, all effects that arise from relativistic states, and only these
effects, are incorporated into renormalizations of the coupling
constants of the extended lagrangian. Thus, in the cutoff theory,
relativistic and nonrelativistic contributions are automatically
separated. This separation is the basis for our analysis of the
annihilation decays of heavy quarkonia.

The utility of the cutoff theory is greatly enhanced if, as a second
step, a Foldy-Wouthuysen-Tani
transformation \cite{foldy} is used to block-diagonalize
the Dirac theory so as to decouple the heavy quark
and antiquark degrees of freedom. Such a decoupling of particle and
antiparticle is a familiar characteristic of nonrelativistic dynamics
and is quite useful in our study of heavy quarkonium. The net effect is
that the usual relativistic field theory of four-component Dirac spinor
fields is replaced by a nonrelativistic Schr\"odinger field theory, with
separate two-component Pauli spinor fields for the heavy quarks and for
the heavy antiquarks. This field theory is NRQCD \cite{caswell-lepage}.
The lagrangian for NRQCD is
\begin{equation}
{\cal L}_{\rm NRQCD} \;=\; {\cal L}_{\rm light} \;+\; {\cal L}_{\rm heavy}
\;+\;
\delta {\cal L} .
\label{LNRQCD}
\end{equation}
The gluons and the $n_f$ flavors of light quarks are described by the fully
relativistic lagrangian
\begin{equation}
{\cal L}_{\rm light} \;=\; - {1 \over 2} {\rm tr} \, G_{\mu \nu} G^{\mu \nu}
	\;+\; \sum \bar q \; i {\not \! \! D} q ,
\label{Llight}
\end{equation}
where $G_{\mu \nu}$ is the gluon field-strength tensor expressed
in the form of an SU(3) matrix,
and $q$ is the Dirac spinor field for a light quark.
The gauge-covariant derivative is $D^\mu = \partial^\mu + i g A^\mu$,
where $A^\mu = (\phi,{\bf A})$ is the SU(3) matrix-valued gauge field
and $g$ is the QCD coupling constant.
The sum in (\ref{Llight}) is over the $n_f$ flavors of light quarks.
The heavy quarks and antiquarks are described by the term
\begin{equation}
{\cal L}_{\rm heavy}
\;=\; \psi^\dagger \, \left( iD_t + \frac{{\bf D}^2}{2M} \right)\, \psi
\;+\; \chi^\dagger \, \left( iD_t - \frac{{\bf D}^2}{2M} \right)\, \chi ,
\label{Lheavy}
\end{equation}
where $\psi$ is the Pauli spinor field that annihilates a heavy quark,
$\chi$ is the Pauli spinor field that creates a heavy antiquark, and
$D_t$ and ${\bf D}$ are the time and space components of the
gauge-covariant derivative $D^\mu$. Color and spin indices
on the fields $\psi$ and $\chi$ have been suppressed.  The lagrangian
${\cal L}_{\rm light} + {\cal L}_{\rm heavy}$ describes ordinary QCD
coupled to a Schr\"odinger field theory for the heavy quarks and antiquarks.
The relativistic effects of full QCD are reproduced through
the correction term $\delta {\cal L}$ in the lagrangian (\ref{LNRQCD}).

The correction terms in the effective lagrangian for NRQCD
that are most important for heavy quarkonium are
bilinear in the quark field or the antiquark field:
\begin{eqnarray}
\delta{\cal L}_{\rm bilinear}
&=& \frac{c_1}{8M^3}
\left( \psi^\dagger ({\bf D}^2)^2 \psi \;-\; \chi^\dagger ({\bf D}^2)^2 \chi
\right)  \nonumber \\
&+& \frac{c_2}{8M^2}
\left( \psi^\dagger ({\bf D} \cdot g {\bf E} - g {\bf E} \cdot {\bf D}) \psi
	\;+\; \chi^\dagger ({\bf D} \cdot g {\bf E} - g {\bf E} \cdot {\bf D}) \chi
\right) \nonumber \\
&+& \frac{c_3}{8M^2}
\left( \psi^\dagger (i {\bf D} \times g {\bf E} - g {\bf E} \times i {\bf D})
\cdot \mbox{\boldmath $\sigma$} \psi
	\;+\; \chi^\dagger (i {\bf D} \times g {\bf E} - g {\bf E} \times i {\bf D})
	\cdot \mbox{\boldmath $\sigma$} \chi \right) \nonumber \\
&+& \frac{c_4}{2M}
\left( \psi^\dagger (g {\bf B} \cdot \mbox{\boldmath $\sigma$}) \psi
	\;-\; \chi^\dagger (g {\bf B} \cdot \mbox{\boldmath $\sigma$}) \chi \right) ,
\label{Lbilinear}
\end{eqnarray}
where $E^i = G^{0i}$ and $B^i = \mbox{$\frac{1}{2}$} \epsilon^{ijk} G^{jk}$
are the electric and magnetic components of the gluon field strength
tensor $G^{\mu \nu}$.
By charge conjugation symmetry, for every term in (\ref{Lbilinear})
involving $\psi$, there is a corresponding term involving the antiquark
field $\chi$, with the same coefficient $c_i$, up to a sign. The
operators in (\ref{Lbilinear}) must be regularized, and they therefore
depend on the ultraviolet cutoff or renormalization scale $\Lambda$ of
NRQCD. The coefficients $c_i(\Lambda)$ also depend on $\Lambda$ in such
a way as to cancel the $\Lambda$-dependence of the operators.
Renormalization theory tells us that NRQCD can be made to reproduce QCD
results as accurately as desired by adding correction terms to the
lagrangian like those in (\ref{Lbilinear}) and tuning the couplings to
appropriate values \cite{lepage-tasi}.

Mixed 2-fermion operators  involving $\chi^\dagger$ and $\psi$
(or $\psi^\dagger$ and $\chi$) correspond to the annihilation
(or the creation) of a $Q \overline{Q}$ pair.
Such terms are excluded from the lagrangian
as part of the definition of NRQCD.  If  such an operator annihilates
a $Q \overline{Q}$ pair, it would, by energy conservation, have to create
gluons (or light quarks) with energies of order $M$. The amplitude for
annihilation of a $Q \overline{Q}$ pair into such high energy gluons cannot be
described accurately in a nonrelativistic theory such as NRQCD.
Nevertheless, as is discussed in Section~\ref{sec:space-time}, the
effects of such annihilation processes on low energy amplitudes can be
reproduced by adding 4-fermion operators such as $\psi^\dagger \chi
\chi^\dagger
\psi$ to the effective lagrangian.

Operators containing higher-order
time derivatives, such as $\psi^\dagger D_t^2 \psi$, are also omitted
from the effective lagrangian as part of the definition of NRQCD.
These operators can be eliminated by field redefinitions that
vanish upon use of the equations of motion.  Because of these
field redefinitions, the off-shell Green's functions of NRQCD
need not agree with those of full QCD, but the two theories
are equivalent for on-shell physical quantities.

The coefficients $c_i$ in (\ref{Lbilinear})
must be tuned as functions of the coupling constant $\alpha_s$,
the heavy-quark mass parameter in full QCD,
and the ultraviolet cutoff $\Lambda$ of NRQCD, so that physical
observables are the same as in full QCD.
The coefficients are conveniently determined
by matching low-energy scattering amplitudes of
heavy quarks and antiquarks in NRQCD, calculated in perturbation
theory in $\alpha_s$ and to a given precision in $v$, with the
corresponding perturbative scattering amplitudes in full QCD.
It is necessary to use on-shell scattering amplitudes
for this purpose, because the equations of motion have been used to simplify
the effective lagrangian for NRQCD by eliminating terms with more than one
power of $D_t$.  The scattering amplitudes can be calculated
using perturbation theory in $\alpha_s$, since the radiative
corrections to the coefficients in the NRQCD lagrangian are dominated by
relativistic momenta. These coefficients therefore have perturbative
expansions in powers of $\alpha_s(M)$ \cite{caswell-lepage,lmnmh}. The
coefficients in (\ref{Lbilinear}) are  defined so that $c_i = 1 +
O(\alpha_s)$.

The explicit factors of $M$ in (\ref{Lbilinear}) were introduced in
order that the coefficients $c_i$ be dimensionless. These coefficients
therefore depend on the definition of the heavy-quark mass parameter
$M$.  Our definition of $M$ is specified by the lagrangian
(\ref{Lheavy}): $1/(2M)$ is the coefficient of the operator
$\psi^\dagger {\bf D}^2 \psi$.  If a different prescription is adopted
for $M$, then all the $c_i$'s must be changed accordingly. The simplest
way to  determine the mass parameter $M$ is to match the location of the
pole in the perturbative propagator for a heavy quark in NRQCD with that
in full QCD.  In both NRQCD and full QCD, the kinetic energy for a heavy
quark of momentum $p$ in perturbation theory has the form $E =
p^2/(2M_{\rm pole}) - p^4/(8M_{\rm pole}^3) + \ldots$, where $M_{\rm
pole}$ is the perturbative pole mass. In Appendix~\ref{app:renorm}, the
self-energy of the heavy quark is calculated in NRQCD to order
$\alpha_s$ and to leading order in $v$. If we use a regularization
scheme in which power divergences are subtracted, then the
energy-momentum relation gives $M = M_{\rm pole}(1 + O(\alpha_s^2))$.
The corresponding calculation using a lattice regularization has been
carried out by Morningstar \cite{cjm}. The perturbative pole mass can be
related to any other definition of the heavy-quark mass by a calculation
in full QCD. 

\subsection{Velocity-scaling Rules}
\label{sec:vscaling}

In principle, infinitely many terms are required in the NRQCD lagrangian
in order to reproduce full QCD, but
in practice only a finite number of these  is needed for precision to
any given order in the typical heavy-quark velocity~$v$.  We can assess the
relative importance of various terms by using velocity-scaling
rules that were derived in Ref. \cite{lmnmh} and are summarized in
Table~\ref{tab}.
This table lists the fields and operators from which terms in the NRQCD
action are built, together with the approximate magnitude of each for
matrix elements between heavy quarkonium states that are localized in space.
The scaling rules were derived in Ref. \cite{lmnmh} by analyzing the
equations of motion for the quantum field operators of NRQCD.
The typical heavy-quark velocity $v$ is determined dynamically
by a balance between the kinetic and potential terms in the equation
of motion for the heavy-quark field, and $v$ can be used as an expansion
parameter in order to analyze the importance of other terms.
The scaling rules are certainly correct within perturbation theory
in $\alpha_s$, but, since they are based on the self-consistency of the field
equations, they should also be valid in the presence of
nonperturbative effects.

There is an important {\it caveat} to the velocity-scaling rules that
involves ultraviolet-divergent loop corrections. Loop corrections to an
operator give rise to power ultraviolet divergences, as well as to
logarithmic divergences. The logarithmic divergences modify the scaling
rules by factors of $\log(\Lambda/Mv)$.  The power divergences can
contribute factors of $1/v^n$, and the scaling rules apply only after
such $1/v^n$ divergences have been subtracted. The subtracted expression
is the relevant one for the following reason. The power-divergent
contributions to a given operator ${\cal O}$ that yield factors of
$1/v^n$ have the form of
renormalizations of lower-dimension operators. When the coefficients of
NRQCD are tuned so as to reproduce full QCD,
the coefficients of the lower-dimension operators are adjusted so that
their contributions to physical quantities cancel the contributions of
the $1/v^n$ power-divergent loop corrections to the operator ${\cal O}$.
Consequently, the inclusion of a given operator in the NRQCD lagrangian
yields a net correction to any physical quantity that is in accordance
with the velocity-scaling rules, up to logarithmic corrections.

The estimates for the magnitudes of $g \phi$ and $g {\bf A}$ in
Table~\ref{tab} hold in Coulomb gauge.  Coulomb gauge is a natural gauge
for analyzing heavy quarkonium, because it avoids spurious retardation
effects that are present in covariant gauges, but cancel out in physical
quantities \cite{love}.  Coulomb gauge is also a physical gauge, that
is, a gauge with no negative norm states. Thus, it allows a sensible
Fock-state expansion for the meson. The dominant Fock state is of course
${| Q \overline{Q} \rangle}$, but the meson also contains the Fock state ${| Q
\overline{Q} g \rangle}$,
which includes a dynamical gluon, and higher Fock states as well.

The estimates in Table~\ref{tab} were derived
assuming that one can do
perturbation theory in the typical heavy-quark velocity.
This perturbation theory relies on the fact that soft
gluons have a weak coupling to heavy quarks, not because the
coupling constant $\alpha_s$ is small, but because the interaction is
proportional to the heavy-quark velocity $v$.
In the derivation of the magnitude of $g {\bf A}$ in Ref. \cite{lmnmh},
dynamical gluons were assumed to have typical momenta of order $Mv$,
which is the inverse size of the quarkonium.
The perturbative estimate for the magnitude of the operator
$g {\bf A}$ is $\alpha_s(k)v k$ for a dynamical gluon of momentum $k$.
For $k$ of order $Mv$,
we can set $\alpha_s \sim v$ and recover the estimate $Mv^3$ given
in Table~\ref{tab}.  For $k$ of order $Mv^2$,
we can set $\alpha_s \sim 1$,
and we again obtain the estimate $Mv^3$.  This estimate relies
on perturbation theory, which may be suspect because of the
strong coupling between gluons with momenta on the order of $Mv^2$.
However such gluons necessarily have wavelengths of order $1/(Mv^2)$ or
larger, which is much larger than the typical size $1/(Mv)$ of the
quarkonium.  For such long-wavelength gluons, the multipole expansion,
whose validity transcends that of perturbation theory in the coupling
constant, can be used to justify the estimate for $g {\bf A}$ in
Table~\ref{tab} \cite{gpl}.

The velocity-scaling rules in Table~\ref{tab} show that the terms
in~$\delta{\cal L}_{\rm bilinear}$ in (\ref{Lbilinear}) all give
contributions that are suppressed by $O(v^2)$ relative to those from the
leading  lagrangian~${\cal L}_{\rm heavy}$. Recalling that mixed
2-fermion operators, such as $\psi^\dagger ({\bf D}^2)^2 \chi$, and operators
involving higher time derivatives, such as $\psi^\dagger D_t^2 \psi$, are
omitted as part of the definition of NRQCD, we see that $\delta {\cal
L}_{\rm bilinear}$ contains all the 2-fermion NRQCD operators of
relative order $v^2$.  The lagrangian ${\cal L}_{\rm light} + {\cal
L}_{\rm heavy} + \delta{\cal L}_{\rm bilinear}$ can therefore be used to
calculate NRQCD matrix elements between heavy-quarkonium states with an
error of order~$v^4$. If an error of order $v^2$ is sufficiently
accurate, then the matrix elements can be calculated by using the
lagrangian ${\cal L}_{\rm light} + {\cal L}_{\rm heavy}$.

It is instructive to contrast the relative magnitudes of the NRQCD
operators in the case of a heavy quarkonium with the relative magnitudes
of the same operators in the case of a heavy-light meson. (In the meson
rest frame, the lagrangian for NRQCD is identical to that for Heavy
Quark Effective Theory, which is the standard formalism for treating
heavy-light mesons \cite{HQET}.) In a heavy-light meson, the typical
3-momentum of the heavy quark is of order $\Lambda_{QCD}$, and is
independent of the heavy-quark mass.  The binding energy is also of
order $\Lambda_{QCD}$,  and is much larger than the heavy-quark kinetic
energy, which is of order $\Lambda_{QCD}^2/M$. Thus, in a heavy-light
meson, the 3-momentum and the energy of the heavy-quark are both of
order $\Lambda_{QCD}$, in contrast with the situation in a heavy
quarkonium, in which the 3-momentum is of order $Mv$ and the energy is
of order $Mv^2$.   Consequently, in a heavy-light meson, the effects of
operators of dimension $d$ are of order $(\Lambda_{QCD}/M)^{d-4}$
relative to the effects of the dimension-4 operator $\psi^\dagger i D_t
\psi$. The leading term $\psi^\dagger i D_t \psi$ describes a static heavy
quark acting as a source of gluon fields. All effects of relative order
$\Lambda_{QCD}/M$ can be taken into account by adding the dimension-5
operators $\psi^\dagger {\bf D}^2 \psi$ and $\psi^\dagger g {\bf B} \cdot
\mbox{\boldmath $\sigma$} \psi$.

\subsection{Quarkonium in NRQCD}

Several qualitative features of heavy
quarkonium can be inferred directly from the NRQCD lagrangian by
exploiting the heavy-quark velocity~$v$ as  an expansion parameter.
Expansions in powers of~$v$ are possible in ordinary QCD, but they are
complicated by the need to make a nonrelativistic expansion
of each individual Lorentz-invariant
operator in order to separate the various powers of~$v$.
Relativistic effects have been unraveled to a large extent in NRQCD,
with the leading $v$-dependence of each operator being
specified by the velocity-scaling rules in Table~\ref{tab}.

The most distinctive phenomenological feature of heavy quarkonium is
that, for many purposes, it is accurately described by the quark
potential model, in which the heavy quark and antiquark are bound by an
instantaneous potential. This model is a tuned phenomenology, rather
than a theory, but it is far simpler than a full field-theoretic
description based on NRQCD or QCD. Its validity  rests upon two
essential ingredients of heavy-quarkonium physics. The first is that the
dominant effect of the exchange of gluons between the heavy quark and
antiquark is to produce an instantaneous interaction.  The reason for
this is that the most important gluons have momenta of order $Mv$ and
energies of order $Mv^2$.  Such gluons are off their energy shells by
amounts of order $Mv$, which are much greater than the typical kinetic
energy $Mv^2$ of the heavy quark.  Consequently, the interaction times
of the gluons are shorter by a factor of $1/v$ than the time scale
associated with the motion of the heavy quarks, and the gluons'
interactions are, therefore, instantaneous as far as the heavy quarks are 
concerned.

The second essential ingredient underlying the quark potential model is
that  the probability of finding dynamical gluons (those that are not
part of the potential) in the meson is small. This is important because
dynamical gluons with very low energy produce effects that are not
instantaneous and are not readily incorporated into the quark potential
model.  In particular, gluons with energies of order $Mv^2$ have
interaction times comparable to that of the heavy quarks, and their
exchange therefore leads to significant retardation effects. The
probability for the Fock state ${| Q \overline{Q} g \rangle}$ of the meson can
be
estimated by considering the energy shift of a quarkonium state
${| H \rangle}$ that is due to the presence of a Fock-state component
${| Q \overline{Q} g \rangle}$. In Coulomb gauge, the only terms that connect
the dominant
Fock state ${| Q \overline{Q} \rangle}$ to the Fock state ${| Q \overline{Q} g
\rangle}$ are terms that
involve the vector potential~${\bf A}$.  At leading order in $v$, the
contributions to the energy shift come from the term
$i g {\bf A} \cdot \psi^\dagger \mbox{\boldmath $\nabla$} \psi/M$ in ${\cal
L}_{\rm heavy}$:
\begin{equation}
\Delta E\;=\; - {1 \over M} \;
{\langle H |} \, \int d^3x \; i g {\bf A} \cdot \psi^\dagger \mbox{\boldmath
$\nabla$} \psi \,{| H \rangle} .
\label{deltaE}
\end{equation}
Using the velocity-scaling rules in Table~\ref{tab} and taking into
account the relevant integration volume $1/(Mv)^3$, we obtain the
estimate $\Delta E \sim M v^4$. This energy shift can  be written in a
different way --- as the product of the probability $P_{Q \overline{Q}
g}$ for the $Q \overline{Q} g$ state multiplied by the energy $E_{Q
\overline{Q} g}$ of that state.  For gluons with momenta $k$ of order
$Mv$, the energy of the $Q \overline{Q} g$ state is dominated by the
energy of the gluon, and we find that $P_{Q \overline{Q} g} \sim v^3$.
For dynamical gluons with very low energies of order $Mv^2$ or less, the
energy of the $Q \overline{Q} g$ state is of order $Mv^2$ and we obtain
the estimate $P_{Q \overline{Q} g} \sim v^2$. For heavy quarkonium, $Q
\overline{Q} g$ states are therefore suppressed relative to the dominant
$Q \overline{Q}$ state by a factor of order $v^2$ in the probability.
Hence, for most quantities, effects due to Fock states  like ${| Q
\overline{Q} g \rangle}$ that contain dynamical gluons are suppressed by
powers of $v$. This might be expected from the phenomenological
successes of the quark potential model.  However, there are quantities,
such as the decay rates of P-wave states into light hadrons \cite{bbl},
for which the effects of the Fock state ${| Q \overline{Q} g \rangle}$
are of leading order in $v$ and the quark potential model fails
completely. 

The above estimates for the probabilities of $| Q \overline{Q}
g \rangle$ Fock states apply if the spin state of the $Q \overline{Q}$
pair is the same as in the dominant $| Q \overline{Q} \rangle$ Fock
state. If the spin state is different, we must replace $g {\bf A} \cdot
\mbox{\boldmath $\nabla$}$ in (2.6) with $g {\bf B} \cdot \mbox{\boldmath
$\sigma$}$ to obtain a nonzero matrix element. Using the
velocity-scaling rules  of Table I, we again obtain an estimate $\Delta
E \sim M v^4$ for the energy shift, implying that the probability for a
$| Q \overline{Q} g \rangle$ state containing a gluon with momentum on
the order of $Mv$ is $P_{Q \overline{Q} g} \sim v^3$. However, in the
derivation of the velocity-scaling rules in Ref.~[14], it was assumed
that dynamical gluons have momenta of order $Mv$. If the gluon has a
much smaller momentum $k$, then the estimate $M^2 v^4$ for the operator
$g {\bf B}$ in Table I should be replaced with $k^2 v^2$. Using this to
estimate the energy shift from a $| Q \overline{Q} g \rangle$ Fock state
containing a gluon with momentum of order $Mv^2$, we obtain $\Delta E
\sim M v^6$ and $P_{Q \overline Q g} \sim v^4$. Thus, gluons with very
low momenta exhibit the suppression that is characteristic of the
multipole expansion. We conclude that a $| Q \overline{Q} g \rangle$
Fock state that can be reached from the dominant $| Q \overline{Q}
\rangle$ Fock state by a spin-flip transition is dominated by dynamical
gluons with momenta of order $Mv$ and that the probability of such a
Fock state is $P_{Q \overline{Q} g} \sim v^3$.

Another  important feature of quarkonium structure is its approximate
independence of the heavy-quark spin. This feature follows immediately
from the structure of the NRQCD lagrangian, which exhibits an approximate
heavy-quark spin symmetry.  The leading term~${\cal L}_{\rm heavy}$ is
completely independent of the heavy-quark spin.
With just this term, states that differ only in the spins of the heavy quark
and antiquark have identical properties; heavy-quark spin is conserved
and can be used to label the  energy eigenstates. Spin-dependence enters
first through the bilinear terms in (\ref{Lbilinear}) that contain Pauli
matrices, and they give corrections that are of relative order
$v^2$.\footnote{In perturbation theory, ladder-like Coulomb-gluon
exchanges between the quark and antiquark give a factor of order
$\alpha_s/v$ for each ladder rung.  The spin-flip contribution is down
by $v^2$ relative to this Coulomb-ladder contribution. For example, in a
two-loop calculation, the Coulomb ladder gives a factor of order
$(\alpha_s/v)^2$, while the ladder with one Coulomb exchange and one
spin-flip exchange gives a factor of order $v^2
(\alpha_s/v)^2=\alpha_s^2$.}
Thus, spin splittings for quarkonia should be
smaller than splittings between radial and orbital-angular-momentum
excitations, with the ratios of these splittings
scaling roughly as $v^2$.  This familiar feature of
the spectra of charmonium and bottomonium
reinforces our confidence in the power-counting rules  and in
the utility of a nonrelativistic framework for studying quarkonium.

The total angular momentum $J$, the parity $P$, and the charge
conjugation $C$ are exactly conserved quantum numbers in NRQCD, as well
as in full QCD. Thus, the energy eigenstates ${| H \rangle}$ of heavy
quarkonium can be labelled by the quantum numbers $J^{PC}$.  By the
arguments given above, the dominant component in the Fock state
expansion of ${| H \rangle}$ is a pure quark-antiquark state ${| Q
\overline{Q} \rangle}$.  The Fock state ${| Q \overline{Q} g \rangle}$,
in which a dynamical gluon is present, has a probability of order $v^2$,
and higher Fock states have probabilities of order $v^4$ or higher.
Since our primary interest is in processes in which the $Q$ and
$\overline{Q}$ in the quarkonium annihilate, we concentrate on the state
of the $Q \overline{Q}$ pair in the various Fock-state components.  For
a general Fock state, the $Q \overline{Q}$ pair can be in either a
color-singlet state or a color-octet state.  Its angular-momentum state
can be denoted by the spectroscopic notation ${}^{2S+1}L_J$, where
$S=0,1$ is the total spin of the quark and antiquark, $L = 0,1,2,\ldots$
(or $L = S,P,D,\ldots$) is the orbital angular momentum, and $J$ is the
total angular momentum.  A $Q \overline{Q}$ pair in a ${}^{2S+1}L_J$
state has parity $P = (-1)^{L+1}$; if it is in a color-singlet state, it
has charge-conjugation number $C = (-1)^{L+S}$. 

In the Fock state ${| Q \overline{Q} \rangle}$, the $Q \overline{Q}$
pair must be in a 
color-singlet state and in an angular-momentum state ${}^{2S+1}L_J$
that is consistent with the quantum numbers $J^{PC}$ of the meson.
Conservation of $J^{PC}$ implies that mixing is allowed only
between the angular-momentum states ${}^3(J-1)_J$ and ${}^3(J+1)_J$.
For example, a ${}^3S_1$ $Q \overline{Q}$ state can mix with a ${}^3D_{1}$
state.
However, such mixing is suppressed because operators that change the
orbital angular momentum must contain at least one power of 
$\mbox{\boldmath $\nabla$}$.  In
general, up to corrections of order $v^2$, we can regard the $Q
\overline{Q}$ component of the meson as being in a definite
angular-momentum state ${}^{2S+1}L_J$. Of course, if the contribution of 
the dominant angular-momentum state is suppressed in a given process, 
then the contribution of the subdominant states takes on increased 
importance. We will present examples of this phenomenon in the 
discussions of the decay and production of P-wave states.

We turn next to the Fock state ${| Q \overline{Q} g \rangle}$ of the
meson, which includes a dynamical gluon and has a component whose
probability is of order $v^2$. In spite of the fact that the dynamics of
the soft gluon is nontrivial, NRQCD tells us much about the quantum
numbers of the $Q \overline{Q}$ pair in the $Q \overline{Q} g$ component
whose probability is of order $v^2$. The pair must of course be in a
color-octet state. Heavy-quark spin symmetry implies that the total spin
quantum number $S$ for the $Q \overline{Q}$ pair is the same as in the
dominant Fock state ${| Q \overline{Q} \rangle}$. But NRQCD also tells
us that the orbital state of the $Q \overline{Q}$ pair is closely
related to that in the Fock state ${| Q \overline{Q} \rangle}$. The
reason for this is that the coupling of the soft gluon can be analyzed
using a multipole expansion, and the usual selection rules for multipole
expansions apply. The leading interaction that couples the dominant Fock
state ${| Q \overline{Q} \rangle}$ to the state ${| Q \overline{Q} g
\rangle}$ is the electric-dipole part of the operator $\psi^\dagger g
{\bf A} \cdot \mbox{\boldmath $\nabla$}\psi$ in ${\cal L}_{\rm heavy}$,
and this changes the orbital-angular-momentum quantum number $L$ of the
$Q \overline{Q}$ pair by $\pm 1$. Higher multipoles bring in additional
powers of $v$, as does second-order perturbation theory. Thus, if the $Q
\overline{Q}$ pair in the dominant Fock state ${| Q \overline{Q}
\rangle}$ has angular-momentum quantum numbers ${}^{2S+1}L_J$, then the
Fock state ${| Q \overline{Q} g \rangle}$ has a probability of order
$v^2$ only if the $Q \overline{Q}$ pair has total spin $S$ and orbital
angular momentum $L+1$ or $L-1$.  For example, if the dominant Fock
state consists of a $Q \overline{Q}$ pair in a ${}^3S_1$ state, then the
Fock state ${| Q \overline{Q} g \rangle}$ has a probability of order
$v^2$ only if the $Q \overline{Q}$ pair is in a color-octet state with
angular-momentum quantum numbers ${}^3P_{0}$, ${}^3P_{1}$, or
${}^3P_{2}$. If the dominant Fock state consists of a $Q \overline{Q}$
pair in a ${}^1P_1$ state, then the Fock state ${| Q \overline{Q} g
\rangle}$ has a probability of order $v^2$ only if the $Q \overline{Q}$
pair is in a color-octet ${}^1S_0$ or ${}^1D_2$ state. 

The above discussion applies to Fock states $| Q \overline{Q} g
\rangle$ in which the $Q \overline{Q}$ pair has the same total spin
quantum number $S$ as in the dominant $| Q \overline{Q} \rangle$ state.
The probabilities for Fock states $| Q \overline{Q} g \rangle$ that can
be reached from the dominant Fock state by a spin-flip transition also
scale in a definite way with $v$. The probability for such a Fock state
to contain a dynamical gluon with momentum of order $Mv$ is of order
$v^3$, just as in the case of a non-spin-flip transition.  However, in
the case of a spin-flip transition, this momentum region dominates
because, as we have seen, gluons with softer momenta, on the order of
$Mv^2$, are suppressed by the multipole expansion. Thus, if the $Q
\overline{Q}$ pair in the dominant Fock state has angular-momentum
quantum numbers ${}^{2S+1}L_J$, then the Fock state $|Q \overline{Q} g
\rangle$,  with the $Q \overline{Q}$ pair in a color-octet state with
the same value of $L$ but different total spin quantum number, has a
probability of order $v^3$. For example, if the dominant Fock state
consists of a $Q \overline{Q}$ pair in a ${}^3S_1$ state, then the Fock
state $| Q \overline{Q} g \rangle$ with the $Q \overline{Q}$ pair in a
color-octet ${}^1S_0$ state has a probability of order $v^3$. If the
dominant Fock state consists of a $Q \overline{Q}$ pair in a ${}^1P_1$
state, then the Fock state $| Q \overline{Q} g \rangle$ with the $Q
\overline{Q}$ pair in a color-octet ${}^3P_J$ state has probability of
order $v^3$.

\subsection{Space-Time Structure of Annihilation}
\label{sec:space-time}

As we will explain in this subsection, the annihilation of a heavy
$Q \overline{Q}$ pair into gluons (or light quarks) occurs at distances that
are
typically of order~$1/M$, that is, at momentum scales of order~$M$.
Because of the large momentum scales involved, the details of the
annihilation process cannot be described accurately within a
nonrelativistic effective theory such as NRQCD.  Nevertheless,
as we will argue in the next subsection, the effects
of annihilation can be incorporated into NRQCD through 4-fermion operators
in the term $\delta {\cal L}$ in the NRQCD lagrangian.
To show that the required operators are local,
it is sufficient to show that the interactions they account for occur
over short distances of order $1/M$.  Strictly local operators are then
obtained by expanding the short-distance interaction
in a Taylor series in the 3-momentum ${\bf p}$ of the heavy quark multiplied
by the characteristic size $1/M$.

Now we wish to argue that the annihilation process is indeed local,
{\it i.e.} that the annihilation does occur within a
distance of order $1/M$.  We note that any annihilation
must result in at least two hard gluons (or light quarks), each with
momentum of order~$M$.  This has two consequences. First, the heavy
quark and antiquark must come within a distance of order $1/M$ in order
to annihilate.  That is because the emission of a hard  gluon from, say,
the heavy quark puts it into a highly virtual state, which can propagate
only a short distance before the quark must annihilate with the antiquark.
Thus, the total annihilation amplitude can be expressed as the sum
of point-like annihilation amplitudes, where the sum extends over the
possible annihilation points inside the meson.  The annihilation rate
is the square  of the total annihilation amplitude, summed over all
possible final states.
The second consequence of the hard gluons is that there is no overlap
between one annihilation amplitude and the complex conjugate of another
if the two annihilation points are separated by a
distance greater than about $1/M$.  This might seem
surprising, since the gluons are, in effect, on their mass shells (that
is, they fragment into jets with invariant masses much less than $M$).
There is no highly virtual state to constrain the distance between the
annihilation points for two amplitudes that produce the same final-state
jets.  Nevertheless, the annihilation points must be in close proximity
to each other in order for there to be an overlap between the final
states. In order to see why this is so, we note that, in classical
mechanics, we could trace the two final-state jets back to the
annihilation vertex, and there would be no ambiguity whatsoever as to
its space-time position. In quantum mechanics, the uncertainty principle
tells us that we can know the position of the annihilation vertex only
to a precision of order $1/M$, since the jet momenta are of order $M$.
Hence, in quantum mechanics, $Q \overline{Q}$ annihilation is not a point-like
process, but it {\it is} a localized process, with a size of order
$1/M$.

In a field-theoretic calculation of the annihilation rate at leading
order in $\alpha_s$, the localization of the annihilation process would
manifest itself as follows. The annihilation rate involves the imaginary
part $\Gamma(P,p,p')$ of the scattering amplitude for a $Q \overline{Q}$ pair
with
total momentum $P$ and initial and final relative momenta $p$ and $p'$.
Consider the Fourier transform of $\Gamma(P,p,p')$ with respect to all
three momentum variables:
\begin{equation}
\int d^4P \, d^4p \, d^4p' \; e^{i P \cdot (X-X')} e^{i p \cdot (x_1-x_2)}
	e^{i p' \cdot (x'_1-x'_2)} \;\Gamma(P,p,p').
\end{equation}
Here, $x_1$ and $x_2$ correspond to quark and antiquark interaction points
in one annihilation amplitude, $x'_1$ and $x'_2$ correspond to
quark and antiquark interaction points in the complex conjugate of a
second annihilation amplitude, and
$X=(x_1+x_2)/2$ and $X'=(x'_1+x'_2)/2$ are
average annihilation points for the first and second amplitudes.
The fact that $\Gamma(P,p,p')$ is insensitive to changes in $p$ and $p'$
that are much less than $M$ implies that, in the Fourier transform, $x_1$
($x'_1$) is localized to within a distance of order $1/M$ of $x_2$
($x'_2$). Similarly, the fact that $\Gamma$ is insensitive to changes in
$P$ that are much less than $M$ implies that the first and second
amplitudes have significant overlap only if $X$ and $X'$ are separated
by a distance of order $1/M$ or less. Note that, if one puts a
restriction in the annihilation rate on one of the components of the jet
momentum, then $\Gamma$ becomes sensitive to that component of $P$, and
the annihilation vertices are no longer localized along that direction.
This is a consequence of the uncertainty principle, which says that
knowledge of a component of the jet momentum along a given direction
reduces our potential knowledge of the position of the annihilation
vertex along that direction.

The radiation of soft or collinear gluons
might seem to violate this simple localization picture
that appears at leading order in the coupling constant.
Gluon radiation from the initial $Q \overline{Q}$ pair
is not a problem, since infrared divergences can be factored into the
long-distance matrix elements of the 4-fermion operators that mediate
the annihilation process in NRQCD,
and collinear divergences are controlled by the heavy-quark mass.
We must, however, worry about infrared or collinear divergences from the
radiation of gluons from the final-state hard gluons. In the presence of
such soft or collinear radiation, the hard gluon can propagate almost on
its mass shell from the annihilation point to the emission vertex. The
energetic final-state gluon jet points back to the emission vertex,
rather than to the annihilation point, which may be far away. In
perturbation theory, infrared and collinear divergences occur in
individual Feynman diagrams and produce a sensitivity to the heavy-quark
momenta in $\Gamma$.   However, the Kinoshita-Lee-Nauenberg (KLN)
theorem \cite{kln} guarantees that, when one sums over the contributions
of all nearly degenerate final states, as is done in forming the
inclusive annihilation rate, the infrared divergences cancel between
diagrams involving real and virtual gluon emission. We can think of this
KLN cancellation as a consequence of a generalized
form of the uncertainty principle: we can localize the annihilation
point, provided that we do not require too much knowledge about the
final state---that is, provided that we do not distinguish between the
various states that contribute to the inclusive cross section.

The locality of the annihilation process is spoiled if the final-state
gluons form a narrow resonance, such as a glueball.
This is because the jets produced by
the decay of the resonance point back to the place where the resonance
decayed.  If the resonance is narrow, this may be far from the point
where the heavy quark and antiquark annihilated.  That is why
perturbation theory cannot be applied directly to the cross section
for $e^+e^-$ annihilation into hadrons in the region of the
charmonium or bottomonium resonances.  In a field-theoretic
calculation, the resonance partially spoils the KLN cancellation of
infrared and collinear divergences.  While contributions from gluons that
have exactly zero momentum or are exactly collinear
still cancel, the real and virtual contributions
no longer cancel for soft gluons whose energy is comparable to the
resonance width or collinear gluons whose transverse momentum
is comparable to the resonance width.
In the case of $e^+ e^-$ annihilation, one can deal with this problem
by forming a suitable average of the cross section over the
resonance region \cite{pqr}.  In perturbation theory,
the effect of this smearing is to allow virtual soft or collinear
emission at one value of the $e^+ e^-$ center-of-mass energy
$\sqrt{s}$ to cancel real soft or collinear emission at a slightly
higher value of $\sqrt{s}$, but the same value of the energy of the
resonating $Q \overline{Q}$ pair.
This solution of smearing in the energy
is not available to us in the case of quarkonium annihilation.
Fortunately, there are no known narrow
glueball resonances in the charmonium or bottomonium region,
so we do not expect the resonance issue to be a problem in practice.

\subsection{Annihilation into Light Hadrons}

Since the annihilation of a $Q \overline{Q}$ pair necessarily produces gluons
or
light quarks with energies of order $M$, the annihilation amplitude
cannot be described accurately within NRQCD.  Nevertheless, the
annihilation rate, which is the square of the amplitude summed over
final states, can be accounted for in NRQCD. Since the annihilation rate
of the $Q \overline{Q}$ pair is localized within a distance of order $1/M$, the
annihilation contribution to a low-energy $Q \overline{Q} \to Q \overline{Q}$
scattering
amplitude can be reproduced in NRQCD by local 4-fermion operators in
$\delta {\cal L}$ involving $\psi$, $\chi^\dagger$, $\chi$, and $\psi^\dagger$.
The optical theorem relates $Q \overline{Q}$ annihilation rates to the
imaginary
parts of $Q \overline{Q} \to Q \overline{Q}$ scattering amplitudes.  This
relation implies
that the coefficients of the 4-fermion operators
in $\delta {\cal L}$ must have imaginary parts.  These imaginary parts
are the manifestation of annihilation in NRQCD.

The 4-fermion interactions that represent the effects of $Q \overline{Q}$
annihilation
in NRQCD have the general form
\begin{equation}
\delta{\cal L}_{\rm 4-fermion}
\;=\; \sum_n {f_n(\Lambda) \over M^{d_n-4}} {\cal O}_n(\Lambda) ,
\label{Lcontact}
\end{equation}
where the ${\cal O}_n$ are local 4-fermion operators, such as $\psi^\dagger
\chi \chi^\dagger \psi$. The naive scaling dimensions $d_n$ of the operators
can be obtained by counting the powers of $M$ using Table~\ref{tab}.
The factors of $M^{d_n-4}$ in (\ref{Lcontact}) have been introduced so
as to make the coefficients $f_n$ dimensionless. The operators ${\cal
O}_n$ must be regularized, and they therefore depend on the ultraviolet
cutoff or renormalization scale $\Lambda$ of the effective theory.  The
natural scale for this cutoff is
$M$, since $1/M$ is the distance scale of the
annihilation process.  However, all results are independent of $\Lambda$,
since the coefficients $f_n(\Lambda)$
depend on $\Lambda$ in such a way as to cancel the
$\Lambda$-dependence of the operators.  The
coefficients can be computed in full QCD
as perturbation series in $\alpha_s(M)$, in which individual terms
may depend on $\log(M/\Lambda)$.

If the analysis of annihilation rates were carried out completely within
full QCD, then the scale $\Lambda$ would arise as an arbitrary
factorization scale that must be introduced in order to separate the
momentum scale $M$ from smaller momentum scales of order $Mv$ or less.
The factorization scale $\Lambda$ should not be confused with the
renormalization scale $\mu$ of the full theory. The coefficients
$f_n(\Lambda)$ are independent of $\mu$ if they are computed to all
orders in $\alpha_s(\mu)$, although some $\mu$-dependence is introduced
as usual by the truncation of the perturbation series.  Unless we
explicitly specify otherwise, we always make the choice $\mu = M$ in
this paper.

The dimension-6 4-fermion terms in $\delta {\cal L}$ are
\begin{eqnarray}
\left(\delta{\cal L}_{\rm 4-fermion}\right)_{d=6}
&=& {f_1({}^1S_0) \over M^2} \, {\cal O}_1({}^1S_0)
	\;+\; {f_1({}^3S_1) \over M^2} \,{\cal O}_1({}^3S_1) \nonumber \\
&& \;+\; {f_8({}^1S_0) \over M^2} \, {\cal O}_8({}^1S_0)
	\;+\; {f_8({}^3S_1) \over M^2} \, {\cal O}_8({}^3S_1) ,
\label{Lcontact6}
\end{eqnarray}
where the dimension-6 operators are
\begin{mathletters}
\label{Odim6}
\begin{eqnarray}
{\cal O}_1({}^1S_0) &=& \psi^\dagger \chi \, \chi^\dagger \psi ,
\label{O1singS} \\
{\cal O}_1({}^3S_1) &=& \psi^\dagger \mbox{\boldmath $\sigma$} \chi \cdot
\chi^\dagger \mbox{\boldmath $\sigma$} \psi ,
\label{O1tripS} \\
{\cal O}_8({}^1S_0) &=& \psi^\dagger T^a \chi \, \chi^\dagger T^a \psi ,
\label{O8singS} \\
{\cal O}_8({}^3S_1) &=& \psi^\dagger \mbox{\boldmath $\sigma$} T^a \chi \cdot
\chi^\dagger \mbox{\boldmath $\sigma$} T^a \psi .
\label{O8tripS}
\end{eqnarray}
\end{mathletters}
The subscript 1 or 8 on the operators and on their coefficients
indicates the color structure of the operator. The arguments
${}^{2S+1}L_J$ indicate the angular-momentum state of the $Q \overline{Q}$ pair
which is annihilated or created by the operator. Normal-ordering of the
4-fermion operators ${\cal O}_n$ will always be understood, so that
matrix elements of ${\cal O}_n$ receive contributions only from
annihilation of the $Q$ and $\overline{Q}$. The dimension-8 terms in the
lagrangian for NRQCD include
\begin{eqnarray}
\left(\delta{\cal L}_{\rm 4-fermion}\right)_{d=8} &=&
{f_1({}^1P_1) \over M^4} \; {\cal O}_1({}^1P_1)
\;+\; {f_1({}^3P_{0}) \over M^4} \; {\cal O}_1({}^3P_{0})
\;+\; {f_1({}^3P_{1}) \over M^4} \; {\cal O}_1({}^3P_{1}) \nonumber \\
&& \;+\; {f_1({}^3P_{2}) \over M^4} \; {\cal O}_1({}^3P_{2})
\;+\; {g_1({}^1S_0) \over M^4} \; {\cal P}_1({}^1S_0)
\;+\; {g_1({}^3S_1) \over M^4} \; {\cal P}_1({}^3S_1) \nonumber \\
&& \;+\; {g_1({}^3S_1,{}^3D_{1}) \over M^4} \, {\cal P}_1({}^3S_1,{}^3D_{1})
\;+\;  \ldots \; .
\label{Lcontact8}
\end{eqnarray}
The dimension-8 operators included explicitly in (\ref{Lcontact8}) are
\begin{mathletters}
\label{Odim8}
\begin{eqnarray}
{\cal O}_1({}^1P_1) &=& \psi^\dagger (-\mbox{$\frac{i}{2}$} \tensor{\bf D})
\chi
	\cdot \chi^\dagger (-\mbox{$\frac{i}{2}$} \tensor{\bf D}) \psi ,
\label{OsingP}
\\
{\cal O}_1({}^3P_{0}) &=&  {1 \over 3} \;
\psi^\dagger (-\mbox{$\frac{i}{2}$} \tensor{\bf D} \cdot \mbox{\boldmath
$\sigma$}) \chi
	\, \chi^\dagger (-\mbox{$\frac{i}{2}$} \tensor{\bf D} \cdot \mbox{\boldmath
$\sigma$}) \psi ,
\label{OtripP0}
\\
{\cal O}_1({}^3P_{1}) &=&  {1 \over 2} \;
\psi^\dagger (-\mbox{$\frac{i}{2}$} \tensor{\bf D} \times \mbox{\boldmath
$\sigma$}) \chi
	\cdot \chi^\dagger (-\mbox{$\frac{i}{2}$} \tensor{\bf D} \times
\mbox{\boldmath $\sigma$}) \psi ,
\label{OtripP1}
\\
{\cal O}_1({}^3P_{2}) &=& \psi^\dagger (-\mbox{$\frac{i}{2}$} \tensor{D}{}^{(i}
\sigma^{j)}) \chi
	\, \chi^\dagger (-\mbox{$\frac{i}{2}$} \tensor{D}{}^{(i} \sigma^{j)}) \psi ,
\label{OtripP2}
\\
{\cal P}_1({}^1S_0) &=& {1\over 2}
\left[\psi^\dagger \chi \, \chi^\dagger (-\mbox{$\frac{i}{2}$} \tensor{\bf
D})^2 \psi \;+\; {\rm h.c.}\right] ,
\label{PsingS}
\\
{\cal P}_1({}^3S_1) &=& {1\over 2}\left[\psi^\dagger \mbox{\boldmath $\sigma$}
\chi
	\cdot \chi^\dagger \mbox{\boldmath $\sigma$} (-\mbox{$\frac{i}{2}$}
\tensor{\bf D})^2 \psi \;+\; {\rm h.c.}\right] ,
\label{PtripS}
\\
{\cal P}_1({}^3S_1,{}^3D_{1}) &=& {1\over 2}\left[\psi^\dagger \sigma^i \chi \,
	\chi^\dagger \sigma^j (-\mbox{$\frac{i}{2}$})^2 \tensor{D}{}^{(i}
\tensor{D}{}^{j)} \psi
\;+\;{\rm h.c.}\right],
\label{PtripSD}
\end{eqnarray}
\end{mathletters}
where $\tensor{\bf D}$ is the difference between the covariant derivative
acting
on the spinor to the right and on the spinor to the left: $\chi^\dagger
\tensor{\bf D}
\psi \equiv \chi^\dagger ({\bf D} \psi) - ({\bf D} \chi)^\dagger \psi$. We have
used
the notation $T^{(ij)}$ for the symmetric traceless
component of a tensor:
$T^{(ij)} = (T^{ij} + T^{ji})/2 - T^{kk} \delta^{ij}/3$.
For each of the operators
shown explicitly in (\ref{Lcontact8}), there is a
corresponding color-octet operator ${\cal O}_8$ or ${\cal P}_8$,
which contains color matrices $T^a$ inserted
between $\psi^\dagger$ and $\chi$ and between $\chi^\dagger$ and $\psi$.
This exhausts the list of the dimension-8 operators that contribute at
tree level to $Q \overline{Q}$ scattering in the center of momentum frame.
There are other dimension-8 operators, such as
$\mbox{\boldmath $\nabla$} (\psi^\dagger \chi) \cdot (\chi^\dagger \tensor{\bf
D} \psi)$ or
${\bf D}(\psi^\dagger T^a \chi) \cdot {\bf D}(\chi^\dagger T^a \psi)$,
in which a derivative acts on the product of $\psi^\dagger$
and $\chi$ or on the product of $\chi^\dagger$ and $\psi$.
Matrix elements of operators such as these are proportional to the
total momentum of the $Q \overline{Q}$ pair, and therefore do not receive any
contributions from the dominant Fock state ${| Q \overline{Q} \rangle}$ in the
meson
rest frame.  They do, however, receive contributions from
higher Fock states, such as ${| Q \overline{Q} g \rangle}$, in which the total
momentum of
the $Q \overline{Q}$ pair is nonzero.

According to the velocity-scaling rules in Table~\ref{tab}, the
dimension-6 terms in (\ref{Lcontact6}) scale as
$v$ relative to the leading
term ${\cal L}_{\rm heavy}$ in the NRQCD Lagrangian.
Thus, if we consider only the dependence on $v$, the terms in
(\ref{Lcontact6}) appear to be more important than the terms
in $\delta {\cal L}_{\rm bilinear}$, which scale as $v^2$ relative
to the terms in ${\cal L}_{\rm heavy}$.
However, the contributions from 4-fermion operators contain
extra suppression factors, owing to the operator coefficients,
whose imaginary parts are of order $\alpha_s^2(M)$ or smaller.
Thus, the contributions to annihilation widths from (\ref{Lcontact6})
are of order $\alpha_s^2(M) v$ or smaller relative to the scale $Mv^2$
of the splittings between radial excitations and between
orbital-angular-momentum excitations.
Similarly, the contributions to annihilation widths from the
dimension-8 operators in (\ref{Lcontact8}) are at most of order
$\alpha_s^2(M) v^3$ relative to the scale $Mv^2$
of splittings between energy levels.
Thus, the annihilation decay rates for heavy-quarkonium states are tiny
perturbations on the energy levels. This is certainly true empirically.
In the  charmonium system,
the ground state $\eta_c$ has the largest annihilation width,
but it is less than $3 \%$ of the splitting between the $\eta_c$
and the first radial or orbital-angular-momentum excitations.
For bottomonium, the annihilation widths are always less than $1 \%$
of the corresponding splittings.

In order to obtain an expression for the annihilation rate, we recall
that the decay rate is $-2$ times the imaginary part of the energy of
the state. The contribution to the imaginary part of the energy that
corresponds to annihilation into light hadrons comes from the
expectation value of $-\delta{\cal L}_{\rm 4-fermion}$, whose
coefficients have imaginary parts.  Thus, we see that the annihilation
rate of a heavy-quarkonium state $H$ into light hadrons is
\begin{equation}
\Gamma(H \to {\rm LH}) \;=\; 2 \; {\rm Im \,} {\langle H |}
	\delta{\cal L}_{\rm 4-fermion} {| H \rangle} ,
\label{Gamlh}
\end{equation}
where ${\rm LH}$ represents all possible light-hadronic final states.
The expectation value is taken in the rest frame of the quarkonium,
where its total momentum ${\bf P}$ vanishes.  The state ${| H \rangle} \equiv
{| H({\bf P}=0) \rangle}$ is an eigenstate of the NRQCD
hamiltonian.\footnote{Radial and orbital-angular-momentum excitations of
a quarkonium may decay through the hermitian part of the NRQCD lagrangian
to lower-lying quarkonium states plus light hadrons.  An example is the
decay of $\psi(2S)$ into $\psi\pi\pi$. In this example, the
spectrum of states in NRQCD contains a continuum of $\psi\pi\pi$
scattering states, each of which includes a small admixture of the bare
$\psi(2S)$ state, and a discrete state, which is mostly the bare
$\psi(2S)$ state, but which also contains a small admixture of bare
$\psi\pi\pi$ scattering states. The $\psi(2S)$ Breit-Wigner
resonance in, for example, the amplitude for $e^+e^-\rightarrow
\mu^+\mu^-$ results from the contributions of the complete spectrum of
states.  However, the resonance in the amplitude can be reproduced by a
single state, with complex energy, that is an eigenstate of the nonlocal
effective Hamiltonian that one would obtain by integrating out the
light-hadron states in NRQCD.  One should identify the state ${| H \rangle}$
in (\ref{master}) with such an eigenstate in applying (\ref{master}) to
an excited quarkonium state that decays through the hermitian part of the
NRQCD lagrangian into a lower-lying quarkonium state.}
It has the standard nonrelativistic normalization: $\langle
H({\bf P}') | H({\bf P}) \rangle = (2 \pi)^3 \delta^3({\bf P} - {\bf P}')$.
Inserting
the expansion (\ref{Lcontact}) into (\ref{Gamlh}), we obtain
\begin{equation}
\Gamma(H \to {\rm LH})
\;=\; \sum_n {2 \; {\rm Im \,} f_n(\Lambda) \over M^{d_n-4}} \;
	{\langle H |} {\cal O}_n(\Lambda) {| H \rangle} .
\label{master}
\end{equation}
The equation (\ref{master}) is our central result for the annihilation
decays into light hadrons. It expresses the decay rate as a sum of
terms, each of which factors into a short-distance coefficient~${\rm Im
\,} f_n$ and a long-distance matrix element~${\langle H |}{\cal O}_n{| H
\rangle}$.
The coefficients ${\rm Im \,} f_n$ in (\ref{master}) are proportional to
the rates for on-shell heavy quarks and antiquarks to annihilate from
appropriate initial configurations into hard gluons and light quarks,
and can be computed as perturbation series in $\alpha_s(M)$. The matrix
elements ${\langle H |} {\cal O}_n {| H \rangle}$ give the probability for
finding
the heavy quark and antiquark in a configuration within the meson that
is suitable for annihilation, and can be evaluated nonperturbatively
using, for example, lattice simulations. The dependence on the arbitrary
factorization scale $\Lambda$ in (\ref{master}) cancels between the
coefficients and the operators.

In some calculations of the matrix elements in (\ref{master}), such as
lattice simulations, it may be useful to approximate the states
${| H \rangle}$ by eigenstates of the hermitian part of the NRQCD Hamiltonian.
We note that corrections to this approximation first appear at
third-order in perturbation theory in ${\rm Im \,} \delta{\cal L}_{\rm
4-fermion}$, since second-order perturbation theory does not give an
imaginary contribution to the energy. The corrections are therefore of
order $(\Gamma/M v^2)^2 \Gamma$. This is of relative order
$\alpha_s^4(M) v^2$ or smaller, since the leading terms in $\Gamma$
scale like $Mv^3$ and are multiplied by short-distance coefficients of
order $\alpha_s^2(M)$ or smaller.  This level of accuracy is sufficient
for most practical purposes.

Applying the velocity-scaling rules of Table~\ref{tab} to the matrix
elements ${\langle H |} {\cal O}_n(\Lambda) {| H \rangle}$, one finds
that the expression (\ref{master}) for the annihilation decay rate can
be organized into an expansion in powers of $v$. Only a finite number of
operators contribute to any given order in $v$. The coefficients
$f_n(\Lambda)$ can be calculated as perturbation series in
$\alpha_s(M)$, so (\ref{master}) is really a double expansion in
$\alpha_s(M)$ and $v$. The simultaneous expansion in $\alpha_s(M)$ and
$v$ is useful to the extent that these two parameters are both small. Of
course, $\alpha_s(M)$ and $v$ are not independent for heavy quarkonium.
According to (\ref{valpha}), $v$ can be identified with $\alpha_s(Mv)$,
which is larger than $\alpha_s(M)$. This implies that it would be futile
to consider corrections to the coefficients ${\rm Im \,} f_n$ of
relative order $\alpha_s^n(M)$ unless matrix elements ${\langle H |}
{\cal O}_n {| H \rangle}$ of relative order $v^n$ have already been
included. 

The relation between $v$ and $\alpha_s(M)$ implied by (\ref{valpha}) follows
from the dynamics of heavy quarkonium.  The factorization formula
(\ref{master}) is actually an operator equation, and it can equally well
be applied to other problems in which the relation between $v$ and
$\alpha_s(M)$ is different.  An example in which $v$ and $\alpha_s(M)$ are
independent is the
annihilation  of a pair of heavy-light mesons, such as $D$
and $\bar D$ mesons, at small relative velocity $v \ll 1$.
As long as $v$ is much larger than $\Lambda_{QCD}/M$, which is the
typical relative velocity of a heavy quark in the heavy-light meson,
it can be identified with the velocity of the heavy quark
and the scaling rules of Table~\ref{tab} apply.

\subsection{Electromagnetic Annihilation}

In addition to annihilating into light hadrons, heavy-quarkonium states
can also annihilate into purely electromagnetic final states containing
only photons and lepton pairs.  The energies of the final-state photons
and leptons are of order $M$.  In NRQCD, the effects of electromagnetic
annihilation can be accounted for in the same way as the effects of
annihilation into light hadrons: by adding 4-fermion terms $\delta{\cal
L}^{\rm EM}_{\rm 4-fermion}$ to the effective lagrangian.  The primary
difference is that in the case of electromagnetic annihilation, the
final state, as far as the strong interactions are concerned, is the QCD
vacuum state ${| 0 \rangle}$. The 4-fermion operators that reproduce the
effects of electromagnetic annihilation therefore differ from those in
(\ref{Lcontact6}) and (\ref{Lcontact8}) by the insertion of an operator
${| 0 \rangle}{\langle 0 |}$ that projects onto the QCD vacuum state.  The
dimension-6 terms that must be added to the lagrangian are
\begin{equation}
\left(\delta{\cal L}^{\rm EM}_{\rm 4-fermion}\right)_{d=6}
\;=\; {f_{{\rm EM}}({}^1S_0) \over M^2} \;
	\psi^\dagger \chi {| 0 \rangle}{\langle 0 |} \chi^\dagger \psi
\;+\; {f_{{\rm EM}}({}^3S_1) \over M^2} \;
	\psi^\dagger \mbox{\boldmath $\sigma$} \chi {| 0 \rangle} \cdot {\langle 0 |}
\chi^\dagger \mbox{\boldmath $\sigma$} \psi .
\label{LEM6}
\end{equation}
Note that color-octet operators, such as
$\psi^\dagger T^a \chi {| 0 \rangle}{\langle 0 |} \chi^\dagger T^a \psi$,
are omitted because they cannot
contribute to matrix elements between color-singlet heavy-quarkonium states.
The dimension-8 terms that must be added to the lagrangian include
\begin{eqnarray}
\left(\delta{\cal L}^{\rm EM}_{\rm 4-fermion}\right)_{d=8}
&=& {f_{{\rm EM}}({}^3P_{0}) \over M^4} \;
{1 \over 3} \; \psi^\dagger (-\mbox{$\frac{i}{2}$} \tensor{\bf D}) \cdot
\mbox{\boldmath $\sigma$} \chi {| 0 \rangle} \;
	{\langle 0 |} \chi^\dagger (-\mbox{$\frac{i}{2}$} \tensor{\bf D}) \cdot
\mbox{\boldmath $\sigma$} \psi \nonumber \\
&& \;+\; {f_{{\rm EM}}({}^3P_{2}) \over M^4} \;
\psi^\dagger (-\mbox{$\frac{i}{2}$} \tensor{D}{}^{(i} \sigma^{j)}) \chi {| 0
\rangle} \;
	{\langle 0 |} \chi^\dagger (-\mbox{$\frac{i}{2}$} \tensor{D}{}^{(i}
\sigma^{j)}) \psi \nonumber \\
&& \;+\; {g_{{\rm EM}}({}^1S_0) \over M^4} {1\over 2}
	\left[ \psi^\dagger \chi {| 0 \rangle}
	{\langle 0 |} \chi^\dagger (-\mbox{$\frac{i}{2}$} \tensor{\bf D})^2 \psi +
{\rm h.c.} \right] \nonumber \\
&& \;+\; {g_{{\rm EM}}({}^3S_1) \over M^4} {1\over 2}
	\left[ \psi^\dagger \mbox{\boldmath $\sigma$} \chi {| 0 \rangle} \cdot
	{\langle 0 |} \chi^\dagger \mbox{\boldmath $\sigma$} (-\mbox{$\frac{i}{2}$}
\tensor{\bf D})^2 \psi + {\rm h.c.} \right]
 \;+\;  \ldots \; .
\label{LEM8}
\end{eqnarray}
We have shown only four of the possible dimension-8 terms.  In particular,
there are terms corresponding to each of the operators shown explicitly in
(\ref{Lcontact8}).  The coefficients
of the operators in (\ref{LEM6}) and (\ref{LEM8})
can be computed as perturbation expansions in $\alpha_s(M)$.

The decay rate of a heavy quarkonium state $H$ into electromagnetic final
states (${\rm EM}$) can be expressed in a factored form that is analogous
to that given in (\ref{master}) for decays into light hadrons:
\begin{equation}
\Gamma(H \to {\rm EM})
\;=\; \sum_n {2 \; {\rm Im \,} f_{{\rm EM},n}(\Lambda) \over M^{d_n-4}} \,
	{\langle H |} \psi^\dagger {\cal K}_n' \chi(\Lambda) {| 0 \rangle} \;
	{\langle 0 |} \chi^\dagger {\cal K}_n \psi(\Lambda) {| H \rangle} ,
\label{masterEM}
\end{equation}
where ${\cal K}_n$ and ${\cal K}_n'$ are products of the unit color matrix,
a spin matrix (the unit matrix or $\sigma^i$), and a polynomial in the
covariant derivative ${\bf D}$ and other fields,
as in (\ref{LEM6}) and (\ref{LEM8}).
The possible electromagnetic final states ${\rm EM}$ include the multiphoton
states $\gamma \gamma$ and $3 \gamma$ and the lepton pairs
$\ell^+ \ell^-$, where $\ell = e, \mu, \tau$.

\subsection{Computation of the Coefficients of the 4-Fermion Operators}
\label{sec:coeffs}

The nonperturbative long-distance dynamics of QCD is described equally
well by full QCD and by NRQCD. The perturbation expansions for full QCD
and NRQCD also give equivalent descriptions of the long-distance
dynamics, although the description is incorrect.  For example,
perturbation theory allows quarks and antiquarks to appear as asymptotic
states. However, because the coefficients of the NRQCD operators are
insensitive to the long-distance dynamics, we can exploit the
equivalence of perturbative QCD and perturbative NRQCD at long distances
as a device to calculate the coefficients of the four-fermion operators.
We compute in perturbation theory in full QCD the annihilation part
$A(Q \overline{Q}\rightarrow Q \overline{Q})$ of the scattering amplitude for
an on-shell
quark and antiquark with small relative momenta. Then we use
perturbation theory in NRQCD to compute the matrix elements of 4-fermion
operators ${\cal O}_n$ between on-shell $Q \overline{Q}$ states.  The
short-distance coefficients are determined by the matching condition
\begin{equation}
A(Q \overline{Q}\rightarrow Q \overline{Q})\Bigg|_{\rm pert.~QCD}
\;=\; \sum_n {f_n(\Lambda) \over M^{d_n-4}} \;
  {\langle Q \overline{Q} |} {\cal O}_n(\Lambda) {| Q \overline{Q}
\rangle}\Bigg|_{\rm pert.~NRQCD}.
\label{matching}
\end{equation}
By expanding the left and right sides of (\ref{matching}) as Taylor
series in the relative momenta ${\bf p}$ and ${\bf p}'$ of the
initial and final $Q \overline{Q}$ pairs, we
can identify the coefficients of the individual operators. These
correspond to the infrared-finite parts of the parton-level amplitudes
for $Q \overline{Q}$ scattering.  Because of the equivalence of NRQCD and full
QCD
at long distances, all of the infrared divergences contained in
$A(Q \overline{Q}\rightarrow Q \overline{Q})$ on the left side of
(\ref{matching})
reside on the right side in the NRQCD
matrix elements ${\langle Q \overline{Q} |} {\cal O}_n(\Lambda) {| Q
\overline{Q} \rangle}$.

The application of the matching condition (\ref{matching}) is
illustrated in Appendix~\ref{app:coeffs}. The imaginary parts of the
coefficients $f_n$ that enter into the annihilation rates of S-wave
states through relative order $v^2$ and the annihilation rates of P-wave
states at leading order in $v$ are computed to order $\alpha_s^2$. In
order to illustrate the use of the matching condition (\ref{matching})
beyond leading order in $\alpha_s$, we also calculate the coefficient
${\rm Im \,} f_1({}^1S_0)$ at next-to-leading order in $\alpha_s$. 

\vfill \eject

\section{Matrix Elements for Heavy Quarkonium}
\label{sec:matrix-el}

The factorization formula (\ref{master}) expresses the decay rate of an
arbitrary heavy quarkonium state $H$ into light hadrons as a sum over
all 4-fermion operators ${\cal O}_n$.  If we truncate the expansion at a
given order in the heavy-quark velocity $v$, then only finitely many of
the operators contribute.  In this section, we show how the number of
independent matrix elements can be reduced further by exploiting
heavy-quark spin symmetry and by using the vacuum-saturation
approximation.  We identify the matrix elements that contribute to the
decays of S-wave states through relative order $v^2$
and the matrix elements
that contribute to the decays of P-wave states at leading order in $v$.
We also discuss the relation between these matrix elements and
Coulomb-gauge wavefunctions, as well as the dependence of the matrix
elements on the factorization scale. For the sake of clarity, we use the
lowest S-wave and P-wave states of charmonium for the purpose of
illustration. However, our results apply equally well to other sets of
S-wave and P-wave states, and they can be extended readily to higher
orbital-angular-momentum states as well.  The lowest-lying S-wave states
in the charmonium system are the $J^{PC} = 0^{-+}$ state $\eta_c$ and
the $1^{--}$ state $J/\psi$ (henceforth referred to simply as $\psi$).
The lowest-lying P-wave states are the $1^{+-}$ state $h_c$ and the
$J^{++}$ states $\chi_{cJ}$, $J=0,1,2$. 

\subsection{Powers of Velocity}
\label{sec:vcounting}

We wish to determine the relative importance of the matrix elements
${\langle H |} {\cal O}_n {| H \rangle}$ of 4-fermion operators ${\cal
O}_n$ for a heavy quarkonium state ${| H \rangle}$. The velocity-scaling
rules in Table~\ref{tab} suggest that ${\langle H |} {\cal O}_n {| H
\rangle}$ is of the same order in $v$ for all the dimension-6 operators
in (\ref{Lcontact6}), and that all the dimension-8 operators in
(\ref{Lcontact8}) are down by 
a power of $v^2$.
There can, however, be
additional suppression by powers of $v$, depending on the quantum
numbers of the state $H$. The velocity-scaling rules in Table~I give the
correct result only if the operator ${\cal O}_n$ annihilates and creates
a color-singlet $Q \overline{Q}$ pair with the same angular-momentum
${}^{2S+1}L_J$ as the $Q \overline{Q}$ pair in the dominant Fock state
${| Q \overline{Q} \rangle}$ of the state ${| H \rangle}$. (In the
notation for 4-fermion operators used in (\ref{Lcontact6}) and
(\ref{Lcontact8}), the subscript 1 or 8 and the argument ${}^{2S+1}L_J$
indicate the color and angular-momentum state of the $Q \overline{Q}$
pair that is annihilated and created by the operator.) The matrix
element ${\langle H |} {\cal O}_n {| H \rangle}$ is suppressed by only
one additional power of $v^2$, relative to the velocity-scaling rules in
Table~1, if ${\cal O}_n$ annihilates and creates $Q \overline{Q}$ pairs
in the same color-spin-orbital state as appears in one of the Fock
states ${| Q \overline{Q} g \rangle}$ whose probability is of order
$v^2$. In particular, if the dominant $Q \overline{Q}$ component is
${}^{2S+1}L_J$,  the $Q \overline{Q}$ pair in the component ${| Q
\overline{Q} g \rangle}$ must be in a color-octet state with spin
quantum number $S$ and orbital-angular-momentum quantum number $L \pm
1$. The matrix element is suppressed by $v^3$ relative to the
velocity-scaling rules in Table~I if ${\cal O}_n$ annihilates and
creates $Q \overline{Q}$ pairs in the same color-spin-orbital state as
appears in one of the Fock states $| Q \overline{Q} g \rangle$ that can
be obtained from the dominant Fock state by a spin-flip transition. In
such a Fock state, the $Q \overline{Q}$ pair must be in a color-octet
state with the same orbital-angular-momentum quantum number $L$ as in
the dominant $| Q \overline{Q} \rangle$ state, but with different total
spin quantum number. In all other cases, the matrix element is down by
$v^4$ or more relative to the velocity-scaling result from Table~I. 

If perturbation theory remained accurate down to the scale
$Mv$, then the spin-flip matrix elements would be suppressed by an
additional power of $v$. The reason for this is that the contribution to
a spin-flip matrix element that is suppressed by only $v^3$ relative to
the velocity-scaling rules is power ultraviolet divergent. Therefore,
one could carry out a renormalization of the matrix element in which
this contribution is subtracted. The corresponding contribution to the
decay rate would then reside in the short-distance coefficient of the
matrix element that is associated with the dominant Fock state. (Such a
subtraction is carried out automatically if dimensional regularization
is used to cut off the ultraviolet divergences in the matrix element.)
Once the subtraction has been made, the leading contribution to the
spin-flip matrix element comes from the scale $Mv^2$. It is subject to
the usual multipole suppression and scales as $v^4$ relative to the
velocity-scaling rules. In practice, one usually makes such subtractions
perturbatively. It is not clear, in the charmonium and bottomonium
systems, that perturbation theory is sufficiently accurate at the scale
$Mv$ to remove the $v^3$ contribution completely. Therefore, we assume
in the  error estimates below that the spin-flip matrix elements scale
as $v^3$ relative to the velocity-scaling rules.

We can now identify the operators that contribute to the annihilation of
the $\eta_c$ into light hadrons through relative order $v^2$. The
$J^{PC} = 0^{-+}$ state ${| \eta_c \rangle}$ consists predominantly of
the Fock state ${| Q \overline{Q} \rangle}$, 
with the $Q \overline{Q}$
pair in a color-singlet ${}^1S_0$ state, but it also contains, with
probability of order $v^2$, the Fock state ${| Q \overline{Q} g
\rangle}$, with the $Q \overline{Q}$ pair in a color-octet ${}^1P_1$
state. 
The $Q \overline{Q}$ pair in the dominant Fock state ${| Q
\overline{Q} \rangle}$ is annihilated and created by the leading
operator ${\cal O}_1({}^1S_0) = \psi^\dagger \chi \chi^\dagger \psi$,
and by the operator ${\cal P}_1({}^1S_0) = \psi^\dagger \chi
\chi^\dagger (-\mbox{$\frac{i}{2}$}\tensor{\bf D})^2 \psi + {\rm h.c.}$,
which is down by $v^2$.  All other operators are suppressed by $v^3$ or
more relative to ${\cal O}_1({}^1S_0)$. For example, the operator $
{\cal O}_8({}^1P_1) = \psi^\dagger (-\mbox{$\frac{i}{2}$} \tensor{\bf
D}) T^a \chi \cdot \chi^\dagger (-\mbox{$\frac{i}{2}$} \tensor{\bf D})
T^a \psi$ scales as $v^2$ relative to ${\cal O}_1({}^1S_0)$, but it
contributes through the Fock state ${| Q \overline{Q} g \rangle}$, which
gives an additional suppression by $v^2$. The operator ${\cal
O}_1({}^3S_1) = \psi^\dagger \mbox{\boldmath $\sigma$} \chi \cdot
\chi^\dagger \mbox{\boldmath $\sigma$} \psi$ scales as ${\cal
O}_1({}^1S_0)$, but it contributes through the Fock state ${| Q
\overline{Q} g g \rangle}$, and is therefore suppressed by an additional
probability factor of $v^4$. As a final example, the operator
$\mbox{\boldmath $\nabla$} (\psi^\dagger \chi) \cdot \mbox{\boldmath
$\nabla$} (\chi^\dagger \psi)$ scales as $v^2$ relative to ${\cal
O}_1({}^1S_0)$, but its matrix element is proportional to the total
momentum of the $Q \overline{Q}$ pair, which vanishes for the Fock state
${| Q \overline{Q} \rangle}$ in the meson rest frame. Thus, there are
only two operators that contribute to the decay rate of the $\eta_c$
into light hadrons through relative  order $v^2$: 
\begin{eqnarray}
\Gamma(\eta_c \to {\rm LH})
&=& {2 \; {\rm Im \,} f_1({}^1S_0) \over M^2} \;
	{\langle \eta_c |} {\cal O}_1({}^1S_0) {| \eta_c \rangle} \nonumber \\
&& \;+\; {2 \; {\rm Im \,} g_1({}^1S_0) \over M^4} \;
	{\langle \eta_c |} {\cal P}_1({}^1S_0) {| \eta_c \rangle}
\;+\; O(v^3 \Gamma).
\label{Geta} \end{eqnarray}

The analysis of the operators that contribute to the decays of the
$\psi$ is similar to that for the $\eta_c$.  The $1^{--}$ state ${| \psi
\rangle}$ consists predominantly of the Fock state ${| Q \overline{Q}
\rangle}$, 
with the $Q \overline{Q}$ pair 
in a color-singlet ${}^3S_1$
state, but it contains, with a probability of order $v^2$, the Fock state 
${| Q \overline{Q} g \rangle}$, with the $Q \overline{Q}$ pair in a
color-octet ${}^3P_{0}$, ${}^3P_{1}$, or ${}^3P_{2}$ state.
At leading
order in $v$, only the operator ${\cal O}_1({}^3S_1)$ contributes to the
decay rate of the $\psi$ into light hadrons. At relative  order $v^2$,
the only additional contribution comes from the operator ${\cal
P}_1({}^3S_1)$. As in the case of the $\eta_c$, all contributions from
Fock states containing dynamical gluons, such as ${| Q \overline{Q} g
\rangle}$, are of order $v^3 \Gamma$ or higher. 

We next determine the operators that contribute to the annihilation
decays of the P-wave states at leading order in $v$. In contrast to the
S-wave states, we find that Fock states containing dynamical gluons play
an important role.  The $1^{+-}$ state ${| h_c \rangle}$ consists
predominantly of the Fock state ${| Q \overline{Q} \rangle}$, with the
$Q \overline{Q}$ pair in a color-singlet ${}^1P_1$ state, but it has 
a probability of order $v^2$ for the Fock state ${| Q \overline{Q} g
\rangle}$, with the $Q \overline{Q}$ pair in a color-octet ${}^1S_0$ or
${}^1D_2$ state.  The Fock state ${| Q \overline{Q} \rangle}$ is created
and annihilated by the dimension-8 operator ${\cal O}_1({}^1P_1)  =
\psi^\dagger (-\mbox{$\frac{i}{2}$} \tensor{\bf D}) \chi \cdot
\chi^\dagger (-\mbox{$\frac{i}{2}$} \tensor{\bf D}) \psi$. The Fock
state ${| Q \overline{Q} g \rangle}$, with the $Q \overline{Q}$ pair in
a color-octet ${}^1S_0$ state, also contributes to the decay at the same
order in $v$ through the operator ${\cal O}_8({}^1S_0) = \psi^\dagger
T^a \chi \chi^\dagger T^a \psi$. The operator scales as $v^{-2}$
relative to ${\cal O}_1({}^1P_1)$, but there is also a suppression
factor of $v^2$ from the probability for the $Q \overline{Q} g$ state.
Thus, the Fock state ${| Q \overline{Q} g \rangle}$, which contains a
dynamical gluon, contributes to the decay rate into light hadrons at the
same order in $v$ as the dominant Fock state ${| Q \overline{Q}
\rangle}$. The resulting expression for the decay rate is 
\begin{eqnarray}
\Gamma(h_c \to {\rm LH})
&=& {2 \; {\rm Im \,} f_1({}^1P_1) \over M^4} \;
{\langle h_c |} {\cal O}_1({}^1P_1) {| h_c \rangle} \nonumber \\
&& \;+\; {2 \; {\rm Im \,} f_8({}^1S_0) \over M^2} \;
	{\langle h_c |} {\cal O}_8({}^1S_0) {| h_c \rangle}
\;+\; O(v^2 \Gamma) .
\label{Gh}
\end{eqnarray}
The analysis of the operators that contribute to the decays of the
$\chi_{c0}$, $\chi_{c1}$, and $\chi_{c2}$ is similar to that for $h_c$.
The $J^{++}$ state ${| \chi_{cJ} \rangle}$ consists predominantly of the
Fock state ${| Q \overline{Q} \rangle}$, with the $Q \overline{Q}$ pair
in a color-singlet ${}^3P_{J}$ state. It also 
contains, with a probability of order
$v^2$, the Fock state ${| Q \overline{Q} g \rangle}$, with the $Q
\overline{Q}$ pair in a color-octet ${}^3S_1$, ${}^3D_{1}$, ${}^3D_{2}$,
or ${}^3D_{3}$ state.
The Fock state ${| Q \overline{Q} \rangle}$
contributes to the annihilation at leading order in $v$ through the
dimension-8 operator $O_1({}^3P_{J})$. The Fock state ${| Q \overline{Q}
g \rangle}$, with the $Q \overline{Q}$ pair in a color-octet ${}^3S_1$
state, also contributes to the annihilation rate at the same order in
$v$, through the operator ${\cal O}_8({}^3S_1) = \psi^\dagger
\mbox{\boldmath $\sigma$} T^a \chi \cdot \chi^\dagger \mbox{\boldmath
$\sigma$} T^a \psi$. 

The analysis of the electromagnetic operators that contribute to the
decays of S-wave and P-wave states is identical to that of the operators
that contribute to the decays into light hadrons, except that there are
no color-octet operators.  We find, therefore, that there are two
operators that contribute to the decay of the $\eta_c$ into two photons
through relative order $v^2$: $\psi^\dagger \chi {| 0 \rangle} {\langle
0 |} \chi^\dagger \psi$ and $\psi^\dagger \chi {| 0 \rangle} {\langle 0
|} \chi^\dagger (-\mbox{$\frac{i}{2}$} \tensor{\bf D})^2 \psi + {\rm
h.c.}$. Thus, the decay rate for $\eta_c \to \gamma \gamma$ is 
\begin{eqnarray}
\Gamma(\eta_c \to \gamma \gamma)
&=& {2 \; {\rm Im \,} f_{\gamma \gamma}({}^1S_0) \over M^2} \;
	\Big| {\langle 0 |} \chi^\dagger \psi {| \eta_c \rangle} \Big|^2 
        \nonumber \\
&& \;+\; {2 \; {\rm Im \,} g_{\gamma \gamma}({}^1S_0) \over M^4} \;
	{\rm Re \,}\left( {\langle \eta_c |} \psi^\dagger \chi {| 0 \rangle}
		{\langle 0 |} \chi^\dagger (-\mbox{$\frac{i}{2}$} 
                \tensor{\bf D})^2 \psi {|
\eta_c \rangle} \right)
\;+\; O(v^4 \Gamma) .
\label{GetaEM}
\end{eqnarray}
Here, it is expressed in terms of the vacuum-to-$\eta_c$ matrix elements
${\langle 0 |} \chi^\dagger \psi {| \eta_c \rangle}$ and ${\langle 0 |}
\chi^\dagger (-\mbox{$\frac{i}{2}$} \tensor{\bf D})^2 \psi {| \eta_c
\rangle}$.  Similarly the decay rate for $\psi \to e^+ e^-$ can be
computed at relative order $v^2$ in terms of ${\langle 0 |} \chi^\dagger
\mbox{\boldmath $\sigma$} \psi {| \psi \rangle}$ and ${\langle 0 |}
\chi^\dagger \mbox{\boldmath $\sigma$} (-\mbox{$\frac{i}{2}$}
\tensor{\bf D})^2 \psi {| \psi \rangle}$. For the decay $\chi_{c0} \to
\gamma \gamma$, the only operator that contributes at leading order in
$v$ is $\psi^\dagger (-\mbox{$\frac{i}{2}$} \tensor{\bf D} \cdot
\mbox{\boldmath $\sigma$}) \chi {| 0 \rangle} 	{\langle 0 |}
\chi^\dagger (-\mbox{$\frac{i}{2}$} \tensor{\bf D} \cdot \mbox{\boldmath
$\sigma$}) \psi$, so the decay rate can be expressed in terms of the
single vacuum-to-$\chi_{c0}$ matrix element ${\langle 0 |} \chi^\dagger
(-\mbox{$\frac{i}{2}$} \tensor{\bf D} \cdot \mbox{\boldmath $\sigma$})
\psi {| \chi_{c0} \rangle}$. Similarly the decay rate for $\chi_{c2} \to
\gamma \gamma$ can be calculated to leading order in $v$ in terms of the
matrix element ${\langle 0 |} \chi^\dagger (-\mbox{$\frac{i}{2}$}
\tensor{D}{}^{(i} \sigma^{j)}) \psi {| \chi_{c2} \rangle}$ only. 

\subsection{Heavy-Quark Spin Symmetry}

Heavy-quark spin symmetry provides approximate relations between matrix
elements for the various spin states of a given radial and orbital
excitation of heavy quarkonium. The leading violations of heavy-quark
spin symmetry come from the spin-flip terms in (\ref{Lbilinear}), whose
effects are of relative order $v^2$. Consequently, the equalities that
follow from heavy-quark spin symmetry hold only through relative order
$v^2$. Nevertheless, these relations can significantly reduce the number
of independent matrix elements that contribute to the decays of the
various spin states. 

When applied to the S-wave states, heavy-quark spin symmetry relates the
$\eta_c$ and the 3 spin states of the $\psi$. For the matrix elements
that contribute to their decays into light hadrons through relative 
order $v^2$, the spin-symmetry relations are 
\begin{mathletters}
\begin{eqnarray}
{\langle \psi |} {\cal O}_1({}^3S_1) {| \psi \rangle}
\;=\; {\langle \eta_c |} {\cal O}_1({}^1S_0) {| \eta_c \rangle}
	\; \Bigg(1 + O(v^2) \Bigg) ,
\label{HQS1}
\\
{\langle \psi |} {\cal P}_1({}^3S_1) {| \psi \rangle}
\;=\; {\langle \eta_c |} {\cal P}_1({}^1S_0) {| \eta_c \rangle}
	\; \Bigg(1 + O(v^2) \Bigg) .
\label{HQS2}
\end{eqnarray}
\end{mathletters}
The 12 spin states of the P-wave states $h_c$, $\chi_{c0}$, $\chi_{c1}$,
and $\chi_{c2}$ form a spin-symmetry multiplet.  The spin-symmetry
relations between the matrix elements that contribute to the decays of
the P-wave states into light hadrons at leading order in $v$ are 
\begin{mathletters}
\begin{eqnarray}
{\langle \chi_{cJ} |} {\cal O}_1({}^3P_{J}) {| \chi_{cJ} \rangle}
\;=\; {\langle h_c |} {\cal O}_1({}^1P_1) {| h_c \rangle}
	\; \Bigg(1 + O(v^2) \Bigg) , \;\; J=0,1,2,
\label{HQP1}
\\
{\langle \chi_{cJ} |} {\cal O}_8({}^3S_1) {| \chi_{cJ} \rangle}
\;=\; {\langle h_c |} {\cal O}_8({}^1S_0) {| h_c \rangle}
	\; \Bigg(1 + O(v^2) \Bigg),\;\; J = 0,1,2.
\label{HQP2}
\end{eqnarray}
\end{mathletters}

Heavy-quark spin symmetry also relates the matrix elements that
contribute to the electromagnetic annihilation decay rates.
For the matrix elements that contribute to the decays of $\eta_c$
and $\psi$, the spin-symmetry relations are
\begin{mathletters}
\label{HQSEM}
\begin{eqnarray}
\mbox{\boldmath$\epsilon$}^* \cdot {\langle 0 |} \chi^\dagger \mbox{\boldmath
$\sigma$} \psi {| \psi(\mbox{\boldmath$\epsilon$}) \rangle}
&=& {\langle 0 |} \chi^\dagger \psi {| \eta_c \rangle} \; \Bigg( 1 \;+\; O(v^2)
\Bigg) ,
\label{HQSEM6} \\
\mbox{\boldmath$\epsilon$}^* \cdot
{\langle 0 |} \chi^\dagger \mbox{\boldmath $\sigma$} (-\mbox{$\frac{i}{2}$}
\tensor{\bf D})^2 \psi {| \psi(\mbox{\boldmath$\epsilon$}) \rangle}
&=& {\langle 0 |} \chi^\dagger (-\mbox{$\frac{i}{2}$} \tensor{\bf D})^2 \psi {|
\eta_c \rangle}
\; \Bigg( 1 \;+\; O(v^2) \Bigg) ,
\label{HQSEM8}
\end{eqnarray}
\end{mathletters}
where the polarization vector $\mbox{\boldmath$\epsilon$}$ satisfies
$\mbox{\boldmath$\epsilon$}^* \cdot \mbox{\boldmath$\epsilon$} = 1$.
For the matrix elements that contribute to the decays of the
$\chi_{c0}$ and $\chi_{c2}$ into two photons,
the spin-symmetry relations are
\begin{equation}
{\epsilon^{ij}}^*
{\langle 0 |} \chi^\dagger (\mbox{$\frac{1}{2}$} \tensor{D}{}^{(i} \sigma^{j)})
\psi {| \chi_{c2}(\epsilon) \rangle}
\;=\; {1 \over \sqrt{3}} \;
{\langle 0 |} \chi^\dagger (\mbox{$\frac{1}{2}$} \tensor{\bf D} \cdot
\mbox{\boldmath $\sigma$}) \psi {| \chi_{c0} \rangle}
\; \Bigg( 1 \;+\; O(v^2) \Bigg) ,
\label{HQPEM}
\end{equation}
where the symmetric polarization tensor $\epsilon^{ij}$
satisfies ${\rm tr}(\epsilon) = 0$ and
${\rm tr}(\epsilon^\dagger \epsilon) = 1$.

\subsection{Vacuum-Saturation Approximation}
\label{sec:vacuum-sat}

The 4-fermion operators in (\ref{Lcontact6}) and (\ref{Lcontact8})
that contribute to the decays of heavy quarkonium into light hadrons
are distinct from those in (\ref{LEM6}) and (\ref{LEM8}) that
contribute to electromagnetic annihilation.  The electromagnetic
matrix elements can be obtained from the corresponding light hadronic matrix
elements by making use of the ``vacuum-saturation
approximation'':  insert a complete set of light hadronic states
$\sum_X {| X \rangle} {\langle X |}$ between $\chi^\dagger$ and $\chi$ and
assume
that the sum is ``saturated'' by the lowest-energy state, the QCD
vacuum ${| 0 \rangle}$.
The vacuum-saturation approximation has been used
in many other contexts in particle physics, but it is usually just a
simplifying assumption without any rigorous basis.  In the case of
heavy quarkonium, we can show that the vacuum-saturation approximation
is actually a controlled approximation.

Consider the matrix element of a color-singlet operator of the form
${\cal O}_n = \psi^\dagger {\cal K}_n' \chi \chi^\dagger {\cal K}_n \psi$,
where ${\cal K}_n$ and ${\cal K}_n'$ are products of a unit color matrix,
a spin matrix (the unit matrix or $\sigma^i$), and a polynomial in ${\bf D}$
and other fields.  The vacuum-saturation approximation to the matrix
element ${\langle H |} {\cal O}_n {| H \rangle}$ is obtained by
inserting a complete set of states ${| X \rangle}$ between $\chi$ and
$\chi^\dagger$,
and assuming that the sum is well approximated by the term involving the
vacuum state ${| 0 \rangle}$:
\begin{eqnarray}
{\langle H |} {\cal O}_n {| H \rangle}
&=& \sum_X {\langle H |} \psi^\dagger {\cal K}_n' \chi {| X \rangle}
	{\langle X |} \chi^\dagger {\cal K}_n \psi {| H \rangle}
\nonumber \\
&\approx&{\langle H |} \psi^\dagger {\cal K}_n' \chi {| 0 \rangle}
	{\langle 0 |} \chi^\dagger {\cal K}_n \psi {| H \rangle}
\label{VSA}
\end{eqnarray}
If the last step in (\ref{VSA}) is to be a controlled approximation, we
must show that the contributions from all other states, such as
multigluon states, are suppressed by powers of $v$. One example for
which this can be done is the matrix element ${\langle \eta_c |}
\psi^\dagger \chi \chi^\dagger \psi {| \eta_c \rangle}$. In the
vacuum-saturation approximation, the last line of (\ref{VSA}) reduces to
$|{\langle 0 |} \chi^\dagger \psi {| \eta_c \rangle}|^2$. This
approximation would be exact if the $\eta_c$ were a pure ${}^1S_0$ $Q
\overline{Q}$ state.  The point-like operator $\chi^\dagger \psi$ would
then annihilate the $\eta_c$ completely, leaving the vacuum state.
However, the $\eta_c$ also has Fock-state components, such as ${| Q
\overline{Q} g \rangle}$ and ${| Q \overline{Q} g g \rangle}$, which
include dynamical gluons. Corrections to the vacuum-saturation
approximation can be attributed to contributions from intermediate
states ${| X \rangle}$ containing such dynamical gluons. For the matrix
element ${\langle \eta_c |} \psi^\dagger \chi \chi^\dagger \psi {|
\eta_c \rangle}$, a single-gluon intermediate state is forbidden by
color conservation, so the leading corrections to the vacuum-saturation
approximation come from two-gluon intermediate states ${| gg \rangle}$. 
The leading contribution to ${\langle \eta_c |} \psi^\dagger \chi{| gg
\rangle}{\langle gg |} \chi^\dagger \psi {| \eta_c \rangle}$ comes from
the ${| Q \overline{Q} g g \rangle}$ component of the $\eta_c$, which
has a probability of order $v^4$.  Thus, the vacuum-saturation
approximation for ${\langle \eta_c |} \psi^\dagger \chi \chi^\dagger
\psi {| \eta_c \rangle}$ holds up to corrections of relative order
$v^4$. 

The vacuum-saturation approximation holds up to corrections of relative
order $v^4$ for any matrix element in which the operator creates and
annihilates the dominant $Q \overline{Q}$ component of the quarkonium
state. For the matrix elements that contribute to the decay rates of the
S-wave states into light hadrons through relative order $v^2$, the
vacuum-saturation approximation gives 
\begin{mathletters}
\label{VSSwave}
\begin{eqnarray}
{\langle \eta_c |} {\cal O}_1({}^1S_0) {| \eta_c \rangle}
&=& \Big| {\langle 0 |} \chi^\dagger \psi{| \eta_c \rangle} \Big|^2
	\; \Bigg(1 + O(v^4) \Bigg) ,
\label{VSeta}
\\
{\langle \psi |} {\cal O}_1({}^3S_1) {| \psi \rangle}
&=& \Big| {\langle 0 |} \chi^\dagger \mbox{\boldmath $\sigma$} \psi {| \psi
\rangle} \Big|^2
	\; \Bigg(1 + O(v^4) \Bigg) ,
\label{VSpsi}
\\
{\langle \eta_c |} {\cal P}_1({}^1S_0) {| \eta_c \rangle}
&=& \; {\rm Re \,}\left( {\langle \eta_c |} \psi^\dagger \chi {| 0 \rangle}
	{\langle 0 |} \chi^\dagger (-\mbox{$\frac{i}{2}$} \tensor{\bf D})^2 \psi {|
\eta_c \rangle} \right)
	\; \Bigg(1 + O(v^4) \Bigg) ,
\label{VSeta2}
\\
{\langle \psi |}
{\cal P}_1({}^3S_1) {| \psi \rangle}
&=& \; {\rm Re \,}\left( {\langle \psi |} \psi^\dagger \mbox{\boldmath
$\sigma$} \chi {| 0 \rangle} \cdot
	{\langle 0 |} \chi^\dagger \mbox{\boldmath $\sigma$} (-\mbox{$\frac{i}{2}$}
\tensor{\bf D})^2 \psi {| \psi \rangle} \right)
	\; \Bigg(1 + O(v^4) \Bigg) .
\label{VSpsi2}
\end{eqnarray}
\end{mathletters}
In the case of P-wave states, the vacuum-saturation approximation
can be applied to the
matrix elements of the color-singlet 4-fermion operators of dimension 8:
\begin{mathletters}
\label{VSPwave}
\begin{eqnarray}
{\langle h_c |} {\cal O}_1({}^1P_1) {| h_c \rangle}
&=& \Big| {\langle 0 |} \chi^\dagger (-\mbox{$\frac{i}{2}$} \tensor{\bf D})
\psi {| h_c \rangle} \Big|^2 \;
	\Bigg( 1 + O(v^4) \Bigg) ,
\label{VSh}
\\
{\langle \chi_{c0} |} {\cal O}_1({}^3P_{0}) {| \chi_{c0} \rangle}
&=& {1 \over 3} \; \Big| {\langle 0 |} \chi^\dagger (-\mbox{$\frac{i}{2}$}
\tensor{\bf D} \cdot \mbox{\boldmath $\sigma$}) \psi
	{| \chi_{c0} \rangle} \Big|^2 \; \Bigg(1 + O(v^4) \Bigg) ,
\label{VSchi0}
\\
{\langle \chi_{c1} |} {\cal O}_1({}^3P_{1}) {| \chi_{c1} \rangle}
&=& {1 \over 2} \; \Big| {\langle 0 |} \chi^\dagger (-\mbox{$\frac{i}{2}$}
\tensor{\bf D} \times \mbox{\boldmath $\sigma$}) \psi
	{| \chi_{c1} \rangle} \Big|^2 \; \Bigg(1 + O(v^4) \Bigg) ,
\label{VSchi1}
\\
{\langle \chi_{c2} |} {\cal O}_1({}^3P_{2}) {| \chi_{c2} \rangle}
&=& \sum_{ij}
	\Big| {\langle 0 |} \chi^\dagger (-\mbox{$\frac{i}{2}$} \tensor{D}{}^{(i}
\sigma^{j)}) \psi
	{| \chi_{c2} \rangle} \Big|^2 \; \Bigg(1 + O(v^4) \Bigg) .
\label{VSchi2}
\end{eqnarray}
\end{mathletters}
The vacuum-saturation approximation cannot be applied to matrix elements
of color-octet operators, such as
${\langle h_c |} \psi^\dagger T^a \chi \chi^\dagger T^a \psi {| h_c \rangle}$,
because
the matrix element ${\langle X |} \chi^\dagger T^a \psi {| h_c \rangle}$
vanishes
if ${\langle X |}$ is the vacuum or any other color-singlet state.

\subsection{Relation to Wavefunctions}
\label{sec:wave-fn}

In most previous work on the annihilation decays of heavy quarkonium,
the nonperturbative factors in the decay rates were expressed in terms of
wavefunctions, or their derivatives, evaluated at the origin.
These ``wavefunctions'' were often identified with the Schr\"odinger
wavefunctions calculated in potential models for heavy quarkonium.
The wavefunction factors were never given rigorous field-theoretic
definitions, so the accuracy of the approximations that were involved
was always vague.  By expressing the decay rates in terms of matrix
elements of NRQCD, we have provided a rigorous field-theoretic definition
of the nonperturbative factors in the decay rates.
Since heavy quarks and antiquarks are described in NRQCD by a
Schr\"odinger field theory, the nonrelativistic wavefunctions can
also be given rigorous field-theoretic definitions, and their relations
to the nonperturbative factors in the decay rates can be clarified.

Nonrelativistic Coulomb-gauge wavefunctions can be defined naturally
as NRQCD Bethe-Salpeter $Q \overline{Q}$ wavefunctions, evaluated at equal
time. For
example, the radial wavefunction $R_{\eta_c}(r)$ for the $\eta_c$ can be
defined as
\begin{equation}
R_{\eta_c}(r) \; {1 \over \sqrt{4 \pi}}
\;\equiv\; {1 \over \sqrt{2 N_c}} \;
{\langle 0 |} \chi^\dagger(- {\bf r}/2) \; \psi(+ {\bf r}/2)
	{| \eta_c \rangle} \Bigg|_{\rm Coulomb} .
\label{psieta}
\end{equation}
The Pauli spinor fields $\psi({\bf r}/2)$ and $\chi^\dagger(- {\bf r}/2)$
are understood to be evaluated at the same time $t=0$.
The factor $1/\sqrt{4 \pi}$ on the left is the spherical harmonic
$Y_{00}({\hat {\bf r}})$, while
the factor of $\sqrt{2 N_c}$ on the left takes into account the
traces of the spin wavefunction $\delta^{m+m',0}/\sqrt{2}$ and the
color wavefunction $\delta^{ij}/\sqrt{N_c}$
of the ${| Q \overline{Q} \rangle}$ component of the $\eta_c$.  In the absence
of a
regulator, the wave function or its derivatives may be singular as ${\bf r}
\rightarrow 0$.  We can
define regularized ``radial wavefunctions at the origin''
${\overline {R_{\eta_c}}}(\Lambda)$ for $\eta_c$ and ${\overline
{R_\psi}}(\Lambda)$ for $\psi$ by
\begin{mathletters}
\label{RSwave}
\begin{eqnarray}
{\overline {R_{\eta_c}}}(\Lambda) &\equiv&
\sqrt{{2 \pi \over N_c}} \; {\langle 0 |} \chi^\dagger \psi(\Lambda) {| \eta_c
\rangle} ,
\label{Reta}
\\
{\overline {R_\psi}}(\Lambda)\; \mbox{\boldmath$\epsilon$} &\equiv& \sqrt{{2
\pi \over N_c}} \;
{\langle 0 |} \chi^\dagger \mbox{\boldmath $\sigma$} \psi(\Lambda) {|
\psi(\mbox{\boldmath$\epsilon$}) \rangle} \;,
\label{Rpsi}
\end{eqnarray}
\end{mathletters}
where $\mbox{\boldmath$\epsilon$}$ is the polarization vector of the
$\psi$. The local operators $\chi^\dagger \psi(\Lambda)$ and
$\chi^\dagger \mbox{\boldmath $\sigma$} \psi(\Lambda)$ can be defined by
dimensional regularization with scale $\Lambda$, together with minimal
subtraction. They can also be defined by a lattice regulator, or any
other convenient regularization scheme. As is suggested by the overline,
the intuitive interpretation of ${\overline {R_{\eta_c}}}(\Lambda)$ and
${\overline {R_\psi}}(\Lambda)$ is that they are the radial
wavefunctions averaged over a region of size $1/\Lambda$ centered at the
origin. Note that the regularized matrix elements in (\ref{RSwave}) are
precisely those that enter into the decay rates for for $\eta_c \to
\gamma \gamma$ and $\psi \to e^+ e^-$ at leading order in $v$. 

The matrix element ${\langle 0 |} \chi^\dagger (-\mbox{$\frac{i}{2}$}
\tensor{\bf D})^2 \psi {| \eta_c \rangle}$ that contributes to the decay
rate for $\eta_c \to \gamma \gamma$ at relative order  $v^2$ can also be
related to the Coulomb-gauge wavefunction defined in (\ref{psieta}). By
the velocity-scaling rules of Table~\ref{tab}, it differs from the
matrix element ${\langle 0 |} \chi^\dagger (-\mbox{$\frac{i}{2}$}
\tensor{\mbox{\boldmath $\nabla$}})^2 	\psi {| \eta_c \rangle}$ in
Coulomb gauge only at relative order $v^2$. With appropriate
regularization, the latter matrix element can be identified with the
limit as $r \to 0$ of $-\mbox{\boldmath $\nabla$}^2 R(r)$, where $R(r)$
is the radial wavefunction defined in (\ref{psieta}). The operator
$\chi^\dagger (\mbox{$\frac{1}{2}$} \tensor{\mbox{\boldmath
$\nabla$}})^2  \psi$ contains a linear ultraviolet divergence
proportional to $\chi^\dagger \psi(\Lambda)$, which we subtract, and a
logarithmic divergence that is cut off at the scale $\Lambda$.  This
subtraction and cutoff define a renormalized laplacian of the radial
wavefunction at the origin, which we denote by ${\overline {\nabla^2
R_{\eta_c}}}$: 
\begin{equation}
{\overline {\nabla^2 R_{\eta_c}}}(\Lambda)
\;\equiv\; {\sqrt{2 \pi \over N_c}} \;
{\langle 0 |} \chi^\dagger (\mbox{$\frac{1}{2}$} \tensor{\mbox{\boldmath
$\nabla$}})^2
        \psi(\Lambda) {| \eta_c \rangle} \Bigg|_{\rm Coulomb}.
\label{R2etaC}
\end{equation}
The analogous quantity ${\overline {\nabla^2 R_\psi}}$ for the $\psi$
can be defined in a similar way. The corresponding gauge-invariant
matrix elements differ from ${\overline {\nabla^2 R_{\eta_c}}}$ and ${\overline
{\nabla^2 R_\psi}}$
only at relative order $v^2$:
\begin{mathletters}
\label{R2Swave}
\begin{eqnarray}
{\langle 0 |} \chi^\dagger (-\mbox{$\frac{i}{2}$} \tensor{\bf D})^2
\psi(\Lambda) {| \eta_c \rangle}
&=& - \sqrt{{N_c \over 2 \pi}} \; {\overline {\nabla^2 R_{\eta_c}}}(\Lambda)
\; \Bigg(1 \;+\; O(v^2) \Bigg) .
\label{R2eta}
\\
\mbox{\boldmath$\epsilon$}^* \cdot
{\langle 0 |} \chi^\dagger \mbox{\boldmath $\sigma$} (-\mbox{$\frac{i}{2}$}
\tensor{\bf D})^2 \psi(\Lambda) {| \psi(\mbox{\boldmath$\epsilon$}) \rangle}
&=& - \; \sqrt{{N_c \over 2 \pi}} \; {\overline {\nabla^2 R_\psi}}(\Lambda) \;
	\Bigg( 1 \;+\; O(v^2) \Bigg).
\label{R2psi}
\end{eqnarray}
\end{mathletters}
The intuitive interpretations of ${\overline {\nabla^2 R_{\eta_c}}}(\Lambda)$
and ${\overline {\nabla^2 R_\psi}}(\Lambda)$ are
somewhat obscured by the subtractions needed to define the
renormalized matrix elements.

Heavy-quark spin symmetry implies that the wavefunctions of the $\eta_c$
and $\psi$ are identical up to corrections of relative order $v^2$: 
\begin{equation}
R_\psi(r) \;=\; R_{\eta_c}(r) \; \Bigg(1 \;+\; O(v^2) \Bigg) .
\label{Retapsi}
\end{equation}
It is convenient to introduce an average radial wavefunction $R_S(r)$
for the $1S$ states $\eta_c$ and $\psi$, which can be used when the
differences of relative order $v^2$ can be neglected: 
\begin{equation}
R_S(r) \;\equiv\; {R_{\eta_c}(r) + 3 R_\psi(r) \over 4} .
\label{RS}
\end{equation}
Because of the heavy-quark spin symmetry, the regularized quantities
${\overline {R_{\eta_c}}}(\Lambda)$ and ${\overline {R_\psi}}(\Lambda)$ differ
only at relative order
$v^2$, as do the renormalized quantities ${\overline {\nabla^2
R_{\eta_c}}}(\Lambda)$ and
${\overline {\nabla^2 R_\psi}}(\Lambda)$. Weighted averages ${\overline
{R_S}}(\Lambda)$ and
${\overline {\nabla^2 R_S}}(\Lambda)$ for the S-wave states can be defined as
in (\ref{RS}).
The S-wave radial wavefunction that is computed in nonrelativistic
potential models can be interpreted as a phenomenological estimate of
the wavefunction (\ref{RS}). Thus, the value $R_S(r=0)$ that is obtained
from potential models can be used as an estimate of the regularized
quantity ${\overline {R_S}}(\Lambda)$ at a scale $\Lambda$ of order $Mv$.  The
relation between ${\overline {\nabla^2 R_S}}(\Lambda)$ and $\mbox{\boldmath
$\nabla$}^2 R_S(r=0)$ in potential
models is more obscure, because of the subtraction that is required to
define the renormalized matrix element in (\ref{R2etaC}), and because
$\mbox{\boldmath $\nabla$}^2 R_S(r)$ diverges linearly as $r \to 0$ if the
potential is
Coulombic at short distances.

Nonrelativistic wavefunctions for the P-wave states can be defined
through matrix elements in Coulomb gauge that are analogous to
(\ref{psieta}).  For example, the radial wavefunction $R_{h_c}(r)$ for
the $h_c$ can be defined as
\begin{equation}
R_{h_c}(r) \; \left( \sqrt{3 \over 4 \pi} \; {\hat {\bf r}} \cdot
\mbox{\boldmath$\epsilon$} \right)
\;\equiv\; {1 \over \sqrt{2 N_c}} \;
{\langle 0 |} \chi^\dagger(- {\bf r}/2) \; \psi(+ {\bf r}/2)
	{| h_c(\mbox{\boldmath$\epsilon$}) \rangle} \Bigg|_{\rm Coulomb},
\label{psih}
\end{equation}
where the polarization vector satisfies $\mbox{\boldmath$\epsilon$} \cdot
\mbox{\boldmath$\epsilon$}^* = 1$.
A regularized derivative of the radial wavefunction at
the origin ${\overline {R_{h_c}'}}(\Lambda)$ can be defined by
\begin{equation}
{\overline {R_{h_c}'}}(\Lambda)\; \mbox{\boldmath$\epsilon$}
\;\equiv\; \sqrt{2 \pi \over 3 N_c} \;
{\langle 0 |} \chi^\dagger (\mbox{$\frac{1}{2}$} \tensor{\mbox{\boldmath
$\nabla$}})
	\psi(\Lambda) {| h_c(\mbox{\boldmath$\epsilon$}) \rangle} \Bigg|_{\rm
Coulomb}.
\label{RhC}
\end{equation}
Analogous quantities ${\overline {R_{\chi_{cJ}}'}}(\Lambda)$ can be defined for
the $\chi_{cJ}$ states.  Since no subtractions are required in order to
define the operator on the right side of (\ref{RhC}),
the quantity ${\overline {R_{h_c}'}}(\Lambda)$ has a straightforward
intuitive interpretation as the derivative of the radial wavefunction
averaged over a region of size $1/\Lambda$ centered at the origin.
In the gauge-invariant analog of the matrix element (\ref{RhC}),
the derivative $\tensor{\mbox{\boldmath $\nabla$}}$ is replaced by the
covariant derivative $\tensor{\bf D}$.
 From the velocity-scaling rules of Table~\ref{tab},
we see that these matrix elements differ only at relative order $v^2$:
\begin{mathletters}
\label{Rhchi}
\begin{eqnarray}
{\langle 0 |} \chi^\dagger (\mbox{$\frac{1}{2}$} \tensor{\bf D}) \psi(\Lambda)
{| h_c(\mbox{\boldmath$\epsilon$}) \rangle}
&=& \sqrt{3 N_c \over 2 \pi} \; {\overline {R_{h_c}'}}(\Lambda) \;
\mbox{\boldmath$\epsilon$}
\; \Bigg(1 \;+\; O(v^2) \Bigg),
\label{Rh}
\\
{1 \over \sqrt{3}} \; {\langle 0 |} \chi^\dagger (\mbox{$\frac{1}{2}$}
\tensor{\bf D} \cdot \mbox{\boldmath $\sigma$}) \psi(\Lambda)
	{| \chi_{c0} \rangle} &=&
\sqrt{3 N_c \over 2 \pi} \; {\overline {R_{\chi_{c0}}'}}(\Lambda)
\; \Bigg(1 \;+\; O(v^2) \Bigg),
\label{Rchi0}
\\
{1 \over \sqrt{2}} \;
{\langle 0 |} \chi^\dagger (-\mbox{$\frac{i}{2}$} \tensor{\bf D} \times
\mbox{\boldmath $\sigma$}) \psi(\Lambda)
	{| \chi_{c1}(\mbox{\boldmath$\epsilon$}) \rangle} &=&
\sqrt{3 N_c \over 2 \pi} \; {\overline {R_{\chi_{c1}}'}}(\Lambda) \;
\mbox{\boldmath$\epsilon$}
\; \Bigg(1 \;+\; O(v^2) \Bigg),
\label{Rchi1}
\\
{\langle 0 |} \chi^\dagger (\mbox{$\frac{1}{2}$} \tensor{D}{}^{(i} \sigma^{j)})
\psi(\Lambda)
	{| \chi_{c2}(\epsilon) \rangle} &=&
\sqrt{3 N_c \over 2 \pi} \; {\overline {R_{\chi_{c2}}'}}(\Lambda) \;
\epsilon^{ij}
\; \Bigg(1 \;+\; O(v^2) \Bigg) .
\label{Rchi2}
\end{eqnarray}
\end{mathletters}
By heavy-quark spin symmetry, ${\overline {R_{h_c}'}}(\Lambda)$ differs
from ${\overline {R_{\chi_{cJ}}'}}(\Lambda)$, $J=0,1,2$, only at
relative order $v^2$.  For applications in which $v^2$ corrections can
be neglected, these wavefunctions can all be replaced by the average
over the 16 P-wave spin states, which we denote by ${\overline
{R_P'}}(\Lambda)$. The value $R_P'(0)$ for the derivative of the radial
wavefunction at the origin that is obtained from nonrelativistic
potential models can be interpreted as a phenomenological estimate of
the regularized quantity ${\overline {R_P'}}(\Lambda)$ at a scale
$\Lambda$ of order $Mv$. 

The vacuum-saturation
approximation discussed in
Section~\ref{sec:vacuum-sat} allows
the matrix elements of some 4-fermion operators to be expressed in terms
of the regularized and renormalized wavefunction parameters defined above.
Combining (\ref{VSSwave}) with (\ref{RSwave}) and (\ref{R2Swave}),
we obtain the following expressions for
the matrix elements that contribute to the
decays of the $\eta_c$ and the $\psi$ into light hadrons:
\begin{mathletters}
\begin{eqnarray}
{\langle \eta_c |} {\cal O}_1({}^1S_0) {| \eta_c \rangle}
&=& {N_c \over 2 \pi} \Big| {\overline {R_{\eta_c}}} \Big|^2 \; \Bigg( 1 \;+\;
O(v^4) \Bigg).
\label{Meta6}
\\
{\langle \psi |} {\cal O}_1({}^3S_1) {| \psi \rangle}
&=& {N_c \over 2 \pi} \Big| {\overline {R_\psi}} \Big|^2 \; \Bigg( 1 \;+\;
O(v^4) \Bigg)
\label{Mpsi6}
\\
{\langle \eta_c |} {\cal P}_1({}^1S_0) {| \eta_c \rangle}
&=& - {N_c \over 2\pi} \; {\rm Re}({\overline {R_S}}{}^*\, {\overline {\nabla^2
R_S}})
	\; \Bigg( 1 \;+\; O(v^2) \Bigg),
\label{Meta8}
\\
{\langle \psi |} {\cal P}_1({}^3S_1) {| \psi \rangle}
&=& - {N_c \over 2\pi} \; {\rm Re}({\overline {R_S}}{}^*\, {\overline {\nabla^2
R_S}})
	\; \Bigg( 1 \;+\; O(v^2) \Bigg).
\label{Mpsi8}
\end{eqnarray}
\end{mathletters}
In (\ref{Meta8}) and (\ref{Mpsi8}), we have used heavy-quark spin
symmetry to replace ${\overline {R_{\eta_c}}}$ and ${\overline
{R_\psi}}$ by their weighted average ${\overline {R_S}}$ and to replace
${\overline {\nabla^2 R_{\eta_c}}}$ and ${\overline {\nabla^2 R_\psi}}$
by ${\overline {\nabla^2 R_S}}$ without any loss of accuracy. If we were
to make the same replacement in (\ref{Meta6}) and (\ref{Mpsi6}), the
relative accuracy would be decreased to $v^2$.  For the decays of the
P-wave states into light hadrons at leading order in $v$, the
vacuum-saturation approximation together with heavy-quark spin symmetry
can be used to express all the color-singlet matrix elements in terms of
the average regularized quantity ${\overline {R_P'}}$. Combining
(\ref{VSPwave}) with  (\ref{Rhchi}), we obtain the approximations 
\begin{mathletters}
\begin{eqnarray}
{\langle h_c |} {\cal O}_1({}^1P_1) {| h_c \rangle}
&=& {3 N_c \over 2 \pi} \Big| {\overline {R_P'}} \Big|^2 \;
	\Bigg( 1 \;+\; O(v^2) \Bigg),
\label{Mh}
\\
{\langle \chi_{cJ} |} {\cal O}_1({}^3P_{J}) {| \chi_{cJ} \rangle}
&=& {3 N_c \over 2 \pi} \Big| {\overline {R_P'}} \Big|^2 \;
	\Bigg( 1 \;+\; O(v^2) \Bigg), \quad J = 0,1,2.
\label{Mchi}
\end{eqnarray}
\end{mathletters}

\subsection{Factorization-Scale Dependence}
\label{sec:fact-scale}

The matrix elements ${\langle H |} {\cal O}_n {| H \rangle}$ that appear in the
factorization formula (\ref{master}) are ultraviolet finite only if the
4-fermion operators ${\cal O}_n$ are properly regularized. The
regularization introduces dependence on the ultraviolet cutoff
$\Lambda$ of NRQCD, and this cutoff-dependence must be understood in
order to make quantitative predictions. We assume that the
operator ${\cal O}_n$ is normal-ordered:  ${\langle 0 |} {\cal O}_n {| 0
\rangle} =
0$. This guarantees that in the matrix element ${\langle H |} {\cal O}_n
{| H \rangle}$, the operator ${\cal O}_n$ annihilates the heavy quark and
antiquark in the initial quarkonium state ${| H \rangle}$.  In addition to
normal-ordering, regularization is needed to control power and
logarithmic divergences.  If a cutoff $\Lambda$ is imposed on loop
momenta, there are power divergences in ${\langle H |} {\cal O}_n {| H
\rangle}$
that are proportional to $\Lambda^p$, where $p = 1,2,\ldots$.  If the
operator ${\cal O}_n$ has dimension $d_n$, then the coefficient of
$\Lambda^p$ is, by dimensional analysis, the sum of matrix elements of
4-fermion operators of dimension $d_n - p$ or larger.  If the dimension
is larger than $d_n-p$, the extra dimensions are balanced by powers of
$1/M$. Similarly the coefficients of logarithmic divergences are
proportional to matrix elements of 4-fermion operators of dimension
$d_n$ or larger.

The power and logarithmic divergences associated with loop corrections
to NRQCD operators can be regularized by a variety of means. A
convenient regularization scheme for analytic calculations is
dimensional regularization with minimal subtraction. The scale
associated with the dimensional regularization then plays the role of
the NRQCD cutoff $\Lambda$. An advantage in using a mass-independent
regulator, such as dimensional regularization, is that power divergences
are automatically discarded.  In other approaches, such as lattice
regularization, the regularized operator may contain divergences that
are proportional to powers of $\Lambda$. These power divergences are
simply artifacts of the regularization scheme and have no physical
content.  Since physical quantities are renormalization-group
invariants, they have no dependence on $\Lambda$. Hence, any power
divergences in NRQCD operator matrix elements must ultimately be
cancelled by power divergences in operator coefficients.

Once one has removed the power divergences, either by employing a
mass-independent regularization scheme or by making explicit
subtractions, the 4-fermion operators satisfy simple evolution equations
of the form
\begin{equation}
\Lambda {d \ \over d \Lambda} {\cal O}_n(\Lambda)
\;=\; \sum_k {\gamma_{nk}(\Lambda) \over M^{d_k - d_n}} \;
{\cal O}_k(\Lambda) ,
\label{evolop}
\end{equation}
where the sum ranges over all 4-fermion operators ${\cal O}_k(\Lambda)$ with
dimensions $d_k \ge d_n$.  The anomalous-dimension coefficients
$\gamma_{nk}(\Lambda)$ are computable as power series in the running
coupling constant $\alpha_s(\Lambda)$.
For $d_n = d_k$, the coefficients $\gamma_{nk}$ are of order $\alpha_s^2$,
because logarithmic ultraviolet divergences in one-loop
diagrams in NRQCD arise only from transverse gluons, whose coupling to the
heavy quark lines brings in a factor of $v$.
The coefficients $\gamma_{nk}$ for $d_n = 6$ and $d_k = 8$
are computed to order $\alpha_s$ in Appendix~\ref{app:evol}.

By taking the matrix elements of (\ref{evolop}) between heavy quarkonium
states ${| H \rangle}$, we obtain the evolution equations for the matrix
elements ${\langle H |} {\cal O}_n(\Lambda) {| H \rangle}$ that appear in the
general factorization formula (\ref{master}):
\begin{equation}
\Lambda {d \ \over d \Lambda} {\langle H |} {\cal O}_n(\Lambda) {| H \rangle}
\;=\; \sum_k {\gamma_{nk}(\Lambda) \over M^{d_k-d_n}}
	{\langle H |} {\cal O}_k(\Lambda) {| H \rangle} .
\label{evolme}
\end{equation}
The leading $v$ behavior of the matrix elements can be determined by
using the velocity-scaling rules developed in the previous sections. At
any given order in $v$, there is only a finite number of terms that
contribute to the evolution equation (\ref{evolme}). The evolution
equations for the dimension-6 4-fermion operators are calculated to
order $\alpha_s$ in Appendix~\ref{app:evol}. The operator evolution
equations for ${\cal O}_1({}^1S_0)$ and ${\cal O}_1({}^3S_1)$ are given
in (\ref{evol1S0}) and (\ref{evol3S1}).  Taking the matrix elements of
these equations and keeping only those terms on the right sides that are
of relative order $v^2$, we find that only the operators ${\cal P}_1$
survive, and we obtain 
\begin{mathletters}
\begin{eqnarray}
\Lambda {d \ \over d \Lambda} {\langle \eta_c |} {\cal O}_1({}^1S_0) {| \eta_c
\rangle}
\;=\; - {8 C_F  \alpha_s(\Lambda) \over 3 \pi M^2}
	{\langle \eta_c |} {\cal P}_1({}^1S_0) {| \eta_c \rangle},
\label{eveta6}
\\
\Lambda {d \ \over d \Lambda} {\langle \psi |} {\cal O}_1({}^3S_1) {| \psi
\rangle}
\;=\; - {8 C_F \alpha_s(\Lambda) \over 3 \pi M^2}
	{\langle \psi |} {\cal P}_1({}^3S_1) {| \psi \rangle},
\label{evpsi6}
\end{eqnarray}
\end{mathletters}
where $C_F = (N_c^2-1)/(2 N_c)$. If the evolution equations are
truncated at leading order in $v$, the right sides of (\ref{eveta6}) and
(\ref{evpsi6}) vanish and we find that the matrix elements ${\langle
\eta_c |} {\cal O}_1({}^1S_0) {| \eta_c \rangle}$ and ${\langle \psi |}
{\cal O}_1({}^3S_1) {| \psi \rangle}$ are renormalization-scale
invariant through order $\alpha_s$. The dimension-8 matrix elements
${\langle \eta_c |} {\cal P}_1({}^1S_0) {| \eta_c \rangle}$ and
${\langle \psi |} {\cal P}_1({}^3S_1) {| \psi \rangle}$ are also
renormalization-scale invariant through order $\alpha_s$ and at leading
order in $v$: 
\begin{mathletters}
\begin{eqnarray}
\Lambda {d \ \over d \Lambda} {\langle \eta_c |} {\cal P}_1({}^1S_0) {| \eta_c
\rangle}
\;=\; 0 ,
\label{eveta8}
\\
\Lambda {d \ \over d \Lambda} {\langle \psi |} {\cal P}_1({}^3S_1) {| \psi
\rangle}
\;=\; 0 .
\label{evpsi8}
\end{eqnarray}
\end{mathletters}
Truncated at order $\alpha_s$, the evolution equations can be solved
analytically for the $\Lambda$-dependence of the matrix elements.
For example, the solution to (\ref{eveta6}) is
\begin{equation}
{\langle \eta_c |} {\cal O}_1({}^1S_0;\Lambda) {| \eta_c \rangle}
\;=\; {\langle \eta_c |} {\cal O}_1({}^1S_0;\Lambda_0) {| \eta_c \rangle}
\;-\; {8 C_F \over 3 \beta_0 M^2}
	\; \log \left(\alpha_s(\Lambda_0) \over \alpha_s(\Lambda)\right)
	\; {\langle \eta_c |} {\cal P}_1({}^1S_0) {| \eta_c \rangle} \;,
\label{soleta6}
\end{equation}
where $\beta_0 = (11 N_c - 2 n_f)/6$ is the first
coefficient in the beta function for QCD with $n_f$ flavors of light quarks:
$\mu (d/d \mu) \alpha_s = - \beta_0 \alpha_s^2/\pi +\ldots$.

We next consider the evolution of the matrix elements that contribute to
P-wave annihilation at leading order in $v$. The color-singlet
dimension-8 matrix elements are renormalization-scale-invariant to this
order in $\alpha_s$: 
\begin{mathletters}
\begin{eqnarray}
\Lambda {d \ \over d \Lambda} {\langle h_c |} {\cal O}_1({}^1P_1) {| h_c
\rangle}
&=& 0 ,
\label{evh}
\\
\Lambda {d \ \over d \Lambda}
{\langle \chi_{cJ} |} {\cal O}_1({}^3P_{J}) {| \chi_{cJ} \rangle}
&=& 0 , \quad J = 0,1,2.
\label{evchi}
\end{eqnarray}
\end{mathletters}
Taking the matrix elements of (\ref{evol1S08}) and (\ref{evol3S18})
in Appendix~\ref{app:evol}, we find that the
color-octet dimension-6 matrix elements have nontrivial scaling behavior
at order $\alpha_s$:
\begin{mathletters}
\label{ev8hchi}
\begin{eqnarray}
\Lambda {d \ \over d \Lambda} {\langle h_c |} {\cal O}_8({}^1S_0) {| h_c
\rangle}
&=& {4 C_F \alpha_s(\Lambda) \over 3 N_c \pi M^2}
	{\langle h_c |} {\cal O}_1({}^1P_1) {| h_c \rangle} ,
\label{ev8h}
\\
\Lambda {d \ \over d \Lambda} {\langle \chi_{cJ} |} {\cal O}_8({}^3S_1)
{| \chi_{cJ} \rangle}
&=& {4 C_F \alpha_s(\Lambda) \over 3 N_c \pi M^2} \;
	{\langle \chi_{cJ} |} {\cal O}_1({}^3P_{J})  {| \chi_{cJ} \rangle} .
\label{ev8chi}
\end{eqnarray}
\end{mathletters}
To this order in $\alpha_s$, we find that the evolution equations can be
solved analytically.  For example, the solution to (\ref{ev8h}) is
\begin{equation}
{\langle h_c |} {\cal O}_8({}^1S_0;\Lambda) {| h_c \rangle}
\;=\; {\langle h_c |} {\cal O}_8({}^1S_0;\Lambda_0) {| h_c \rangle}
\;+\; {4 C_F \over 3 N_c \beta_0 M^2}
\log \left( \alpha_s(\Lambda_0) \over \alpha_s(\Lambda) \right)
	{\langle h_c |} {\cal O}_1({}^1P_1) {| h_c \rangle} .
\label{evsol}
\end{equation}

The solution (\ref{evsol})
to the evolution equation can be used to provide a crude
estimate of the color-octet matrix element
${\langle h_c |} {\cal O}_8({}^1S_0;\Lambda) {| h_c \rangle}$
in terms of the color-singlet matrix element
${\langle h_c |} {\cal O}_1({}^1P_1) {| h_c \rangle}$.
Suppose that we approximate (\ref{evsol}) by the evolution term on the 
right side.  The evolution term is largest, relative to the matrix element
${\langle h_c |} {\cal O}_8({}^1S_0;\Lambda_0) {| h_c \rangle}$, 
when the scales $\Lambda_0$ and $\Lambda$ are as widely 
separated as possible.  However, the logarithmic evolution holds only
down to scales of order $Mv$.  Thus, we choose $\Lambda_0=Mv$. Then,
setting $\Lambda=M$, neglecting the initial matrix element in
(\ref{evsol}), and assuming that $\alpha_s(\Lambda_0)=\alpha_s(Mv)\sim
v$, we find that (\ref{evsol}) reduces to 
\begin{equation}
{\langle h_c |} {\cal O}_8({}^1S_0;M) {| h_c \rangle}
\;\approx\; {4 C_F \over 3 N_c \beta_0 M^2}
\log \left( v\over \alpha_s(M) \right)
	{\langle h_c |} {\cal O}_1({}^1P_1) {| h_c \rangle} \; .
\label{evesth}
\end{equation}
The same method can be used to obtain crude estimates
for the corresponding matrix elements for the $\chi_{cJ}$ states:
\begin{equation}
{\langle \chi_{cJ} |} {\cal O}_8({}^3S_1;M) {| \chi_{cJ} \rangle}
\;\approx\; {4 C_F \over 3 N_c \beta_0 M^2}
	\log \left( v \over \alpha_s(M) \right)
   {\langle \chi_{cJ} |} {\cal O}_1({}^3P_{J}) {| \chi_{cJ} \rangle} \; .
\label{evestchi}
\end{equation}
The terms that we have retained in obtaining these estimates are
enhanced by one power of $\log [v/\alpha_s(M)]$ relative to the terms
that we have neglected.  Since this is not a large enhancement factor,
particularly in the case of charmonium, these estimates should be
regarded as giving only the orders of magnitude of the matrix elements. 
\vfill \eject

\section{Annihilation Decays of Heavy Quarkonium}
\label{sec:annih}

In Section~\ref{sec:matrix-el}, we derived a factorization formula
(\ref{master}) for the decay rates of heavy quarkonium states into light
hadrons. In this section, we apply this formula to the decays of S-wave
states,
up to corrections of relative order $v^3$, 
and to the decays of P-wave states, 
up to corrections of relative order $v^2$.
We also treat the decays into the
electromagnetic final states by using the analogous formula
(\ref{masterEM}). As in Section~\ref{sec:matrix-el}, we use the
lowest-lying S-wave and P-wave states of charmonium for the purpose of
illustration. 

\subsection{S-wave Annihilation}

Most previous treatments of the annihilation rates of the S-wave states
of heavy quarkonium have been restricted to leading order in $v$. In
these analyses, long-distance effects were absorbed into a
nonperturbative factor $|R_S(0)|^2$, where $R_S(0)$ is the radial
wavefunction at the origin. We improve on these previous treatments by
providing a rigorous definition of the nonperturbative factor in terms
of matrix elements of NRQCD.  We also extend the analysis of the decay
rates to relative order $v^2$, and show that 3 independent
nonperturbative factors are sufficient to calculate all the S-wave
annihilation rates through this order. 

We first consider the decays of the $J^{PC}= 0^{-+}$ state $\eta_c$ and
the $1^{--}$ state $\psi$ into light hadrons. As was shown in
Section~\ref{sec:vcounting}, there are only two operators that
contribute to each of these decay rates through relative order $v^2$.
According to (\ref{master}), the decay rates into light hadrons are
therefore 
\begin{mathletters}
\label{GSwavelh}
\begin{eqnarray}
\Gamma(\eta_c \to {\rm LH})
&=& {2 \; {\rm Im \,} f_1({}^1S_0) \over M^2} \;
	{\langle \eta_c |} {\cal O}_1({}^1S_0) {| \eta_c \rangle}
\;+\; {2 \; {\rm Im \,} g_1({}^1S_0) \over M^4}
	\; {\langle \eta_c |} {\cal P}_1({}^1S_0) {| \eta_c \rangle}
\nonumber \\
&& \quad \quad \quad \quad \quad \quad \quad \quad \quad \quad
\;+\; O(v^3 \Gamma) ,
\label{Getalh}
\\
\Gamma(\psi \to {\rm LH})
&=& {2 \; {\rm Im \,} f_1({}^3S_1) \over M^2} \;
	{\langle \psi |} {\cal O}_1({}^3S_1) {| \psi \rangle}
\;+\; {2 \; {\rm Im \,} g_1({}^3S_1) \over M^4} \;
	{\langle \psi |} {\cal P}_1({}^3S_1) {| \psi \rangle}
\nonumber \\
&& \quad \quad \quad \quad \quad \quad \quad \quad \quad \quad
\;+\; O(v^3 \Gamma) .
\label{Gpsilh}
\end{eqnarray}
\end{mathletters}

The imaginary parts of the coefficients in (\ref{GSwavelh}) are
calculated at order $\alpha_s^2$ in Appendix~\ref{app:alphasq}, and
${\rm Im \,} f_1({}^1S_0)$ and ${\rm Im \,} f_1({}^3S_1)$ are given
through next-to-leading order in $\alpha_s$ in
Appendix~\ref{app:higher}. According to the factorization formula
(\ref{masterEM}) for electromagnetic annihilation, the decay rates for
$\eta_c \to \gamma \gamma$ and $\psi \to e^+e^-$ are
\begin{mathletters}
\label{GSwaveEM}
\begin{eqnarray}
&& \Gamma(\eta_c \to \gamma \gamma)
\;=\; {2 \; {\rm Im \,} f_{\gamma \gamma}({}^1S_0) \over M^2} \;
	\Big| {\langle 0 |} \chi^\dagger \psi {| \eta_c \rangle} \Big|^2 \nonumber \\
&& \quad \quad \quad
\;+\; {2 \; {\rm Im \,} g_{\gamma \gamma}({}^1S_0) \over M^4} \;
	{\rm Re \,}\left( {\langle \eta_c |} \psi^\dagger \chi {| 0 \rangle}
	{\langle 0 |} \chi^\dagger (-\mbox{$\frac{i}{2}$} \tensor{\bf D})^2 \psi {|
\eta_c \rangle} \right)
\;+\; O(v^4 \Gamma) ,
\label{Getagg}
\\
&& \Gamma(\psi \to e^+ e^-)
\;=\; {2 \; {\rm Im \,} f_{\rm ee}({}^3S_1) \over M^2} \;
	\Big| {\langle 0 |} \chi^\dagger \mbox{\boldmath $\sigma$} \psi {| \psi
\rangle} \Big|^2 \nonumber \\
&& \quad \quad \quad
\;+\; {2 \; {\rm Im \,} g_{\rm ee}({}^3S_1) \over M^4} \;
	{\rm Re \,}\left( {\langle \psi |} \psi^\dagger \mbox{\boldmath $\sigma$} \chi
{| 0 \rangle} \cdot
	{\langle 0 |} \chi^\dagger \mbox{\boldmath $\sigma$} (-\mbox{$\frac{i}{2}$}
\tensor{\bf D})^2 \psi {| \psi \rangle} \right)
\;+\; O(v^4 \Gamma) .
\label{Gpsiee}
\end{eqnarray}
\end{mathletters}
The decay rate for $\psi \to \gamma \gamma \gamma$ is given by an
expression that is identical to (\ref{Gpsiee}), but with coefficients
$f_{3 \gamma}({}^3S_1)$ and $g_{3 \gamma}({}^3S_1)$. The imaginary parts
of the coefficients in (\ref{GSwaveEM}) are calculated at order
$\alpha^2$ in Appendix~\ref{app:electro}, and order-$\alpha_s$
corrections are given for ${\rm Im \,} f_{\gamma \gamma}({}^1S_0)$,
${\rm Im \,} f_{\rm ee}({}^3S_1)$, and ${\rm Im \,} f_{3
\gamma}({}^3S_1)$. The matrix elements in (\ref{GSwavelh}) and
(\ref{GSwaveEM}) can be computed using lattice simulations of NRQCD.
Since matrix elements of relative order $v^3$ have been omitted, there
is nothing to be gained by computing the dimension-6 matrix elements to
an accuracy of better than $v^2$. Similarly, the dimension-8 matrix
elements need be computed only at leading order in 
$v$.

At the level of accuracy in (\ref{GSwavelh}) and (\ref{GSwaveEM}),
the matrix elements are not all independent.
The vacuum-saturation approximation (\ref{VSSwave}) can be used to express
the 4-fermion matrix elements in (\ref{GSwavelh}) in terms of
the vacuum-to-quarkonium matrix elements in (\ref{GSwaveEM}).
Furthermore, the heavy-quark spin-symmetry relation (\ref{HQSEM})
can be used to equate the matrix elements in the second terms on the
right sides of (\ref{Getagg}) and (\ref{Gpsiee}).  The net result is
that the 8 matrix elements in (\ref{GSwavelh}) and (\ref{GSwaveEM})
can be reduced to 3 independent
nonperturbative quantities, which we can take to be $|{\overline
{R_{\eta_c}}}|^2$, $|{\overline {R_\psi}}|^2$,
and ${\rm Re}({\overline {R_S}}{}^*\, {\overline {\nabla^2 R_S}})$.
The resulting expressions for the decay rates are
\begin{mathletters}
\label{GSwave2}
\begin{eqnarray}
\Gamma(\eta_c \to {\rm LH})
&=& {N_c \; {\rm Im \,} f_1({}^1S_0) \over \pi M^2 } \;
	\Big| {\overline {R_{\eta_c}}} \Big|^2
\;-\; { N_c \; {\rm Im \,} g_1({}^1S_0) \over \pi M^4} \;
	{\rm Re}({\overline {R_S}}{}^*\, {\overline {\nabla^2 R_S}})
\;+\; O(v^3 \Gamma) ,
\label{GetalhR}
\\
\Gamma(\psi \to {\rm LH})
&=& {N_c \; {\rm Im \,} f_1({}^3S_1) \over \pi M^2} \; \Big| {\overline
{R_\psi}} \Big|^2
\;-\; { N_c \; {\rm Im \,} g_1({}^3S_1) \over \pi M^4} \;
	{\rm Re}({\overline {R_S}}{}^*\, {\overline {\nabla^2 R_S}})
\;+\; O(v^3 \Gamma) ,
\label{GpsilhR}
\\
\Gamma(\eta_c \to \gamma \gamma)
&=& {N_c \; {\rm Im \,} f_{\gamma \gamma}({}^1S_0) \over \pi M^2} \; \Big|
{\overline {R_{\eta_c}}} \Big|^2
\;-\; { N_c \; {\rm Im \,} g_{\gamma \gamma}({}^1S_0) \over \pi M^4} \;
	{\rm Re}({\overline {R_S}}{}^*\, {\overline {\nabla^2 R_S}})
\;+\; O(v^4 \Gamma) ,
\label{GetaggR}
\\
\Gamma(\psi \to e^+ e^-)
&=& {N_c \; {\rm Im \,} f_{\rm ee}({}^3S_1) \over \pi M^2} \; \Big| {\overline
{R_\psi}} \Big|^2
\;-\; { N_c \; {\rm Im \,} g_{\rm ee}({}^3S_1) \over \pi M^4} \;
	{\rm Re}({\overline {R_S}}{}^*\, {\overline {\nabla^2 R_S}})
\;+\; O(v^4 \Gamma) .
\label{GpsieeR}
\end{eqnarray}
\end{mathletters}
The quantities ${\overline {R_{\eta_c}}}$, ${\overline {R_\psi}}$, and
${\overline {\nabla^2 R_S}}$
are defined in Section~\ref{sec:wave-fn} in terms of
vacuum-to-quarkonium matrix elements in NRQCD, and can, therefore,
be calculated using nonperturbative methods,
such as lattice simulations.
They can also be estimated using the wavefunctions that are obtained
from nonrelativistic potential models of quarkonium.
Alternatively, since there are more
decay rates than there are parameters, they can be treated as
purely phenomenological parameters, to be determined by experiment.

The approximations to the matrix elements that have been made in
(\ref{GSwave2}) imply restrictions on the order in
$\alpha_s(M)$ to which the coefficients can be included meaningfully.
Because of the identification of $v$ with $\alpha_s(Mv)$ in (\ref{valpha}),
we should consider $v$ to be less than or of order $\alpha_s(M)$.
There is no point in calculating the coefficients
to relative order $\alpha_s^n$ unless we have included all
operators whose matrix elements are of relative order $v^n$ or less.
Hence, there is no gain in accuracy if the coefficients of $|{\overline
{R_{\eta_c}}}|^2$ and
$|{\overline {R_\psi}}|^2$ are calculated beyond relative order $\alpha_s^3$,
or if the coefficients of ${\rm Re}({\overline {R_S}}^* {\overline {\nabla^2
R_S}})$
are calculated beyond relative order $\alpha_s$.

If we require accuracy only to leading order in $v$, then the decay
rates in (\ref{GSwave2}) can be simplified further.  The difference
between ${\overline {R_{\eta_c}}}$ and ${\overline {R_\psi}}$ is of
relative order $v^2$, so both can be replaced by their weighted average
${\overline {R_S}}$. The factor ${\rm Re}({\overline {R_S}}{}^*\,
{\overline {\nabla^2 R_S}})$ is of order $v^2$ relative to $|R_S|^2$ and
can therefore be set to 0. We thereby recover the familiar factorization
formulas assumed in previous work: 
\begin{mathletters}
\label{GSwave0}
\begin{eqnarray}
\Gamma(\eta_c \to {\rm LH})
&=& {N_c \; {\rm Im \,} f_1({}^1S_0) \over \pi M^2} \; \Big| {\overline {R_S}}
\Big|^2
\;+\; O(v^2 \Gamma) ,
\label{GetalhS}
\\
\Gamma(\psi \to {\rm LH})
&=& {N_c \; {\rm Im \,} f_1({}^3S_1) \over \pi M^2} \; \Big| {\overline {R_S}}
\Big|^2
\;+\; O(v^2 \Gamma) ,
\label{GpsilhS}
\\
\Gamma(\eta_c \to \gamma \gamma)
&=& {N_c \; {\rm Im \,} f_{\gamma \gamma}({}^1S_0) \over \pi M^2} \; \Big|
{\overline {R_S}} \Big|^2
\;+\; O(v^2 \Gamma) ,
\label{GetaggS}
\\
\Gamma(\psi \to e^+ e^-)
&=& {N_c \; {\rm Im \,} f_{\rm ee}({}^3S_1) \over \pi M^2} \; \Big| {\overline
{R_S}} \Big|^2
\;+\; O(v^2 \Gamma) .
\label{GpsieeS}
\end{eqnarray}
\end{mathletters}
Because corrections of relative order $v^2$ have been neglected in
(\ref{GSwave0}), there is no point in calculating the regularized
wavefunction at the origin ${\overline {R_S}}$ to an accuracy of
relative order $v^2$. Similarly, because of the identification of $v$
with $\alpha_s(Mv)$, there is no increase in accuracy if the
coefficients of $|{\overline {R_S}}|^2$ in (\ref{GSwave0}) are
calculated beyond next-to-leading order in $\alpha_s$. The effects of
matrix elements of relative order $v^2$ are probably more important than
perturbative corrections to the coefficients that are of relative order
$\alpha_s^2$. 

\subsection{P-wave Annihilation}

In most previous work on the annihilation decays of P-wave states, it
was assumed that long-distance effects could be factored into a single
nonperturbative quantity $|R_P'(0)|^2$, where $R_P'(0)$ is the
derivative of the radial wavefunction at the origin.  By explicit
calculation, Barbieri {\it et al.} \cite{barbe,barbc} found that the
coefficients of $|R_P'(0)|^2$ depend logarithmically on an infrared
cutoff on the energies of the final-state gluons. In subsequent
phenomenological applications of these calculations, the infrared cutoff
has been identified with the binding energy of the quarkonium state,
which is of order $Mv^2$, the inverse of the radius of the bound state,
which is of order $Mv$, or the inverse of the confinement radius, which
is of order $\Lambda_{QCD}$. It should be clear, however, that the
infrared divergence is a signal of the breakdown of the factorization
assumption upon which the calculation is based. The solution to the
problem of infrared divergences in the calculation of the P-wave decay
rates into light hadrons was first presented in Ref. \cite{bbl}.  We will
review the resolution of this problem later in this subsection.

As was shown in Section~\ref{sec:vcounting}, there are two 4-fermion
operators that contribute to the decay rates of any of the P-wave states
into light hadrons at leading order in $v$. According to our
factorization formula (\ref{master}), the decay rates of the four P-wave
states into light hadrons are 
\begin{mathletters}
\label{GPwavelh}
\begin{eqnarray}
\Gamma(h_c \to {\rm LH})
&=& {2 \; {\rm Im \,} f_1({}^1P_1) \over M^4} \;
	{\langle h_c |}
{\cal O}_1({}^1P_1) {| h_c \rangle} \nonumber \\
&& \;+\; {2 \; {\rm Im \,} f_8({}^1S_0) \over M^2} \;
	{\langle h_c |} {\cal O}_8({}^1S_0) {| h_c \rangle}
\;+\; O(v^2 \Gamma) ,
\label{Ghlh}
\\
\Gamma(\chi_{cJ} \to {\rm LH})
&=& {2 \; {\rm Im \,} f_1({}^3P_{J}) \over M^4} \;
	{\langle \chi_{cJ} |} {\cal O}_1({}^3P_{J}) {| \chi_{cJ} \rangle} \nonumber \\
&& \;+\; {2 \; {\rm Im \,} f_8({}^3S_1) \over M^2} \;
	{\langle \chi_{cJ} |} {\cal O}_8({}^3S_1) {| \chi_{cJ} \rangle}
\;+\; O(v^2 \Gamma) , \quad J = 0,1,2.
\label{Gchilh}
\end{eqnarray}
\end{mathletters}
The imaginary parts of the coefficients $f_1({}^3P_{0})$,
$f_1({}^3P_{2})$, $f_8({}^1S_0)$, and $f_8({}^3S_1)$ are calculated in
order $\alpha_s^2$ in Appendix~\ref{app:alphasq}. The color-octet matrix
elements in the factorization formulas (\ref{GPwavelh}) represent
contributions to the annihilation rates from the Fock states ${| Q
\overline{Q} g \rangle}$. Thus, we see that the decays of P-wave (and
higher orbital-angular-momentum states) can probe components of the
meson wavefunction that involve dynamical gluons. For the decays of the
$\chi_{c0}$ and $\chi_{c2}$ into two photons, there is only one operator
that contributes at leading order in $v$: 
\begin{mathletters}
\label{GPwaveEM}
\begin{eqnarray}
\Gamma(\chi_{c0} \to \gamma \gamma)
&=& {2 \; {\rm Im \,} f_{\gamma \gamma}({}^3P_{0}) \over M^4} \; {1 \over 3} \;
\Big| {\langle 0 |} \chi^\dagger (-\mbox{$\frac{i}{2}$} \tensor{\bf D} \cdot
\mbox{\boldmath $\sigma$}) \psi {| \chi_{c0} \rangle} \Big|^2
\;+\; O(v^2 \Gamma) ,
\label{Gchi0gg}
\\
\Gamma(\chi_{c2} \to \gamma \gamma)
&=& {2 \; {\rm Im \,} f_{\gamma \gamma}({}^3P_{2}) \over M^4} \; \sum_{ij}
\Big| {\langle 0 |} \chi^\dagger (-\mbox{$\frac{i}{2}$} \tensor{D}{}^{(i}
\sigma^{j)})
	\psi {| \chi_{c2} \rangle} \Big|^2
\;+\; O(v^2 \Gamma) .
\label{Gchi2gg}
\end{eqnarray}
\end{mathletters}
The coefficients ${\rm Im \,} f_{\gamma \gamma}({}^3P_{0})$ and ${\rm Im
\,} f_{\gamma \gamma}({}^3P_{2})$ are calculated at order $\alpha^2$ in
Appendix~\ref{app:electro}, and the order-$\alpha_s$ corrections are
given as well. There is no increase in accuracy if the matrix elements
in (\ref{GPwavelh}) and (\ref{GPwaveEM}) are calculated to an accuracy
of relative order  $v^2$, since matrix elements of relative order $v^2$
have been omitted. Because of the identification (\ref{valpha}) of $v$
with $\alpha_s(Mv)$, there is no increase in accuracy if the
coefficients in (\ref{GPwavelh}) and (\ref{GPwaveEM}) are calculated
beyond next-to-leading order in $\alpha_s(M)$.  Perturbative corrections
of relative order $\alpha_s^2(M)$ are probably less important than
contributions of other matrix elements of relative order $v^2$. 

To the order in $v$ that is being considered in (\ref{GPwavelh}) and
(\ref{GPwaveEM}), the matrix elements are not all independent. The
vacuum-saturation approximation (\ref{VSPwave}) can be used to express
the matrix elements of ${\cal O}_1({}^1P_1)$ and ${\cal O}_1({}^3P_{J})$
in (\ref{GPwavelh}) in terms of vacuum-to-quarkonium matrix elements.
These matrix elements can be related to regularized derivatives of
radial wavefunctions at the origin by using (\ref{Rhchi}). Because of
the heavy-quark spin symmetry, the derivatives of the radial
wavefunctions at the origin can all be replaced by the average value
${\overline {R_P'}}$ for the 12 spin states of the P-wave system,
without any loss of accuracy. The heavy-quark spin-symmetry relation
(\ref{HQP2}) also implies that the matrix elements of ${\cal
O}_8({}^1S_0)$ and ${\cal O}_8({}^3S_1)$ in (\ref{GPwavelh}) are the
same,
up to corrections of relative order $v^2$.
Thus, the decay rates (\ref{GPwavelh})
and (\ref{GPwaveEM}) can all be expressed in terms of two
nonperturbative quantities $|{\overline {R_P'}}|^2$ and ${\langle h_c
|}{\cal O}_8({}^1S_0){| h_c \rangle}$ (or, alternatively, the average of
the color-octet matrix elements for the 12 P-wave spin states): 
\begin{mathletters}
\label{HPwave}
\begin{eqnarray}
\Gamma(h_c \to {\rm LH})
&=& {3 N_c \; {\rm Im \,} f_1({}^1P_1) \over \pi M^4} \; \Big| {\overline
{R_P'}} \Big|^2
\;+\; {2 \; {\rm Im \,} f_8({}^1S_0) \over M^2} \;
	{\langle h_c |}{\cal O}_8({}^1S_0){| h_c \rangle}
\;+\; O(v^2 \Gamma) ,
\label{Hhlh}
\\
\Gamma(\chi_{cJ} \to {\rm LH})
&=& {3 N_c \; {\rm Im \,} f_1({}^3P_{J}) \over \pi M^4} \; \Big| {\overline
{R_P'}} \Big|^2
\;+\; {2 \; {\rm Im \,} f_8({}^3S_1) \over M^2} \;
	{\langle h_c |}{\cal O}_8({}^1S_0){| h_c \rangle}
\nonumber \\
&& \quad \quad \quad \quad \quad \quad \quad \quad \quad \quad
\;+\; O(v^2 \Gamma) , \quad \; J=0,1,2,
\label{HchiJlh}
\\
\Gamma(\chi_{cJ} \to \gamma \gamma)
&=& {3 N_c \; {\rm Im \,} f_{\gamma \gamma}({}^3P_{J}) \over \pi M^4} \;
	\Big| {\overline {R_P'}} \Big|^2
\;+\; O(v^2 \Gamma) , \quad \; J=0,2.
\label{Hchi0em}
\end{eqnarray}
\end{mathletters}
Since ${\overline {R_P'}}$ is proportional to a vacuum-to-quarkonium matrix
element,
it can be calculated more easily in lattice NRQCD simulations than can
${\langle h_c |} {\cal O}_8({}^1S_0) {| h_c \rangle}$, which is a matrix
element
between quarkonium states.

As we have already mentioned, in the calculations of Barbieri {\it et
al.} of the P-wave decay rates into light hadrons
\cite{barbe,barbc,barbd}, a logarithmic dependence on an infrared cutoff
appeared in the coefficients of  $|R_P'(0)|^2$. We now explain why this
infrared-cutoff dependence is absent in the factorization formulas
(\ref{HPwave}).  The coefficients of $|{\overline {R_P'}}|^2$ in (\ref{HPwave})
depend logarithmically on the NRQCD cutoff $\Lambda$.  In these
coefficients, $\Lambda$ plays the same role as did the infrared cutoff
in the Barbieri {\it et al.} calculations.  According to the evolution
equation (\ref{ev8h}), the matrix element ${\langle h_c |} {\cal O}_8({}^1S_0)
{| h_c \rangle}$, also depends logarithmically on $\Lambda$.  In this case,
$\Lambda$ plays the role of an ultraviolet cutoff.  Because physical
quantities, such as decay rates, are renormalization-group invariants,
the $\Lambda$-dependence in ${\langle h_c |} {\cal O}_8({}^1S_0) {| h_c
\rangle}$
cancels the $\Lambda$-dependence in the coefficients of $|{\overline
{R_P'}}|^2$ in
(\ref{HPwave}).  Thus, we see that the inclusion of the color-octet term
proportional to ${\langle h_c |} {\cal O}_8({}^1S_0) {| h_c \rangle}$ in the
factorization formulas removes the dependence of the decay rate on an
arbitrary infrared cutoff.

The factorization formulas (\ref{HPwave}) for the annihilation decays of
P-waves at leading order in $v$ were first given in Ref. \cite{bbl} in
the form 
\begin{mathletters}
\label{bblP}
\begin{eqnarray}
\Gamma(h_c \to {\rm LH})
&=& H_1 \; {\widehat \Gamma}_1(Q \overline{Q}({}^1P_1) \to {\rm partons})
\;+\; H_8 \; {\widehat \Gamma}_8(Q \overline{Q}({}^1S_0) \to {\rm partons}) ,
\label{bblh}
\\
\Gamma(\chi_{cJ} \to {\rm LH})
&=& H_1 \; {\widehat \Gamma}_1(Q \overline{Q}({}^3P_{J}) \to {\rm partons})
\;+\; H_8 \; {\widehat \Gamma}_8(Q \overline{Q}({}^3S_1) \to {\rm partons}) ,
\nonumber \\
&& \quad \quad \quad \quad \quad \quad \quad \quad \quad \quad
J = 0,1,2,
\label{bblchi}
\\
\Gamma(\chi_{cJ} \to \gamma \gamma)
&=& H_1 \; {\widehat \Gamma}_1(Q \overline{Q}({}^3P_{J}) \to \gamma \gamma), \;
\quad J = 0, 2 .
\label{bblEM}
\end{eqnarray}
\end{mathletters}
The coefficients ${\widehat \Gamma}_1$ and ${\widehat \Gamma}_8$ in
(\ref{bblP}) are proportional to the annihilation rates of on-shell
$Q \overline{Q}$ pairs in color-singlet P-wave and color-octet S-wave states,
respectively.
The nonperturbative parameters $H_1$ and $H_8$ that were introduced in
Ref.~\cite{bbl}
can be defined rigorously in terms of matrix elements in NRQCD:
\begin{mathletters}
\label{H18}
\begin{eqnarray}
H_1 &=& {1 \over M^4} \;
{\langle h_c |} {\cal O}_1({}^1P_1) {| h_c \rangle} \; ,
\label{H1}
\\
H_8(\Lambda) &=& {1 \over M^2} \;
{\langle h_c |} {\cal O}_8({}^1S_0) {| h_c \rangle} \; .
\label{H8}
\end{eqnarray}
\end{mathletters}
The factors of $1/M^4$ and $1/M^2$ in (\ref{H1}) and
(\ref{H8}) were chosen in Ref. \cite{bbl} so that $H_1$ and $H_8$ would
be the combinations of NRQCD matrix elements and quark masses that are
determined in experimental measurements of the P-wave decay rates.

In retrospect, the choice made in Ref. \cite{bbl}  to include factors of
$1/M$ in the definitions of $H_1$ and $H_8$  in (\ref{H18})
was unfortunate.  The factors of $1/M$  are more properly associated with the
coefficients ${\widehat \Gamma}_1$ and ${\widehat \Gamma}_8$, since
they involve short-distance physics at scales of order $1/\Lambda$ or less.
The factorization formulas (\ref{HPwave})
are, therefore, to be preferred over the forms (\ref{bblP}),
because they incorporate all effects of the short distance scale $1/M$
into the coefficients, leaving matrix elements that depend only on
physics at length scales $1/(Mv)$ and longer.

\vfill \eject

\section{Perturbative Factorization}
\label{sec:pert-fac}

In this section, we sketch the connection between the NRQCD
approach and conventional perturbative methods for demonstrating
the factorization of cross sections involving large momentum transfer in QCD.
In perturbative proofs of factorization, the aim
is to demonstrate that, to all orders in perturbation theory,
infrared and collinear divergences either cancel or can be absorbed into
well-defined nonperturbative long-distance quantities.
Some familiar examples of such nonperturbative quantities are parton
distributions in the case of deep-inelastic lepton-hadron scattering
and fragmentation functions in the case of inclusive hadron production
at large transverse momentum in $e^+e^-$ annihilation.
The cross sections can be written as sums of products of long-distance
quantities with infrared-safe ({\it i.e.,} short-distance) parton-level
cross sections.
Our factorization formula for heavy quarkonium annihilation rates
is also of this form, and it is illuminating to see how it could
be derived from a more conventional perturbative analysis.

\subsection{Topological Factorization}

We remind the reader that, in QCD, infrared (or soft) divergences are
logarithmic and arise only from the emission of a gluon for which all
components of the 4-momentum are small. Collinear divergences (or mass
singularities) are also logarithmic, and arise when one parton (gluon or
light quark) splits into two or more partons and all of their 4-momenta
are parallel. Collinear divergences are cut off by quark masses, which
necessarily introduce a non-parallel component into the 4-momenta of the
splitting partons.

Let us focus first on the infrared divergences that arise
in the radiation of gluons from final-state partons
and on the collinear divergences that arise in the splitting of a
final-state parton into collinear partons.
The Kinoshita-Lee-Nauenberg theorem \cite{kln}
guarantees that all such divergences cancel
when one sums over those final-state cuts of a given
diagram that contribute to an inclusive cross section. For example, the
diagram shown in Fig.~\ref{fig:kln} has three cuts that
correspond to gluonic final states, and each cut
contains infrared and collinear divergences.  However, these divergences
cancel when one adds the real-emission cut of Fig.~\ref{fig:kln}(b)
to the virtual-emission cuts of Figs.~\ref{fig:kln}(a) and
\ref{fig:kln}(c).  This  step in the proof of perturbative factorization
is related to the localization of the
annihilation vertex, which was discussed in Section~\ref{sec:space-time}.

Next, let us consider the radiation of gluons from
the heavy-quark lines.  Such contributions are protected from collinear
divergences by the heavy-quark mass, so we need consider only the
possibility of infrared divergences. One key to analyzing the infrared
divergences is the concept of a ``controlling momentum''. The essential
idea is that the infrared divergence associated with an integration over
propagators and vertices in some portion of a Feynman diagram is cut off
by the largest external momentum that enters the propagators.  For
example, an infrared divergence could potentially arise from the square
of the diagram in Fig.~\ref{fig:real}(c) when all components of
the 4-momentum of the middle gluon
become small. However, because of simple kinematics, the other two
final-state gluons must both carry large momenta, some of whose components
are of order $M$. That large momentum  must flow through the heavy-quark
propagator to which the soft gluon attaches, and,
consequently, it cuts off the potential infrared divergence.

In this example, and in general, the concept of a controlling momentum
tells us that an infrared divergence can never arise from
a soft gluon that attaches to a propagator that is off-shell by order $M$.
That means that the infrared-divergent part of a Feynman
diagram can always be separated from the ``short-distance part'' by
cutting through heavy-quark propagators that are off the mass shell by
amounts that are much less than $M$.  (By the short-distance part, we
mean that part of the diagram that includes the hard final-state partons
and all propagators that are off-shell by order $M$.) This ``topological
factorization'' is the crucial step in a perturbative demonstration of
factorization.  It implies that
the infrared divergences can be disentangled from the short-distance
part of the diagram and absorbed into the long-distance part of the
diagram, which also includes the quarkonium wavefunctions.

The topological factorization of the annihilation rate of heavy quarkonium
is represented schematically in Fig.~\ref{fig:top}.
The shaded ovals represent the wavefunction for a quarkonium state.
A typical Fock state contains a $Q \overline{Q}$ pair and zero or more gluons
or light quark pairs.  The short-distance part of the annihilation
rate is represented by the circle labelled {\bf H} (for {\it hard}).
At leading order in $v$,
the only lines that attach to the short-distance
part are the incoming $Q$ and $\overline{Q}$
and the outgoing $Q$ and $\overline{Q}$.  The long-distance part includes
the wavefunction of the initial meson and its complex conjugate.  These
wavefunctions are connected by any extra partons that may be present
in the Fock state, which are represented in Fig.~\ref{fig:top} by gluon lines.
The long-distance part also includes soft-gluon interactions
between the extra partons, which are represented
by the circle labelled {\bf S} (for {\it soft}).

Once topological factorization has been demonstrated, two additional
steps are required in order to complete the proof of perturbative
factorization. First, one must decouple the relative 4-momentum $p$ of
the heavy quark and antiquark from the short-distance part of the
amplitude by expanding the short-distance part as a Taylor series in
$p$.  Second, one must decouple the Dirac indices and color indices that
connect the short-distance part to the long-distance part.
This can be accomplished by making use of Fierz rearrangements. In the
factored decay rate, the long-distance parts correspond to the matrix
elements of the NRQCD 4-fermion operators in the quarkonium state; the
short-distance parts correspond to the imaginary parts of the
coefficients of those operators in the NRQCD lagrangian.

In order to see in more detail how the perturbative analysis leads to
the results that we have obtained from NRQCD, let us consider two
examples: annihilation of S-wave and P-wave quarkonium at leading
nontrivial order in $v$ and through order $\alpha_s^3$ in QCD
perturbation theory. We use the specific example of decays into 2 and 3
gluons in the discussions below. However, the essential ingredients of
the discussion apply also to decays into a light quark-antiquark pair
and decays into a $q {\bar q}$ pair and a gluon.

\subsection{Annihilation of S-wave Quarkonium}
\label{sec:swavepert}

The first step in the analysis of annihilation of S-wave quarkonium
is to identify the short-distance part in the
topological factorization of the amplitude.  The dominant
component of the bound-state wavefunction
consists of a heavy quark and antiquark in a color-singlet state.
We take the $Q \overline{Q}$ pair to have total 4-momentum $P$
and relative 4-momentum $2p$.  We assume that, owing to the bound-state
dynamics, the $Q$ and $\overline{Q}$ have inverse propagators
$(\mbox{$\frac{1}{2}$} P \pm p)^2 - M^2$ of order $M^2 v^2$, with $v^2 \ll 1$.
In the meson rest frame, the energies $\mbox{$\frac{1}{2}$} P_0 \pm p_0$
of the $Q$ and $\overline{Q}$ then differ from the mass $M$ by order $Mv^2$
and their momenta $\mbox{$\frac{1}{2}$} {\bf P} \pm {\bf p}$ are of order $Mv$.
At order $\alpha_s^2$, the $Q \overline{Q}$ pair can annihilate into two gluons
through the diagrams in Fig.~\ref{fig:leading}.
By energy conservation, the two gluons must both have momenta of order $M$.
At this order, the topological factorization of the annihilation rate
is trivial. The $Q \overline{Q}$ annihilation amplitude belongs
entirely to the short-distance part of the annihilation rate in
Fig.~\ref{fig:top}, while the quarkonium wavefunction
belongs to the long-distance part.

We next consider the annihilation rate of the $Q \overline{Q}$ pair
at order $\alpha_s^3$.  This rate has contributions from the
annihilation into three gluons through the diagrams in Fig.~\ref{fig:real},
and also from the annihilation into two gluons, due to
the interference between next-to-leading order
diagrams such as those in Fig.~\ref{fig:virtual}
and the leading-order diagrams in Fig.~\ref{fig:leading}.
We begin by examining the infrared divergences in the
diagrams for the emission of a real gluon of 4-momentum $l$
shown in Fig.~\ref{fig:real}.  As we
have already explained, the diagram in Fig.~\ref{fig:real}(c) contains
no infrared divergence.  For the diagrams in Figs.~\ref{fig:real}(a) and
(b), we identify the infrared contribution that is leading in $v$
by assuming that $P_0 \approx 2 M$, that $l_0 = |{\bf l}|$, ${\bf P}$, and
${\bf p}$
are of order $Mv$, and that $p_0$ is of order $Mv^2$.
The emission vertex for the gluon with momentum $l$ and the
two adjacent heavy-quark propagators can then be approximated as follows:
\begin{eqnarray}
{(\pm \mbox{$\frac{1}{2}$} P + p) \cdot \gamma + M
	\over (\pm \mbox{$\frac{1}{2}$} P + p)^2 - M^2 + i \epsilon}
& \gamma^\mu  &
{(\pm \mbox{$\frac{1}{2}$} P + p \mp l) \cdot \gamma + M
	\over (\pm \mbox{$\frac{1}{2}$} P + p \mp l)^2 - M^2 + i \epsilon}
\nonumber \\
&\approx&
{M (1 \pm \gamma_0) \over (\pm \mbox{$\frac{1}{2}$} P + p)^2 -M^2 + i \epsilon}
\; \gamma^\mu \;
{M (1 \pm \gamma_0) \over -2 M l_0 + i \epsilon}
\nonumber \\
&=& {M (1 \pm \gamma_0) \over (\pm \mbox{$\frac{1}{2}$} P + p)^2 - M^2 + i
\epsilon}
	\left( {\pm g^{\mu 0} \over -|{\bf l}|} \right),
\label{swave-ir}
\end{eqnarray}
where the upper and lower signs correspond to Fig.~\ref{fig:real}(a)
and Fig.~\ref{fig:real}(b), respectively.  (In the case of the lower sign,
the order of the gamma matrices should actually be reversed, but the
last line is unaffected.)
The factor $\pm g^{\mu 0}$ is called
the ``eikonal vertex'', and the factor $1/(-l_0+i\epsilon)=1/(-|{\bf l}|)$
is called the ``eikonal propagator''. Their product is called the
``eikonal factor''.  We see that the eikonal factor
for the contribution of Fig.~\ref{fig:real}(a)
is equal and opposite in sign to the eikonal factor for the contribution
of Fig.~\ref{fig:real}(b).  All other propagator and vertex factors
in the two diagrams are the same.  If the $Q \overline{Q}$ pair is in a
color-singlet
state, then the color factors in the two diagrams are also the same,
and the infrared contributions from the region $|{\bf l}| \to 0$ cancel.
This cancellation is a consequence of the fact
that, in the infrared limit, the soft gluon couples to the
color charges of the quark and antiquark.  Since the quarkonium is a
color singlet, the quark and antiquark have opposite color charges.

Because of the infrared cancellation, the topological factorization of
the real-emission diagrams in Fig.~\ref{fig:real} is trivial.
The amplitudes for $Q \overline{Q} \to ggg$ all belong to the short distance
part
of the annihilation rate in Fig.~\ref{fig:top},
while the quarkonium wavefunction belongs to the long-distance part.

Now let us turn to the virtual-gluon-emission diagrams shown in
Fig.~\ref{fig:virtual}. Once again, we can identify the infrared part by
neglecting $l$ and $p$ compared to $M$.  As in the preceding example,
the eikonal vertices are proportional to $g^{\mu 0}$ times the quark
(or antiquark) charge. For Fig.~\ref{fig:virtual}(a), the eikonal
propagator factors associated with the exchange of the gluon with
momentum $l$ are $[1/(-l_0+i\epsilon)][-1/(l_0+i\epsilon)]$.  Each of
the diagrams of Figs.~\ref{fig:virtual}(b)~and~(c) contains a mass
renormalization, which we subtract.  The remaining contribution is a
wavefunction renormalization, half of which we absorb into the
quarkonium wavefunction.  The other half yields the eikonal
propagator factor $(-1/2)[1/(-l_0+i\epsilon)]^2$. (The squared
propagator appears after subtraction of the mass-renormalization
contribution.)  The eikonal factors from the three diagrams would
cancel, were it not for the $i\epsilon$'s in the propagator
denominators.  Instead, the eikonal factors yield
\begin{equation}
{1 \over -l_0 + i \epsilon} \, {-1 \over l_0 + i \epsilon}
\;-\; {1 \over -l_0 + i \epsilon} \, {1 \over -l_0 + i \epsilon}
\;=\; 2 \pi i \delta(l_0) \, {1 \over -l_0 + i \epsilon}.
\label{pinch}
\end{equation}
The $\delta(l_0)$ contribution arises because of the pinch in the $l_0$
integration contour in the contribution of the diagram of
Fig.~\ref{fig:virtual}(a). This $\delta(l_0)$ contribution corresponds
to the exchange of a space-like gluon with temporal polarization between
the quark and antiquark.  That is, it corresponds to the Coulomb
scattering of the quark and antiquark. Note that the factor multiplying
$\delta(l_0)$ is divergent at $l_0=0$. This somewhat unexpected
divergence has arisen because we have neglected the relative momentum
${\bf p}$ of the heavy quark and antiquark. Had we retained that momentum in
the
propagator denominators, we would have obtained a $1/v$ singularity,
where $v$ is the relative velocity of the quark and antiquark. [This
$1/v$ contribution is calculated in detail in
(\ref{A3bint4})--(\ref{A3b}).]  Ordinarily, in the absence of a
collinear singularity, the phase space for two partons to be moving with
small relative velocity would be unimportant.  Here, that region of
phase space is important by virtue of the quarkonium bound state.  (In
fact, it is the $1/v$ singularity that builds up the bound-state wave
function in a perturbative analysis of the Bethe-Salpeter equation for
positronium.)

At this point, the topological factorization of the virtual-emission
diagrams can be carried out. For the diagrams in Fig.~\ref{fig:virtual},
one factors the following contributions into the long-distance part: the
wavefunctions, with which we associate the square root of each quark or
antiquark wavefunction renormalization, and the $1/v$ singularity that
arises from the diagram of Fig.~\ref{fig:virtual}(a).  The remaining
contributions from these diagrams are factored into the short-distance
part.

Many discussions of perturbative factorization make use of the
Grammer-Yennie technique \cite{gy} for analyzing infrared
divergences.  As an aside,
let us indicate briefly how that technique would apply to
the example at hand. From our previous discussion, we see that, in the
infrared limit, the infrared-gluon  vertex $V^\mu$ is well approximated
by $g^{\mu 0}V_0$.
Then we can write
\begin{equation}
V^\mu \; \approx \; g^{\mu 0} \; {V\cdot l\over l_0-i\epsilon},
\label{g-y}
\end{equation}
provided that $l_0$ is not small compared with the other components of
$l$.  This is always the case for real emission.  For virtual emission,
we can eliminate the region of small $l_0$ by deforming the $l_0$
contour of integration into the lower half of the complex plane.  As can
be seen from an examination of the propagator denominators, all of the
singularities in the lower half of the complex plane are order $M^0$
away from the origin, except in the diagram of
Fig.~\ref{fig:virtual}(a).  This is the momentum-space manifestation of
the fact that a space-like gluon (with small $l_0$) can be exchanged causally
only between co-moving particles.  In carrying out the contour
deformation for the diagram of Fig.~\ref{fig:virtual}(a), and only in
this case, we unavoidably pick  up the residue of a pole at $l_0\approx 0$.
This residue yields the $1/v$ singularity in (\ref{pinch}) that was
noted earlier. Along the deformed contours, the Grammer-Yennie
approximation (\ref{g-y}) is valid. Substituting (\ref{g-y}) for the
infrared vertices, we can make use of Ward identities (current
conservation) to show that the contributions of the deformed contours
cancel. In order to obtain this Ward-identity cancellation, one needs,
in addition to the Grammer-Yennie contributions of
Figs.~\ref{fig:real}~and~\ref{fig:virtual}, Grammer-Yennie contributions
in which the infrared gluon attaches to the short-distance part of the
process. But, as we have already argued, these diagrams give
contributions that vanish in the infrared region, so there is no harm in
 applying the Grammer-Yennie approximation to them.

After topological factorization, the short-distance and long-distance
parts of the annihilation rate are still tied together by
integrations over the relative 4-momenta $p$ and $p'$ of the $Q \overline{Q}$
pairs
entering and leaving the short distance part
and by sums over the color and Dirac indices associated with the
heavy-quark propagators.
To complete the factorization, we must decouple these integrals and sums.

The decoupling of the integration over $p$ and $p'$ is accomplished simply by
expanding the short-distance contribution in a Taylor series in $p$ and $p'$.
Taking ${\bf p}$ and ${\bf p}'$ to be of order $Mv$ and $p_0$ and $p_0'$
to be of order $Mv^2$,
we see that the Taylor expansion of the short-distance part
corresponds to an expansion in powers of $v$.
All of the dependence on $p$ and $p'$ is now in the long-distance
part and in the explicit powers of $p$ and $p'$ from the Taylor expansion.
To analyze S-wave decays at leading order in $v$, we need keep only
the zeroth order terms in the Taylor expansion, which  amounts to
setting $p = p' = 0$.   In the meson rest frame, the $Q \overline{Q}$ pair has
total energy $P_0$ which differs from $2 M$ by an amount of order $Mv^2$
, and total momentum ${\bf P}$ of order $Mv$.  At leading order in $v$, we
can set $P_0 = 2M$ and ${\bf P} = 0$ in the short-distance part of the
annihilation rate. Thus the incoming quark and antiquark can be taken to
be on their mass shells and at threshold.

The decoupling of the color indices connecting the short-distance and
long-distance parts of the annihilation rate is straightforward.
The short-distance part is a color tensor $C_{ij,kl}$, with
color indices $i$ and $j$ for the incoming $Q$ and $\overline{Q}$
and $j$ and $k$ for the outgoing $Q$ and $\overline{Q}$.
The indices $i$ and $j$ can be decoupled from the tensor by using the
Fierz rearrangement
\begin{equation}
\delta_{i'i} \; \delta_{jj'}
\;=\; {1 \over N_c} \; \delta_{ji} \; \delta_{i'j'}
\;+\; 2 \; T^a_{ji} \; T^a_{i'j'}.
\label{colorFierz}
\end{equation}
A similar rearrangement can be used for the indices $k$ and $l$.
By color symmetry, $T^a_{ij} C_{ij,kk}$ must vanish and
$T^a_{ij} C_{ij,kl} T^b_{lk}$ must be proportional to the unit tensor
$\delta^{ab}$.  The resulting rearrangement formula is
\begin{equation}
C_{ij,kl}
\;=\; {1 \over N_c^2} \; \delta_{ji} \;
	\left( C_{i'i',j'j'} \right) \; \delta_{kl}
\;+\; {4 \over N_c^2-1} \; T^a_{ji} \;
	\left( T^b_{i'j'} C_{i'j',k'l'} T^b_{l'k'} \right) T^a_{kl} .
\label{colorRe}
\end{equation}
The indices have been decoupled from the tensor by decomposing it into a
term in which both pairs of indices are projected onto a color-singlet
state and a term in which both pairs of indices are projected onto a
color-octet state. For S-wave quarkonium, the dominant Fock state
contains a color-singlet $Q \overline{Q}$ pair, so only the first term on the
right
side of (\ref{colorRe}) contributes at leading order in $v$.

The Dirac indices connecting the short-distance and long-distance parts
of the amplitude can be decoupled in a similar way, although the algebra
is a little more cumbersome than it is for the color indices.  Having
set $p = p' = 0$ and $P = (2M,{\bf 0})$ in the short-distance part of
the amplitude, we find that the numerators of the four propagators
connecting it to the long-distance part reduce to $M(\gamma_0+1)$ for
the quarks and $M(-\gamma_0+1)$ for the antiquarks.  The Dirac structure
of the short-distance part of the amplitude is therefore a tensor
$\Gamma_{ij,kl}$, in which the Dirac indices $i$ and $k$ of the quarks
are contracted with projectors $P_+ = (1 + \gamma_0)/2$, and the Dirac
indices $j$ and $l$ of the antiquarks are contracted with projectors
$P_- = (1 - \gamma_0)/2$. The indices $i$ and $j$ can be decoupled from
the Dirac tensor by using the Fierz rearrangement
\begin{equation}
\left( P_+ \right)_{i'i} \left( P_- \right)_{jj'}
\;=\; {1 \over 2} \; \left( \gamma_5 P_+ \right)_{ji}
		\; \left( \gamma_5 P_- \right)_{i'j'}
\;+\; {1 \over 2} \; \left( \sigma^a \gamma_5 P_+ \right)_{ji}
		\; \left( \sigma^a \gamma_5 P_- \right)_{i'j'} ,
\label{DiracFierz}
\end{equation}
where $\sigma^i = (i/2) \epsilon^{0ijk} [\gamma^j, \gamma^k]$.
A similar Fierz rearrangement can be used to decouple the indices
$k$ and $l$ from the Dirac tensor.
Since all 3-momenta have been set to 0 in the Dirac tensor
$\Gamma^{ij,kl}$, there is no 3-vector on which $\Gamma^{ij,kl}$ can depend.
Its transformation properties under rotations then imply that the vector
$(\sigma^a \gamma_5 P_-)_{ij} \Gamma_{ij,kl} (\gamma_5 P_+)_{lk}$
must vanish, while the tensor $(\sigma^a \gamma_5 P_-)_{ij} \Gamma_{ij,kl}
(\sigma^b \gamma_5 P_+)_{lk}$ must be proportional to
the Cartesian unit tensor $\delta^{ab}$.  Consequently, one obtains the
rearrangement formula
\begin{eqnarray}
&& \left( P_+ \right)_{i'i} \left( P_- \right)_{jj'}
	\; \Gamma_{i'j',k'l'} \;
	\left( P_+ \right)_{kk'} \left( P_- \right)_{l'l}
\nonumber \\
&& \quad \quad
\;=\; {1 \over 4} \; \left( \gamma_5 P_+ \right)_{ji}
	\; \left[ \left( \gamma_5 P_- \right)_{i'j'} \Gamma_{i'j',k'l'}
		\left( \gamma_5 P_+ \right)_{l'k'} \right]
	\; \left( \gamma_5 P_- \right)_{kl}
\nonumber \\
&& \quad \quad
\;+\; {1 \over 12} \; \left( \sigma^m \gamma_5 P_+ \right)_{ji}
	\; \left[ \left( \sigma^a \gamma_5 P_- \right)_{i'j'}
		\Gamma_{i'j',k'l'}
		\left( \sigma^a \gamma_5 P_+ \right)_{l'k'} \right]
	\; \left( \sigma^m \gamma_5 P_- \right)_{kl} .
\label{DiracS}
\end{eqnarray}
This rearrangement of the Dirac indices corresponds to the decomposition
of the Dirac tensor into spin-singlet and spin-triplet pieces. The Dirac
matrix $\gamma_5 P_- = P_+ \gamma_5$ projects a $Q \overline{Q}$ pair at rest
onto
a state with total spin 0, as can be seen from the identity
\begin{equation}
\sum_{m m'} \langle 0,0 | \mbox{$\frac{1}{2}$},m;\mbox{$\frac{1}{2}$},m'
\rangle \; u_m {\bar v}_{m'}
\;=\; {1 \over \sqrt{2}} \left( \gamma_5 P_- \right) ,
\label{projzero}
\end{equation}
where $u_m$ and ${\bar v}_{m'}$ are Dirac spinors evaluated at zero
3-momentum. Similarly, the Dirac matrix $\sigma^i \gamma_5 P_- = P_+
\sigma^i \gamma_5$ projects a $Q \overline{Q}$ pair at rest onto a state of
total
spin 1:
\begin{equation}
\sum_{m m'} \langle 1,M | \mbox{$\frac{1}{2}$},m;\mbox{$\frac{1}{2}$},m'
\rangle \; u_m {\bar v}_{m'}
\;=\; {1 \over \sqrt{2}} U^{Mi} \left( \sigma^i \gamma_5 P_- \right) ,
\label{projone}
\end{equation}
where $U^{Mi}$ is the unitary matrix that transforms from the Cartesian
basis to the basis of angular-momentum eigenstates.

Now that we have decoupled the integrations over $p$ and $p'$ and
the sums over color
and Dirac indices, the factorization of the annihilation rate is complete.
In the rearrangement identity (\ref{DiracS}), the factors on the right
side that are enclosed in square brackets belong to the short-distance
part of the annihilation rate.  They correspond to the operator
coefficients in the NRQCD approach.  The remaining factors to the right and to
the left of the square brackets belong to the long-distance part.
It is evident that the long-distance parts of the annihilation rate can be
reproduced by matrix elements  of local operators in the quarkonium state.
The two operators that contribute to the annihilation of S-wave states
at leading order in $v$ may be identified as
\begin{mathletters}
\label{OS}
\begin{eqnarray}
\omit $- \overline{\Psi} \gamma_5 P_+ \Psi \overline{\Psi} \gamma_5 P_-
\Psi$\hfil&
\; \approx \;& {\cal O}_1({}^1S_0) ,
\label{OsingS}
\\
\omit $- \overline{\Psi} \mbox{\boldmath $\sigma$} \gamma_5 P_+ \Psi \cdot
\overline{\Psi} \mbox{\boldmath $\sigma$} \gamma_5 P_-
\Psi$\hfil
&\; \approx \;& {\cal O}_1({}^3S_1) ,
\label{OtripS}
\end{eqnarray}
\end{mathletters}
where $\Psi$ is the Dirac field for the heavy quark.
For matrix elements between quarkonium states, these operators reduce at
leading order in $v$ to the NRQCD operators
${\cal O}_1({}^1S_0) = \psi^\dagger \chi \chi^\dagger \psi$ and
${\cal O}_1({}^3S_1)
	= \psi^\dagger \mbox{\boldmath $\sigma$} \chi \cdot \chi^\dagger
\mbox{\boldmath $\sigma$} \psi$,
respectively.   Thus, perturbative factorization  yields the same
operator matrix elements as appear in the NRQCD analysis.
It should be noted, however, that the identifications (\ref{OS})
are not unique.  For example, the operator
$- \overline{\Psi} \gamma_5 \Psi \overline{\Psi} \gamma_5 \Psi$,
when sandwiched between quarkonium states, also reduces at
leading order in $v$ to ${\cal O}_1({}^1S_0)$ and both
$- \overline{\Psi} \mbox{\boldmath $\sigma$} \gamma_5 \Psi \cdot
\overline{\Psi} \mbox{\boldmath $\sigma$} \gamma_5 \Psi$ and
$- \overline{\Psi} \mbox{\boldmath $\gamma$} \Psi \cdot \overline{\Psi}
\mbox{\boldmath $\gamma$} \Psi$
reduce to ${\cal O}_1({}^3S_1)$.

\subsection{Annihilation of P-wave Quarkonium}

Now let us analyze the annihilation of P-wave quarkonium at leading
nontrivial order in $v$.
First we note that, because the spatial part of the P-wave quarkonium
wavefunction transforms under rotations like a vector, the
${{\bf p}}$-independent part of the $Q \overline{Q}$ annihilation amplitude
vanishes on carrying out the angular part of the
integration over ${\bf p}$.  Thus, we must retain terms with one factor
of ${\bf p}$ in the annihilation amplitude, which means that the leading
amplitude is down by one power of $v$ relative to the S-wave case.

At order $\alpha_s^2$, the annihilation proceeds through the diagrams
in Fig.~\ref{fig:leading}.  In this case, the factor of ${\bf p}$
in the $Q \overline{Q}$ annihilation amplitude can come only from expanding
the propagator of the virtual heavy quark, which is off its mass-shell by
an amount of order $M$.  The topological factorization is therefore trivial.
The amplitude for $Q \overline{Q} \to gg$ belongs to the short-distance part
of Fig.~\ref{fig:top}, and the quarkonium wavefunction belongs to the
long-distance part.

We next consider the annihilation at order $\alpha_s^3$, which receives
contributions from the real-emission diagrams in Fig.~\ref{fig:real}
and from the virtual-emission diagrams in Fig.~\ref{fig:virtual}.
The factor of ${\bf p}$ can come from one of two sources: the purely
short-distance (infrared-safe) part of the diagram, or the potentially
infrared-divergent part, which consists of the soft gluon
and the heavy-quark propagators to which it attaches.

If the factor of ${\bf p}$ comes from the short-distance part of the
diagram, then the analysis of the infrared divergences goes through
exactly as in the S-wave case.  Infrared divergences cancel between the
real-emission diagrams, but the exchange of a virtual gluon between the
$Q$ and $\overline{Q}$ [Fig.~\ref{fig:virtual}(c)] results in a $1/v$
singularity.  Topological factorization is trivial, except for this
$1/v$ singularity. It must be factored into the long-distance part of
the annihilation rate.

We proceed to consider the case in which the factor of ${\bf p}$ comes from
the potentially infrared-divergent part of the diagram.
We consider separately the cases of virtual-gluon emission
and real-gluon emission.

The diagrams for virtual-gluon emission are shown in
Fig.~\ref{fig:virtual}.  The potentially infrared-divergent part of the
amplitude includes the factors in the first line of (\ref{swave-ir}).
The required factor of ${\bf p}$ can come either from a ${\bf p} \cdot
\mbox{\boldmath $\gamma$}$
in the numerator of a propagator or from expanding out the denominator.
The terms with a factor of ${\bf p}$ that comes from a propagator
denominator are easily seen to be suppressed by a power of $v$.  The
terms that contain a ${\bf p} \cdot \mbox{\boldmath $\gamma$}$ in the numerator
are also
suppressed by a factor of $v$ because of the Dirac structure.  To see
this, first consider the case of a soft gluon with temporal
polarization. From the identity $P_+ {\bf p} \cdot \mbox{\boldmath $\gamma$} =
{\bf p} \cdot
\mbox{\boldmath $\gamma$} P_-$, one can see that the factor ${\bf p} \cdot
\mbox{\boldmath $\gamma$}$ connects
``large'' components of Dirac matrices to ``small'' components.  This
gives rise to the suppression by a factor of $v$. Now consider the case
of a virtual gluon with spatial polarization vector
$\mbox{\boldmath$\epsilon$}$.  Both
of the spatial-gluon vertices bring in factors of $\mbox{\boldmath$\epsilon$}
\cdot
\mbox{\boldmath $\gamma$}$.  The combined effect of these two factors and the
factor of
${\bf p} \cdot \mbox{\boldmath $\gamma$}$ is again to connect large and small
components,
which costs a factor of $v$.  Thus for virtual-gluon emission, there are
no infrared divergences at leading order in $v$. The topological
factorization is therefore trivial. The amplitude for $Q \overline{Q} \to gg$
belongs entirely to the short-distance factor in the annihilation rate,
aside from the square root of each heavy-quark or heavy-antiquark
wavefunction renormalization, which we associate with the quarkonium
wavefunction.

Finally, we consider the case of real-gluon emission
through the diagrams in Fig.~\ref{fig:real}.
Let us examine the infrared limit of the diagrams in
Fig.~\ref{fig:real}(a) and \ref{fig:real}(b) for the case in which the
soft gluon with momentum $l$ has spatial polarization vector
$\mbox{\boldmath$\epsilon$}$.
The emission vertex and the adjacent heavy-quark propagators
can be approximated as follows:
\begin{eqnarray}
{(\pm \mbox{$\frac{1}{2}$} P + p) \cdot \gamma + M
	\over (\pm \mbox{$\frac{1}{2}$} P + p)^2 - M^2 + i \epsilon}
\; & \mbox{\boldmath$\epsilon$} \cdot \mbox{\boldmath $\gamma$} & \;
{(\pm \mbox{$\frac{1}{2}$} P + p \mp l) \cdot \gamma + M
	\over (\pm \mbox{$\frac{1}{2}$} P + p \mp l)^2 - M^2 + i \epsilon}
\nonumber \\
&\approx& {M (1 \pm \gamma_0) - {\bf p} \cdot \mbox{\boldmath $\gamma$}
	\over (\pm \mbox{$\frac{1}{2}$} P + p)^2 - M^2 + i \epsilon}
\; \mbox{\boldmath$\epsilon$} \cdot \mbox{\boldmath $\gamma$} \;
{M (1 \pm \gamma_0) - {\bf p} \cdot \mbox{\boldmath $\gamma$} \over -2 M l_0 +
i \epsilon}
\nonumber \\
&\approx& {M (1 \pm \gamma_0) \over (\pm \mbox{$\frac{1}{2}$} P + p)^2 - M^2 +
i \epsilon}
\left({2 {\bf p} \cdot \mbox{\boldmath$\epsilon$} \over -|{\bf l}|} \right) .
\label{pwave-ir}
\end{eqnarray}
The upper and lower signs apply to Figs.~\ref{fig:real}(a) and
\ref{fig:real}(b), respectively. (For the lower sign, the order of the
Dirac matrices should actually be reversed.)  In the last line of
(\ref{pwave-ir}), we have retained only those numerator terms that
contain one power of  ${\bf p}$. The factor $2{{\bf p}}\cdot
\mbox{\boldmath$\epsilon$}/(-|{\bf l}|)$ is the infrared-emission factor. In
contrast with
the S-wave case, the infrared contributions from the two real-emission
diagrams add, rather than cancelling.  Because we have retained one
power of ${\bf p}$, the soft gluon couples to the color
current of the heavy quark, rather than to the color charge. Since the
heavy quark and antiquark have opposite color charges and, in the CM
frame, opposite momenta, their color currents are equal.
Note that, because of the vector ${\bf p}$ in the infrared-emission factor,
the emission of the soft gluon changes the orbital-angular-momentum
quantum number of the $Q \overline{Q}$ pair by one unit, but it does not
flip the  spin of the quark or antiquark.  Thus, it
converts the heavy quark and antiquark from a
color-singlet P-wave state to a color-octet S-wave state.

In the decay rate, we must integrate the infrared emission factors from
the square of the sum of the amplitudes over the phase space of the
gluon. Keeping only the logarithmically divergent part of the integral,
we find the result
\begin{equation}
4\,\int_\lambda^M {d^3l \over (2\pi)^3} \, {1 \over 2 |{\bf l}|}
\left( { 2 {\bf p} \cdot \mbox{\boldmath$\epsilon$} \over - |{\bf l}|} \right)
\left( { 2 {\bf p}' \cdot \mbox{\boldmath$\epsilon$}^* \over - |{\bf l}|}
\right)
\;=\;
{4{\bf p} \cdot \mbox{\boldmath$\epsilon$} \, {{\bf p}}' \cdot
\mbox{\boldmath$\epsilon$}^* \over \pi^2}
\left( \log{M\over\Lambda} \;+\; \log{\Lambda\over\lambda} \right),
\label{pwave-long}
\end{equation}
where $\lambda$ is an infrared cutoff of order $Mv$, and we have
arbitrarily set the upper limit on $|{\bf l}|$ to $M$.  We have introduced a
factorization scale $\Lambda$ to separate the infrared divergence from
the short-distance part of the integral. The long-distance contribution
that is proportional to $\log(\Lambda/\lambda)$ in (\ref{pwave-long})
can be interpreted as the probability for a heavy quark and antiquark in
a color-singlet P-wave state to make a transition to a  color-octet
S-wave state by radiating a soft gluon.

We can now carry out the topological factorization of the diagrams in
Fig.~\ref{fig:real} for real gluon emission.  The square of the
amplitude for $Q \overline{Q} \to g g g$, integrated over phase space, belongs
to the short-distance part of the annihilation rate in
Fig.~\ref{fig:top}, except for the second term on the right side of
(\ref{pwave-long}).  This term, which contains the infrared divergent
contribution that arises from the emission of the soft gluon in
Figs.~\ref{fig:real}(a) and \ref{fig:real}(b), is included in the
long-distance part, along with the quarkonium wavefunction. The soft
gluon is an example of a light parton that connects the initial and
final wavefunctions in Fig.~\ref{fig:top}. Note that, in this
contribution to the annihilation rate, the heavy quark and antiquark
enter the short-distance part in a color-octet S-wave state.  We call
this contribution to the annihilation rate the ``color-octet
contribution''. In all the other contributions to the P-wave
annihilation rate at this order, the heavy quark and antiquark enter the
short-distance part in a color-singlet P-wave state.  We refer to those
contributions as the ``color-singlet contribution''.

At this point, we have identified the long- and short-distance parts in
the topological factorization of the annihilation rate.
It remains only to decouple the integrations over the relative momenta
$p$ and $p'$ of the $Q \overline{Q}$ pairs and the sums over color and Dirac
indices.

We first discuss the color-singlet contribution. The color indices of
the short-distance and long-distance parts are easily decoupled by using
the rearrangement identity (\ref{colorRe}). Only the first term on the
right side of (\ref{colorRe}) contributes, since the $Q \overline{Q}$ pair is
in a
color-singlet state. In order to decouple the integrals over the
relative momenta $p$ and $p'$ of the initial and final $Q \overline{Q}$ pairs,
we
expand the short distance part as a Taylor series in $p$ and $p'$.  At
leading order in $v$, we set $p_0 = {p_0}' = 0$ and keep only those
terms linear in both ${\bf p}$ and ${\bf p}'$. The resulting amplitude has the
structure $p^m \Gamma^{mn}_{ij,kl} {p'}^n$, in which the Dirac indices
$i$ and $k$ of the quark are contracted with projectors $P_+$ and the
Dirac indices $j$ and $l$ of the antiquarks are contracted with
projectors $P_-$. The Dirac indices of the initial and final $Q \overline{Q}$
pair
are decoupled from the short distance factor $\Gamma^{mn}_{ij,kl}$ by
applying the Fierz identity (\ref{DiracFierz}) to both the initial and
final indices. The resulting rearrangement identity can be greatly
simplified by making use of the fact that the Fierz-decoupled
short-distance parts transform like tensors under rotations and the fact
that there are no three-vectors on which $\Gamma^{mn}_{ij,kl}$ can
depend.  For example, the tensor $(\gamma_5 P_+)_{ij}
\Gamma^{mn}_{ij,kl} (\gamma_5 P_+)_{lk}$ must be proportional to the
unit tensor $\delta^{mn}$, and the tensor $(\sigma^a \gamma_5 P_+)_{ij}
\Gamma^{mn}_{ij,kl} (\gamma_5 P_+)_{lk}$ must be proportional to the
totally antisymmetric tensor $\epsilon^{amn}$.
Since $(\sigma^a \gamma_5 P_+)_{ij} \Gamma^{mn}_{ij,kl}
	(\sigma^b \gamma_5 P_+)_{lk}$
is a Cartesian tensor in 3 dimensions with 4 indices, it can be decomposed
into a linear combination of the three tensors
$\delta^{am} \delta^{bn}$, $\epsilon^{amx} \epsilon^{bnx}$, and
$\mbox{$\frac{1}{2}$} (\delta^{ab}\delta^{mn} + \delta^{an}\delta^{mb})
	- {1\over3} \delta^{am}\delta^{bn}$,
which correspond to total angular momentum 0, 1, and 2, respectively.
Consequently, one obtains the rearrangement formula
\begin{eqnarray}
&& \left( P_+ \right)_{i'i} \left( P_- \right)_{jj'}
	\; \left( p^m \; \Gamma^{mn}_{i'j',k'l'} \; {p'}^n \right) \;
	\left( P_+ \right)_{kk'} \left( P_- \right)_{l'l}
\nonumber \\
&&
\;=\; {1 \over 12} \; \left( p^m \gamma_5 P_+ \right)_{ji}
\; \left[ \left( \gamma_5 P_- \right)_{i'j'}
	\Gamma^{aa}_{i'j',k'l'}
	\left( \gamma_5 P_+ \right)_{l'k'} \right]
\; \left( {p'}^m \gamma_5 P_- \right)_{kl}
\nonumber \\
&& \quad \quad
\;+\; {1 \over 36} \; \left( {\bf p} \cdot \mbox{\boldmath $\sigma$} \gamma_5
P_+ \right)_{ji}
\; \left[ \left( \sigma^a \gamma_5 P_- \right)_{i'j'}
	\Gamma^{ab}_{i'j',k'l'}
	\left( \sigma^b \gamma_5 P_+ \right)_{l'k'} \right]
\; \left( {\bf p}' \cdot \mbox{\boldmath $\sigma$} \gamma_5 P_- \right)_{kl}
\nonumber \\
&& \quad \quad
\;+\; {1 \over 24} \; \left( ({\bf p} \times \mbox{\boldmath $\sigma$})^m
\gamma_5 P_+ \right)_{ji}
\; \left[ \left( \sigma^{[a} \gamma_5 P_- \right)_{i'j'}
	\Gamma^{b][b}_{i'j',k'l'}
	\left( \sigma^{a]} \gamma_5 P_+ \right)_{l'k'} \right]
\; \left( ({\bf p}' \times \mbox{\boldmath $\sigma$})^m \gamma_5 P_-
\right)_{kl}
\nonumber \\
&& \quad \quad
\;+\; {1 \over 20} \; \left( p^{(m} \sigma^{n)} \gamma_5 P_+ \right)_{ji}
\; \left[ \left( \sigma^{(a} \gamma_5 P_- \right)_{i'j'}
	\Gamma^{b)(a}_{i'j',k'l'}
	\left( \sigma^{b)} \gamma_5 P_+ \right)_{l'k'} \right]
\; \left( {p'}^{(m} \sigma^{n)} \gamma_5 P_- \right)_{kl}
\nonumber \\
&& \quad \quad
\;-\; {1 \over 24} \; \left( ({\bf p} \times \mbox{\boldmath $\sigma$})^m
\gamma_5 P_+ \right)_{ji}
\; \left[ \epsilon^{abc} \left(  \sigma^{a} \gamma_5 P_- \right)_{i'j'}
	\Gamma^{bc}_{i'j',k'l'}
	\left( \gamma_5 P_+ \right)_{l'k'} \right]
\; \left( {p'}^m \gamma_5 P_- \right)_{kl}
\nonumber \\
&& \quad \quad
\;+\; {1 \over 24} \;  \left( p^m \gamma_5 P_+ \right)_{ji}
\; \left[ \epsilon^{abc} \left( \gamma_5 P_- \right)_{i'j'}
	\Gamma^{ab}_{i'j',k'l'}
	\left( \sigma^c \gamma_5 P_+ \right)_{l'k'} \right]
\; \left( ({\bf p}' \times \mbox{\boldmath $\sigma$})^m \gamma_5 P_-
\right)_{kl} .
\label{DiracP}
\end{eqnarray}
We use the notation $T^{(ab)}$ for the symmetric traceless part of a
tensor $T^{ab}$ and $T^{[ab]} = \mbox{$\frac{1}{2}$}(T^{ab} - T^{ba})$
for the antisymmetric part.

At this point, the factorization of the color-singlet contribution
to the annihilation rate is complete.  The factors in square
brackets on the right side of (\ref{DiracP}) belong to the short-distance
part, while the factors to the right and to the left of the square brackets
belong to the long-distance part.
It is apparent from the rearrangement identity (\ref{DiracP})
that the long-distance parts can be
reproduced by matrix elements  of the following local operators:
\begin{mathletters}
\begin{eqnarray}
\omit $ - \overline{\Psi} (-\mbox{$\frac{i}{2}$} \tensor{\bf D}) \gamma_5 P_+
\Psi
	\cdot \overline{\Psi} (-\mbox{$\frac{i}{2}$} \tensor{\bf D}) \gamma_5 P_-
\Psi$\hfil &
\; \approx \;&{\cal O}_1({}^1P_1) ,
\\
\omit $ - {1\over 3} \overline{\Psi} (-\mbox{$\frac{i}{2}$} \tensor{\bf D}
\cdot \mbox{\boldmath $\sigma$})\gamma_5 P_+ \Psi
  \, \overline{\Psi} (-\mbox{$\frac{i}{2}$} \tensor{\bf D} \cdot
\mbox{\boldmath $\sigma$})\gamma_5 P_-\Psi$\hfil &
\; \approx \; &{\cal O}_1({}^3P_{0}) ,
\\
\omit $- {1\over 2} \overline{\Psi} (-\mbox{$\frac{i}{2}$} \tensor{\bf D}
\times \mbox{\boldmath $\sigma$}) \gamma_5 P_+ \Psi
	\cdot \overline{\Psi} (-\mbox{$\frac{i}{2}$} \tensor{\bf D} \times
\mbox{\boldmath $\sigma$}) \gamma_5 P_-
\Psi$\hfil &
\; \approx \;& {\cal O}_1({}^3P_{1}) ,
\\
\omit $ - \overline{\Psi} (-\mbox{$\frac{i}{2}$} \tensor{D}{}^{(i} \sigma^{j)})
\gamma_5 P_+ \Psi
	\, \overline{\Psi} (-\mbox{$\frac{i}{2}$} \tensor{D}{}^{(i} \sigma^{j)})
\gamma_5 P_-
\Psi$\hfil &
\; \approx \; &{\cal O}_1({}^3P_{2}) ,
\\
\omit $ - \overline{\Psi} (-\mbox{$\frac{i}{2}$} \tensor{\bf D}) \gamma_5 P_+
\Psi
	\cdot \overline{\Psi} (-\mbox{$\frac{i}{2}$} \tensor{\bf D} \times
\mbox{\boldmath $\sigma$}) \gamma_5 P_-
\Psi,$\hfil &&
\\
\omit $- \overline{\Psi} (-\mbox{$\frac{i}{2}$} \tensor{\bf D} \times
\mbox{\boldmath $\sigma$}) \gamma_5 P_- \Psi
	\cdot \overline{\Psi} (-\mbox{$\frac{i}{2}$} \tensor{\bf D}) \gamma_5 P_+
\Psi.$\hfil &&
\end{eqnarray}
\end{mathletters}
The matrix elements of the last two operators vanish for a quarkonium state
that is a charge-conjugation eigenstate.  The other four operators reduce
at leading order in $v$ to the NRQCD operators ${\cal O}({}^1P_1)$
and ${\cal O}({}^3P_{J}), J = 0,1,2$, respectively.

Finally, we consider the factorization of the color-octet contribution
to the annihilation rate, for which the short-distance part involves
the annihilation of a $Q \overline{Q}$ pair in a color-octet S-wave state.
The color indices of the short-distance and long-distance parts
are easily decoupled by using the rearrangement identity (\ref{colorRe}).
Only the second term on the right side of (\ref{colorRe}) is
non-vanishing for the color-octet contribution.
The decoupling of the momentum integrations and the Dirac indices
proceeds along the same lines as for S-wave quarkonium,
which was discussed in subsection \ref{sec:swavepert}.
The momentum integrations are decoupled by
Taylor-expanding the short-distance part in $p$ and $p'$, and setting $p
= p' = 0$. The decoupling of the Dirac indices is accomplished by using
the rearrangement formula (\ref{DiracS}). The factors in square brackets
in (\ref{DiracS}) belong to the short-distance part of the annihilation
rate, while the factors to the right and to the left belong to the
long-distance part. From the rearrangement identities (\ref{colorRe})
and (\ref{DiracS}), it is evident that the long-distance parts are
reproduced by matrix elements of the operators
\begin{mathletters}
\begin{eqnarray}
\omit $-\overline{\Psi} \gamma_5 T^a P_+ \Psi \overline{\Psi} \gamma_5 T^a P_-
\Psi$\hfil &
\; \approx \;& {\cal O}_8({}^1S_0) ,
\\
\omit $-\overline{\Psi} \mbox{\boldmath $\sigma$} \gamma_5 T^a P_+ \Psi
	\cdot \overline{\Psi} \mbox{\boldmath $\sigma$} \gamma_5 T^a P_- \Psi$\hfil &
\; \approx \; &{\cal O}_8({}^3S_1) .
\end{eqnarray}
\end{mathletters}
These operators reduce, at leading order in $v$, to the
operators ${\cal O}_8({}^1S_0) = \psi^\dagger T^a \chi \chi^\dagger T^a \psi$
and
${\cal O}_8({}^3S_1) = \psi^\dagger \mbox{\boldmath $\sigma$} T^a \chi \cdot
\chi^\dagger \mbox{\boldmath $\sigma$} T^a \psi$
in the NRQCD analysis.
The matrix elements of ${\cal O}_8({}^1S_0)$ and ${\cal O}_8({}^3S_1)$
include the probability factor proportional to
$\log(\Lambda/\lambda)$ in (\ref{pwave-long}). The logarithmic dependence on
$\Lambda$ is reflected in the evolution of these operators,
which is given in (\ref{ev8h}) and (\ref{ev8chi}). Thus, the factorization
scale $\Lambda$ in the perturbative approach can be identified
with the  ultraviolet cutoff of NRQCD.

\vfill \eject

\section{Production of Heavy Quarkonium}
\label{sec:prod}

In this section, we present a general factorization formula for
computing inclusive heavy-quarkonium production rates in high-energy
processes that involve a momentum transfer $Q^2$ that is of order $M^2$
or larger. In the case of S-wave quarkonium, our factorization formalism
coincides with the ``color-singlet model'' for quarkonium production 
\cite{schuler} in the nonrelativistic limit, but it also allows the
systematic calculation of relativistic corrections that are suppressed
by powers of $v$. In the case of P-wave quarkonium, our formalism
reveals that the color-singlet model is incomplete, even at leading
order in $v$, and must be supplemented by including the ``color-octet
mechanism'' for P-wave quarkonium production \cite{bbly}.

\subsection{Factorization of the Production Rate}

Our goal, as in the discussion of heavy-quarkonium annihilation,
is to express the
inclusive production rate for a quarkonium state in a factored form.
That is, we wish to write the production rate as a sum of terms, each of
which consists of a short-distance part, which can be calculated in QCD
perturbation theory, multiplied by a long-distance part that can be
expressed as a matrix element in NRQCD.
Our arguments for the factorization of the production rate are
based on the all-orders properties of QCD perturbation theory.
In this sense, the level of rigor of these arguments is comparable to that
in the proofs of factorization for the
Drell-Yan process for lepton pair-production in hadron-hadron
collisions \cite{drell-yan}.
These arguments are less rigorous than those that we
have given for the factorization of the quarkonium annihilation rate.
The latter arguments
rely only on the general space-time structure of the annihilation
process and on the validity of the effective-field-theory approach.
Their level of rigor is comparable to that
in the proofs of factorization in  deep-inelastic lepton-hadron scattering,
which can be formulated in terms of the operator-product expansion.

When a quarkonium state is produced in a process that involves
momentum transfer $Q^2$ of order $M^2$ or larger, the production of the $Q
\overline{Q}$
pair that forms the bound state takes place at short distances of order
$1/M$ or smaller.  A simple example of such a process, which the reader can
keep in mind throughout the following discussion, is
the production in $e^+e^-$ annihilation
at a center-of-mass energy $\sqrt{s} \gg M$ of a heavy quarkonium $H$,
with 4-momentum $P$,
recoiling against two light hadron jets.  At leading order in QCD
perturbation theory, the relevant parton process is
$e^+e^- \to Q \overline{Q} g g$. We take the $Q$ and $\overline{Q}$ to have
momenta
$P/2+p$ and $P/2-p$.  The relative 3-momentum ${\bf p}$ must be
of order $Mv$ in the ${\bf P} = 0$ frame
in order for the $Q \overline{Q}$ pair to have a significant probability
for forming the bound state $H$.
The amplitude for the  production of the $Q \overline{Q}$ pair is insensitive
to changes in the relative 4-momentum $p$
that are much less than $M$, and therefore the quark and antiquark
are produced with a separation of order $1/M$ or less.
Similarly, the square of the amplitude is insensitive to changes in the
the total 4-momentum $P$ of the heavy pair that are much less than
$M$. Thus, the product of one amplitude and the complex conjugate
of a second will contribute significantly to the $Q \overline{Q}$-production
cross section only if the corresponding production
points are separated by a distance of order $1/M$ or less.
We therefore conclude
that the production of the $Q \overline{Q}$ pair is indeed a short-distance
process
that takes place within a distance of order $1/M$.

In the framework of NRQCD, the effect of the short-distance part of a
production amplitude is simply to create a $Q \overline{Q}$ pair at a spacetime
point.
The formation of the quarkonium state $H$ from the $Q \overline{Q}$ pair takes
place over
distances that are of order $1/(Mv)$ or larger in the quarkonium rest frame,
so it is described accurately by NRQCD.  Therefore, in NRQCD,
the production rate (the square of the amplitude summed over final states)
involves the creation of a $Q \overline{Q}$ pair at a spacetime point, its
propagation into the asymptotic future, where the out state includes the
quarkonium $H$, and, finally, the propagation of the $Q \overline{Q}$ pair back
in
time to the creation point. That is, the long-distance part of the
production rate is given in NRQCD by vacuum matrix elements of local
4-fermion operators.  The effects of the short-distance parts of the
production rate are taken into account through the coefficients of the
4-fermion operators. Since the final state must include a quarkonium,
the 4-fermion operators that appear in production cross sections involve
projections onto the space of states that contain, in the asymptotic
future, the quarkonium state $H$ plus anything else.
The generic form of a production operator is
\begin{eqnarray}
{\cal O}^H_n &=&
\chi^\dagger {\cal K}_n \psi \left( \sum_X \sum_{m_J}{| H+X \rangle}
	{\langle H+X |} \right) \psi^\dagger {\cal K}_n' \chi
\nonumber\\
&=& \chi^\dagger {\cal K}_n \psi \left( a_H^\dagger a_H \right) \psi^\dagger
{\cal K}_n' \chi,
\label{prod-op}
\end{eqnarray}
where the sums are over the $2J+1$ spin states of the quarkonium $H$ and
over all other final-state particles $X$.  In the second line of
(\ref{prod-op}), the projection has been expressed compactly in terms of
the operator $a_H^\dagger$ that creates the quarkonium $H$ in the out
state. A sum over the angular-momentum quantum numbers $m_J$ is implicit
in $a_H^\dagger a_H$.
The factors ${\cal K}_n$ and ${\cal K}_n'$ in the operator are products of a
color matrix (either the unit matrix or $T^a$), a spin matrix (either
the unit matrix or $\sigma^i$), and a polynomial in the covariant
derivative ${\bf D}$ and other fields. The overall operator ${\cal O}^H_n$
is invariant under color and spatial rotations.\footnote{Here we
consider explicitly only unpolarized production of heavy quarkonium.
In the case of polarized production, $a_H^\dagger$ would create a state
of definite polarization, and ${\cal K}_n$ and ${\cal K}_n'$ would, in general,
depend on one or more vectors associated with the incoming particles,
such as the directions of their spins and momenta.}
We assume that any
matrix elements of ${\cal O}^H_n$ will be evaluated in the quarkonium
rest frame; otherwise the factors ${\cal K}_n$ and ${\cal K}_n'$ may depend on
the
4-momentum of the quarkonium.

It is convenient to introduce notation
for the production operators that is analogous to that
for the decay operators defined in (\ref{Odim6}) and
(\ref{Odim8}).  The production operators of dimension 6 are
\begin{mathletters}
\begin{eqnarray}
{\cal O}^H_1({}^1S_0)
&=& \chi^\dagger \psi \left( a_H^\dagger a_H \right) \psi^\dagger \chi,
\\
{\cal O}^H_1({}^3S_1) &=& \chi^\dagger \sigma^i \psi \left( a_H^\dagger a_H
\right)
	\psi^\dagger \sigma^i \chi,
\\
{\cal O}^H_8({}^1S_0) &=& \chi^\dagger T^a \psi \left( a_H^\dagger a_H \right)
	\psi^\dagger T^a \chi,
\\
{\cal O}^H_8({}^3S_1)
&=& \chi^\dagger \sigma^i T^a \psi \left( a_H^\dagger a_H \right)
	\psi^\dagger \sigma^i T^a \chi.
\end{eqnarray}
\end{mathletters}
Some of the color-singlet production operators of dimension 8 are
\begin{mathletters}
\label{Odim8-prod}
\begin{eqnarray}
{\cal O}^H_1({}^1P_1)
&=& \chi^\dagger (-\mbox{$\frac{i}{2}$} \tensor{D}^i) \psi \left( a_H^\dagger
a_H \right)
	\psi^\dagger (-\mbox{$\frac{i}{2}$} \tensor{D}^i) \chi ,
\\
{\cal O}^H_1({}^3P_{0})
&=&  {1 \over 3} \; \chi^\dagger (-\mbox{$\frac{i}{2}$} \tensor{\bf D} \cdot
\mbox{\boldmath $\sigma$}) \psi
	\left( a_H^\dagger a_H \right)
	\psi^\dagger (-\mbox{$\frac{i}{2}$} \tensor{\bf D} \cdot 
	\mbox{\boldmath $\sigma$}) \chi ,
\\
{\cal O}^H_1({}^3P_{1}) &=&
	{1 \over 2} \; \chi^\dagger (-\mbox{$\frac{i}{2}$} \tensor{\bf D} \times
	\mbox{\boldmath $\sigma$})^i \psi
	\left( a_H^\dagger a_H \right)
	\psi^\dagger (-\mbox{$\frac{i}{2}$} \tensor{\bf D} \times 
	\mbox{\boldmath $\sigma$})^i \chi ,
\\
{\cal O}^H_1({}^3P_{2}) &=& \chi^\dagger (-\mbox{$\frac{i}{2}$}
\tensor{D}{}^{(i} \sigma^{j)}) \psi
	\left( a_H^\dagger a_H \right)
	\, \psi^\dagger (-\mbox{$\frac{i}{2}$} \tensor{D}{}^{(i} \sigma^{j)}) 
	\chi ,
\\
{\cal P}^H_1({}^1S_0) &=& {1\over 2}\left[
\chi^\dagger \psi \, \left( a_H^\dagger a_H \right)
	\psi^\dagger (-\mbox{$\frac{i}{2}$} \tensor{\bf D})^2 \chi \;+\; {\rm
h.c.}\right] ,
\\
{\cal P}^H_1({}^3S_1) &=&{1\over 2}\left[
\chi^\dagger \sigma^i \psi \left( a_H^\dagger a_H \right)
	\psi^\dagger \sigma^i (-\mbox{$\frac{i}{2}$} \tensor{\bf D})^2 
	\chi
\;+\; {\rm h.c.}\right] .
\end{eqnarray}
\end{mathletters}

Given that the long-distance part of the production rate can be expressed
in terms of vacuum matrix elements of operators of the form given in
(\ref{prod-op}), the inclusive production cross section must have the form
\begin{equation}
\sigma(H)
\;=\; \sum_n {F_n(\Lambda) \over M^{d_n-4}} \;
  {\langle 0 |} {\cal O}^H_n(\Lambda) {| 0 \rangle},
\label{prod-master}
\end{equation}
where it is understood that the
matrix element is to be evaluated in the quarkonium rest frame.
The short-distance coefficients $F_n$ depend on all the kinematic variables
of the production process, but they are independent of the quarkonium
state $H$.
Equation~(\ref{prod-master}) is the equivalent for production of our
factorization formula (\ref{master}) for quarkonium decay.

Beyond leading order in perturbation theory, interactions involving soft
(infrared) gluons and gluons collinear to the final-state jets potentially
spoil this factorization picture, both by making the $Q
\overline{Q}$-production
process long-ranged and by making connections between the outgoing
quarkonium and the final-state jets that destroy the topological
factorization. In the
case of quarkonium decay, we were able to use the KLN theorem to argue that
such final-state soft and collinear interactions cancel in the inclusive
decay rate.  In the case of quarkonium production, the KLN theorem does not
apply directly because we have specified that the final state contain
the quarkonium: some of the cuts in the KLN sum are missing.  Cuts are
missing only for diagrams in which a soft or collinear gluon attaches to
one of the heavy $Q$ or $\overline{Q}$ lines.  If only one end of a gluon
attaches to a $Q$ or $\overline{Q}$ line and the other end attaches to a
final-state jet, then the sum over cuts along the jet line is sufficient
by itself to effect the KLN cancellation. If both ends of a soft gluon
attach to a heavy $Q$ or $\overline{Q}$ line, then there is no KLN
cancellation.  However, this contribution is part of the matrix element
of the NRQCD 4-fermion operator.

In the case that ${\cal O}^H_n$ is a color-octet operator, one might
worry that, because the intermediate states in the first line of
(\ref{prod-op}) carry net color charge, the factorization of the cross
section in (\ref{prod-master}) is not valid.  Owing to the property of
confinement, such colored states have infinite energy. (Their energies
would be finite in a finite volume, however.)  Of course, the complete
final state is color neutral and contains only color-singlet hadrons.
One can picture the color neutralization of the partons in perturbation
theory as a process involving soft-gluon exchanges between the partons.
In particular, there can be color-neutralizing soft-gluon exchanges
between partons that are comoving with the quarkonium and partons in
other hadron jets produced by the short distance process. However, the
KLN argument tells us that, at least in perturbation theory, the
infrared and collinear divergences from such soft interactions cancel in
the inclusive quarkonium production rate. That is, for purposes of
computing the inclusive quarkonium production rate, the colored partons
can be treated as if they were unconfined. Of course, the complete
operator ${\cal O}^H_n$ is invariant under color rotations, and one can
deal with it without referring to the troublesome colored intermediate
states by making use of the form given in the second line of
(\ref{prod-op}). This approach might be useful in lattice measurements
of the production matrix elements.

If we consider production of quarkonium in hadron-induced processes,
then a host of new difficulties arise in proving that the production
rate factors. These include exchanges of soft, collinear, and Glauber
(quasi-elastic) gluons involving spectator partons in the initial state
and exchanges of soft and collinear gluons involving
active partons in the initial state.
Rather than discuss the resolution of these difficulties here,
we will merely assume that the Glauber divergences
cancel, that the only noncancelling infrared divergences
are those associated with the matrix elements of the 4-fermion operators,
and that the noncancelling collinear divergences can be absorbed into initial
parton distributions.  We refer the reader to the proofs of
factorization of the Drell-Yan cross section
\cite{drell-yan,qiu-sterman} for detailed discussions of these points.
Given these assumptions, the factored form (\ref{prod-master})
holds to all orders in perturbation theory.
It should be noted that, in the case of hadron-hadron collisions,
there is a limit to the precision of the factored form of the
cross section.  Generally, because
of soft exchanges between spectators, one can prove only that a factored form
holds through next-to-leading order in an expansion
in inverse powers of the large momentum transfer $Q^2$ \cite{qiu-sterman}.
Beyond that order, factorization is known to fail \cite{doria-frenkel-taylor}.

\subsection{Relation of Production Matrix Elements to Decay Matrix
Elements}

The NRQCD matrix elements that appear in the production rate
(\ref{prod-master}) are related to the NRQCD matrix elements that appear
in decay rates through a crossing of the quarkonium from the final state
to the initial state.  This relation is analogous to the one between
parton distribution functions and parton fragmentation functions
\cite{collins-soper}. In
general, the crossing relation is very complicated.  There are, however,
two instances in which one can obtain simple results.

Through order $\alpha_s$ in QCD perturbation theory, the crossing
relation between ${\langle H |} {\cal O}_n {| H \rangle}$ and the
corresponding production operator ${\langle 0 |} {\cal O}^H_n {| 0
\rangle}$ is a simple equality, up to a factor of $2J+1$ for the number
of spin states. Finite-order perturbation theory is usually of little
help in dealing with long-distance matrix elements.  It does tell us,
though, that, to leading order in $\alpha_s$, the evolution equations
for the production operators are the same as the evolution equations for
the corresponding decay operators. For example, the evolution equation
for the production matrix element ${\langle 0 |} {\cal
O}_8^{h_c}({}^1S_0) {| 0 \rangle}$ in terms of ${\langle 0 |} {\cal
O}_1^{h_c}({}^1P_1) {| 0 \rangle}$ is identical at leading order in
$\alpha_s$ and in $v$ to the evolution equation (\ref{ev8h}) for the
corresponding decay matrix elements: 
\begin{equation}
\Lambda {d \ \over d \Lambda} {\langle 0 |} {\cal O}_8^{h_c}({}^1S_0) {| 0
\rangle}
\;=\; {4 C_F \alpha_s(\Lambda) \over 3 N_c \pi M^2}
{\langle 0 |} {\cal O}_1^{h_c}({}^1P_1) {| 0 \rangle} .
\label{ev8h-prod}
\end{equation}

When ${\cal O}^H_n$ is a color-singlet operator, the
vacuum-saturation approximation can sometimes be used
to simplify the matrix element.  Assuming that the sum over states
in the first line of (\ref{prod-op}) is dominated by the quarkonium state
$H$ plus the vacuum, we obtain
\begin{eqnarray}
{\langle 0 |} {\cal O}^H_n {| 0 \rangle}
&\approx& {\langle 0 |} \chi^\dagger {\cal K}_n \psi
	\Biggl(\sum_{m_J}{| H \rangle} {\langle H |} \Biggr)
	\psi^\dagger {\cal K}_n' \chi {| 0 \rangle}
\nonumber \\
&=& (2J+1){\langle H |} \psi^\dagger {\cal K}_n' \chi {| 0 \rangle}
	{\langle 0 |} \chi^\dagger {\cal K}_n \psi {| H \rangle}
\nonumber \\
&\approx& (2J+1){\langle H |} {\cal O}_n {| H \rangle},
\label{vac-sat-prod}
\end{eqnarray}
where ${\cal O}_n = \psi^\dagger {\cal K}_n' \chi \chi^\dagger {\cal K}_n
\psi$. In the second
line, we have used the rotational invariance of the operator $\psi^\dagger
{\cal K}_n' \chi {| 0 \rangle} {\langle 0 |} \chi^\dagger {\cal K}_n \psi$,
which implies that the
matrix element is identical for each of the $2J+1$ angular-momentum
states $H$ that differ only in the quantum number $m_J$. In the last
line, we have used the vacuum-saturation approximation (\ref{VSA}) for
the decay matrix element ${\langle H |} {\cal O}_n {| H \rangle}$.

For the vacuum-saturation approximation to be a controlled
approximation, we must be able to show that the contributions of all the
other states in the sum in (\ref{prod-op}) are suppressed by powers of
$v$. This is in fact the case if the operator ${\cal O}_n^H$ creates and
annihilates the $Q \overline{Q}$ pair in the angular-momentum state that
corresponds to the dominant Fock state of the meson $H$.  In this case,
the vacuum-saturation approximation result (\ref{vac-sat-prod}) is
correct up to an error of relative order $v^4$.

In the case of a color-octet operator, the states ${| H+X \rangle}$
in the first line of (\ref{prod-op}) have
nonzero color, and the vacuum-saturation approximation is not
applicable.  In perturbation theory, we can approximate the sum by
retaining only the terms involving intermediate states ${| H+g \rangle}$ that
contain a single gluon. Similarly, we can approximate the sum for the
corresponding decay matrix element by retaining the terms that involve
single-gluon intermediate states ${| g \rangle}$. The resulting matrix
elements ${\langle 0 |} \chi^\dagger {\cal K}_n \psi {| H+g \rangle}$
and ${\langle g |} \chi^\dagger {\cal K}_n \psi {| H \rangle}$ are related by
crossing.
Unfortunately the crossing relation is a simple equality
only at leading order in perturbation theory.
In the absence of any rigorous relation between them,
we treat the matrix elements of the color-octet production
operators and the color-octet
decay operators as independent nonperturbative quantities.

\subsection{Computation of the Operator Coefficients}

The short-distance part of the quarkonium production rate is
insensitive to the long-distance $Q \overline{Q}$ dynamics.  Therefore,
following the same reasoning as in Section~\ref{sec:coeffs},
we can exploit the equivalence of perturbative QCD and perturbative NRQCD
at long distances as a device
to calculate the coefficients of the matrix elements
in (\ref{prod-master}).  We compute the production rate for an
on-shell $Q \overline{Q}$ pair with small relative momentum using perturbation
theory
in full QCD. Then we use perturbation theory in NRQCD to compute
the matrix elements of 4-fermion operators ${\cal O}_n^{Q \overline{Q}}$,
which are analogous to those in (\ref{prod-op}) except that the projection
is onto on-shell $Q \overline{Q}$ states.  The short-distance
coefficients are then determined by the matching condition
\begin{equation}
\sigma(Q \overline{Q})\Bigg|_{\rm pert.~QCD}
\;=\; \sum_n {F_n(\Lambda) \over M^{d_n-4}} \;
  {\langle 0 |} {\cal O}_n^{Q \overline{Q}}(\Lambda) {| 0 \rangle}\Bigg|_{\rm
pert.~NRQCD}.
\label{prod-matching}
\end{equation}
By expanding the left and right sides of (\ref{prod-matching})
as Taylor series in the
relative momentum ${\bf p}$ between the $Q$ and $\overline{Q}$,
we can identify the coefficients of the individual operators.
They correspond to the infrared- and collinear-finite parts of
cross sections for $Q \overline{Q}$~production.  One useful
way to evaluate the left side of
(\ref{prod-matching}) is to express the projection of
the product $u(P/2+p){\bar v}(P/2-p)$
of the $Q$ and $\overline{Q}$ spinors onto a particular angular
momentum state in Lorentz-invariant form.  We refer the reader to
Ref.~\cite{kuhn-guberina} for examples.  Then the left side of
(\ref{prod-matching}) can be evaluated in any convenient frame,
such as the CM frame of the overall production process.
It is understood, of course, that the matrix elements on the right side
of (\ref{prod-matching})
are to be evaluated in the rest frame of the quarkonium.

\subsection{S-wave Production}

We now apply the factorization formalism to the production of S-wave
quarkonium through relative order $v^2$. For definiteness, we use the
lowest-lying S-wave levels of charmonium for the purpose of
illustration. Of course, the results that we give generalize immediately
to other S-wave quarkonium systems. According to (\ref{prod-master}),
the cross section for the inclusive production of S-wave charmonium is 
\begin{mathletters}
\label{s-prod}
\begin{eqnarray}
\sigma(\eta_c)
&=& {F_1({}^1S_0) \over M^2} \;
	{\langle 0 |} {\cal O}_1^{\eta_c}({}^1S_0) {| 0 \rangle}
\;+\; {G_1({}^1S_0) \over M^4}
	\; {\langle 0 |} {\cal P}_1^{\eta_c}({}^1S_0) {| 0 \rangle}
\;+\; O(v^3 \sigma) ,
\label{eta-prod}
\\
\sigma(\psi)
&=& {F_1({}^3S_1) \over M^2} \; {\langle 0 |} {\cal O}_1^\psi({}^3S_1)
{| 0 \rangle}
\;+\; {G_1({}^3S_1) \over M^4} {\langle 0 |} {\cal P}_1^\psi({}^3S_1)
{| 0 \rangle}
\;+\; O(v^3 \sigma).
\label{psi-prod}
\end{eqnarray}
\end{mathletters}
The vacuum-saturation approximation (\ref{vac-sat-prod}) can be used to
reduce the 4-fermion matrix elements to products of matrix elements
between the vacuum and the quarkonium state.  These can, in turn, be
related to the quarkonium wavefunctions given in
Section~\ref{sec:wave-fn}.  Finally, heavy-quark spin symmetry can be
used to reduce the matrix elements to the same three nonperturbative
parameters that appear in charmonium decay: $|{\overline
{R_{\eta_c}}}|^2$, $|{\overline {R_\psi}}|^2$, and ${\rm Re}({\overline
{R_S}}{}^*\, {\overline {\nabla^2 R_S}})$. Taking into account factors
of $2J+1$ for the number of spin states, we find that the cross sections
are 
\begin{mathletters}
\label{s-prodR}
\begin{eqnarray}
\sigma(\eta_c)
&=& {N_c \; F_1({}^1S_0) \over 2\pi M^2 } \;
	\Big| {\overline {R_{\eta_c}}} \Big|^2
\;-\; {N_c \; G_1({}^1S_0) \over 2\pi M^4} \; {\rm Re}({\overline {R_S}}{}^*\,
{\overline {\nabla^2 R_S}})
\;+\; O(v^3 \sigma) ,
\label{eta-prodR}
\\
\sigma(\psi)
&=& {3 N_c \; F_1({}^3S_1) \over 2\pi M^2} \; \Big| {\overline {R_\psi}}
\Big|^2
\;-\; {3N_c \; G_1({}^3S_1) \over 2\pi M^4}
	\; {\rm Re}({\overline {R_S}}{}^*\, {\overline {\nabla^2 R_S}})
\;+\; O(v^3 \sigma).
\label{psi-prodR}
\end{eqnarray}
\end{mathletters}

If we require only accuracy to leading order in $v$, then we can
simplify the production rates in (\ref{s-prodR}) further by dropping the
terms proportional to ${\rm Re}({\overline {R_S}}{}^*\, {\overline
{\nabla^2 R_S}})$ and replacing ${\overline {R_{\eta_c}}}$ and
${\overline {R_\psi}}$ by their weighted average ${\overline {R_S}}$. We
then recover the familiar factorization formulas used in most previous
work: 
\begin{mathletters}
\label{s-prodS}
\begin{eqnarray}
\sigma(\eta_c)
&=& {N_c \; F_1({}^1S_0) \over 2\pi M^2} \; \Big| {\overline {R_S}} \Big|^2
\;+\; O(v^2 \sigma) ,
\label{eta-prodS}
\\
\sigma(\psi)
&=& {3 N_c \; F_1({}^3S_1) \over 2\pi M^2} \; \Big| {\overline {R_S}} \Big|^2
\;+\; O(v^2 \sigma).
\label{psi-prodS}
\end{eqnarray}
\end{mathletters}

In applying the factorization formula (\ref{prod-master}), one should
keep in mind that the short-distance coefficients $F_n(\Lambda)$ depend
not only on $\alpha_s(M)$ but also on dimensionless ratios of kinematic
variables. For example, in the case of production of heavy quarkonium at
large transverse momentum $p_T$, the coefficients  $F_n(\Lambda)$ depend
 strongly on $p_T^2/M^2$. In determining the relative importance of the
various terms in (\ref{prod-master}), one must take into account not
only the size of the matrix element and the leading power of
$\alpha_s(M)$ in the short-distance coefficient, but also the dependence
of $F_n(\Lambda)$ on dimensionless ratios of kinematic variables. The
terms given explicitly in (\ref{s-prod}) may not be the dominant
contributions to the cross sections if the coefficients of the matrix
elements are sufficiently suppressed relative to the coefficients of
other matrix elements that are of higher order in $v$. 

\subsection{P-wave Production}

We next apply the factorization formalism to the production of P-wave
quarkonium to leading order in $v$, using the lowest-lying P-wave levels
of charmonium for the purpose of illustration. According to our
factorization formula (\ref{prod-master}), the inclusive production
rates for P-wave charmonium are 
\begin{mathletters}
\label{p-prod}
\begin{eqnarray}
\sigma(h_c)
&=& {F_1({}^1P_1) \over M^4} \;
	{\langle 0 |} {\cal O}_1^{h_c}({}^1P_1) {| 0 \rangle}
\;+\; {F_8({}^1S_0) \over M^2} \;
	{\langle 0 |} {\cal O}_8^{h_c}({}^1S_0) {| 0 \rangle}
\;+\; O(v^2 \sigma) ,
\label{h-prod}
\\
\sigma(\chi_{cJ})
&=& {F_1({}^3P_{J}) \over M^4} \;
	{\langle 0 |} {\cal O}_1^{\chi_{cJ}}({}^3P_{J}) {| 0 \rangle}
\;+\; {F_8({}^3S_1) \over M^2} \;
	{\langle 0 |} {\cal O}_8^{\chi_{cJ}}({}^3S_1) {| 0 \rangle}
\nonumber \\
&&  \quad \quad \quad \quad \quad \quad \quad \quad \quad \quad
\;+\; O(v^2 \sigma), \quad J = 0,1,2.
\label{chi-prod}
\end{eqnarray}
\end{mathletters}
The vacuum-saturation approximation (\ref{vac-sat-prod}) can be applied
to the color-singlet matrix elements to express them in terms of
vacuum-to-quarkonium matrix elements. These matrix elements can be
expressed in terms of regularized derivatives of radial wavefunctions at
the origin by using (\ref{Rhchi}). Because of heavy-quark spin symmetry,
they can all be replaced, without loss of accuracy, by their weighted
average ${\overline {R_P'}}$. Heavy-quark spin symmetry also implies
that the color-octet matrix elements in (\ref{p-prod}) are proportional
to $2J+1$, up to corrections of relative order $v^2$. Thus, the P-wave
charmonium production rates can all be expressed in terms of the two
nonperturbative parameters $|{\overline {R_P'}}|^2$ and ${\langle 0 |}
{\cal O}_8^{h_c}({}^1S_0) {| 0 \rangle}$ (or, alternatively, the average
over the P-wave states of $3/(2J+1)$ times the color-octet matrix
elements): 
\begin{mathletters}
\label{P-prodR}
\begin{eqnarray}
\sigma(h_c)
&=& {9 N_c \; F_1({}^1P_1) \over 2\pi M^4} \; \Big| {\overline {R_P'}} \Big|^2
\;+\; {F_8({}^1S_0) \over M^2} \;
{\langle 0 |} {\cal O}_8^{h_c}({}^1S_0) {| 0 \rangle}
\;+\; O(v^2 \sigma) ,
\label{h-prodR}
\\
\sigma(\chi_{cJ})
&=& {(2J+1) 3 N_c \; F_1({}^3P_{J}) \over 2\pi M^4} \; \Big| {\overline {R_P'}}
\Big|^2
\;+\; {(2J+1) F_8({}^3S_1) \over 3 M^2} \; {\langle 0 |} {\cal
O}_8^{h_c}({}^1S_0)
{| 0 \rangle}
\nonumber \\
&& \quad \quad \quad \quad \quad \quad \quad \quad \quad \quad
\;+\; O(v^2 \sigma),\quad J=0,1,2.
\label{chi-prodR}
\end{eqnarray}
\end{mathletters}
Note that the color-octet matrix element
${\langle 0 |} {\cal O}_8^{h_c}({}^1S_0) {| 0 \rangle}$ in (\ref{P-prodR})
cannot be identified with
the decay matrix element ${\langle h_c |} {\cal O}_8({}^1S_0) {| h_c \rangle}$
in
(\ref{HPwave}).

The first application of the factorization formulas (\ref{P-prodR})
was to the inclusive production of P-wave charmonium states
in $B$-meson decay \cite{bbly}.
The factorization formulas were given in the form
\begin{mathletters}
\label{bbl-bmesonx}
\begin{eqnarray}
\Gamma \left( b \rightarrow h_c + X \right) \; &=& \;
H_1 \; {\widehat \Gamma}_1
\left( b \rightarrow c {\bar c} (^1P_1) + X, \mu\right)\nonumber\\
&&\qquad +\;  3 \; H_8'(\mu) \; {\widehat \Gamma}_8
\left( b \rightarrow c {\bar c} (^1S_0) + X \right) \; ,
\label{bbl-bmesonh}\\
\Gamma \left( b \rightarrow \chi_{cJ} + X \right) \; &=& \;
H_1 \; {\widehat \Gamma}_1
\left( b \rightarrow c {\bar c} (^3P_J) + X, \mu \right) \;\nonumber\\
&&\qquad + \;  (2J + 1) \; H_8'(\mu) \; {\widehat \Gamma}_8
\left( b \rightarrow c {\bar c} (^3S_1) + X \right) \; .
\label{bbl-bmesonchi}
\end{eqnarray}
\end{mathletters}
The coefficients ${\widehat \Gamma}_1$ and ${\widehat \Gamma}_8$ are
proportional to the production rates for on-shell $Q \overline{Q}$ pairs in
color-singlet P-wave and color-octet S-wave states, respectively.
The factors $H_1$ and $H'_8$ can
be expressed in terms of NRQCD matrix elements
divided by appropriate factors of the heavy-quark mass:
\begin{mathletters}
\label{H-prod}
\begin{eqnarray}
H_1 &=& {1 \over 3 M^4} \;
{\langle 0 |} {\cal O}^{h_c}_1({}^1P_1) {| 0 \rangle},
\label{H1-prod}
\\
H'_8(\Lambda) &=& {1 \over 3 M^2} \;
{\langle 0 |} {\cal O}^{h_c}_8({}^1S_0) {| 0 \rangle}.
\label{H8-prod}
\end{eqnarray}
\end{mathletters}
The definitions (\ref{H1-prod}) and (\ref{H8-prod}) were chosen in
Ref.~\cite{bbly} so that $H_1$ and $H'_8$ would coincide as closely as
possible with the decay matrix elements $H_1$ and $H_8$. Using the
vacuum-saturation approximation (\ref{vac-sat-prod}), we see that the
definition of $H_1$ given in (\ref{H1-prod}) is equal to that given in
(\ref{H1}), up to corrections of relative order $v^4$. A crude estimate
for $H_8(M)$ in terms of $H_1$ is given in (\ref{evesth}).  A similar
estimate of $H_8'(M)$ in terms of $H_1$ can be obtained by solving the
evolution equation (\ref{ev8h-prod}) and assuming that ${\langle 0 |}
{\cal O}_8^{h_c}({}^1S_0;\Lambda) {| 0 \rangle}$ can be neglected at
some initial scale $\Lambda = \Lambda_0$.  With the normalizations in
(\ref{H8}) and (\ref{H8-prod}), the resulting estimates for $H_8(M)$ and
$H_8'(M)$ are equal. However, there is no apparent rigorous relation
between these two matrix elements. 

As we have already remarked
in connection with the decay matrix elements, the factors of $1/M$ in
(\ref{H1-prod}) and (\ref{H8-prod}) are more properly associated with
the operator coefficients, since they involve short-distance physics at
distance scales of order $1/\Lambda$ or less. Therefore, the
factorization formulas (\ref{P-prodR}) are preferable to the forms given
in (\ref{bbl-bmesonx}).

In Ref.~\cite{bbly}, which discusses the decay of a B~meson into a
charmonium state, the NRQCD cutoff $\Lambda$ was set equal to the
scale of the large momentum transfer in the process, which is the
bottom-quark mass $m_b$. This choice of cutoff is inappropriate because the
NRQCD evolution equation (\ref{ev8h-prod}) accurately reflects the
behavior of full QCD only for cutoffs $\Lambda$ that are less than $M$.
That is, the NRQCD evolution equation cannot be used to sum logarithms of
$Q^2/M^2$, where $Q^2$ is the large momentum transfer in a production
process.  Therefore, a more appropriate choice of NRQCD cutoff for the
process analyzed in \cite{bbly} is $\Lambda=m_c$, where $m_c=M$ is the
charmed-quark mass.  Note, however, that a change of NRQCD cutoff from
$m_b$ to $m_c$ does not affect the short-distance coefficients in the
leading-order calculation presented in \cite{bbly}, and is, in general,
insignificant numerically.

\vfill \eject

\section{Discussion and Outlook}

The factorization approach that we have developed in this paper
provides a systematic theoretical framework for
understanding the annihilation and production of heavy quarkonium.
In this section, we discuss the relation between our
approach and previous models for quarkonium production and annihilation.
We also summarize the current status of theoretical calculations
of annihilation rates and production cross sections.

\subsection{Comparison with Previous Approaches}

We have presented a rigorous formalism for calculating the
inclusive annihilation rates of heavy quarkonia.
It is based on the use of NRQCD to separate the annihilation rate
into short-distance parts,
involving distance scales on the order of $1/M$, and long-distance parts.
The short-distance parts are identified
with the imaginary parts of coefficients in the NRQCD lagrangian,
and can be computed as perturbation expansions in $\alpha_s(M)$.
The long-distance parts are expressed as matrix elements of 4-fermion
operators in NRQCD  and can be computed nonperturbatively by using
lattice simulations.  We have also developed an analogous formalism
for computing inclusive production rates of heavy quarkonia in processes
involving large momentum transfers.  The cross sections are factored
into short-distance parts, which can be computed perturbatively, and
long-distance parts, which are expressed as NRQCD matrix elements.

The factorization approach provides a firm
theoretical foundation for calculations
of the annihilation and production rates for heavy quarkonium.
It can be used to assess the degree of validity and the limitations
of models used in previous work on heavy quarkonium production
and annihilation.
The most thoroughly developed model for the calculation of production
rates is the ``color-singlet model'' \cite{csm1,csm2,csm3,csm4}.
Most calculations of  annihilation rates have also been carried out
within this model.  In the color-singlet model,
the quarkonium state is modeled by a color-singlet $Q \overline{Q}$ pair
that is in the appropriate angular-momentum state
and has vanishing relative momentum.  Nonperturbative effects
are assumed to factor into a single nonperturbative quantity that is
proportional to the square of the radial wavefunction, or one of its
derivatives, evaluated at the origin.

The factorization formalism represents a significant advance over the
color-singlet model in several respects. First, it provides a systematic
framework for calculating perturbative corrections to the short-distance
factors to arbitrarily high orders in $\alpha_s$. The infrared
divergences that are encountered at any order of the perturbation
expansion can be factored into specific nonperturbative matrix elements.
Perturbative calculations in the color-singlet model are based on the
assumption that all infrared divergences can be factored into a single
nonperturbative quantity.  In the case of S-waves, calculations at
next-to-leading order in $\alpha_s$ (NLO) provide empirical support for
the assumption that long-distance effects can be factored into the
quantity $|R_S(0)|^2$. Our formalism reveals that this assumption is, in
fact, correct for any specific S-wave process in the nonrelativistic
limit to all orders in $\alpha_s$. It has often been assumed, in
addition, that the same quantity $|R_S(0)|^2$ describes processes
involving both the $0^{-+}$ and $1^{--}$ S-wave states. Our formalism
shows that this additional assumption is correct only up to corrections
of relative order $v^2$. The assumption that the same quantity
$|R_S(0)|^2$ describes annihilation into light hadrons and
electromagnetic annihilation also fails at relative order $v^2$, as does
the assumption that the same quantity $|R_S(0)|^2$ describes both
annihilation and production processes. In the case of P-waves, explicit
calculations of the decay rates into light hadrons reveal that the
assumption of a single long-distance factor $|R_P'(0)|^2$ fails at
leading order in $\alpha_s$ (LO) for $h_c$ and $\chi_{c1}$ \cite{barbc}
and at NLO for $\chi_{c0}$ and $\chi_{c2}$ \cite{barbe}. In the context
of our formalism, these results follow simply from the existence of a
second independent matrix element that contributes to the annihilation
rates of P-wave quarkonia in the nonrelativistic limit. 

The factorization formalism also improves upon the color-singlet model
by allowing the systematic calculation of relativistic corrections to
annihilation and production rates.  Relativistic corrections are
incorporated by including nonperturbative matrix elements that scale as
higher powers of $v$.  In the case of S-waves, our formalism for
computing the $v^2$ corrections is similar at leading order in
$\alpha_s$ to a model for relativistic corrections developed by Keung
and Muzinich \cite{keung}.  The major differences are that the
factorization formalism provides nonperturbative definitions for the
long-distance factors, it allows the short-distance coefficients to be
calculated beyond leading order in $\alpha_s$, and it can be used to
treat corrections of order $v^3$ and higher. 

Another advantage of the factorization formalism is that it provides
unambiguous field-theoretic definitions of the long-distance factors in
annihilation and production rates.  This allows one to compute them
nonperturbatively using, for example, lattice simulations of NRQCD.
Previous approaches have relied either on determining the long-distance
factors phenomenologically or on relating them to potential-model
wavefunctions.  Both of these approaches are of limited utility. The
purely phenomenological approach can be applied only in situations in
which the number of accurately-measured experimental observables is
greater than the number of nonperturbative matrix elements.
Potential-model estimates can be used for color-singlet matrix elements
that have simple potential-model analogs, but they cannot be used for
other matrix elements, such as the color-octet matrix elements that
contribute to the annihilation of P-wave states into light hadrons at
leading order in $v$. It is also difficult to gauge the accuracy of
potential-model estimates in the absence of a rigorous connection to
QCD. Since our formalism provides unambiguous definitions of the
long-distance factors in annihilation and production processes, it
allows us to quantify relations between these matrix elements and
Coulomb-gauge wavefunctions in NRQCD.  It also allows us to quantify the
differences between matrix elements for decays into light hadrons and
matrix elements for decays into electromagnetic final states, as well as
the differences between annihilation matrix elements and production
matrix elements. 

A final advantage of the factorization formalism is that it takes into
account the complete Fock-space structure of the quarkonium. In the
color-singlet model, the quarkonium is assumed to be simply a $Q
\overline{Q}$ pair in a color-singlet state with definite
angular-momentum quantum numbers ${}^{2S+1}L_J$. However, a quarkonium
also has a probability of order $v^2$ to be in a $Q \overline{Q} g$ Fock
state, and it has probabilities of order $v^4$ or smaller for the higher
Fock states. In the case of P-waves, the factorization formalism reveals
that the $Q \overline{Q} g$ component can play just as important a role
in annihilation and in production as the dominant $Q \overline{Q}$
component. In the case of S-waves, the higher Fock states can be ignored
in the nonrelativistic limit, and even at relative order $v^2$, but the
factorization formalism indicates that they do contribute at relative
order $v^3$. 

The factorization formalism for describing the annihilation of heavy
quarkonia is in many ways similar to the operator-product-expansion
formalism for calculating the inclusive decay rates of heavy-light
mesons \cite{bigi}.  These decay rates can be factored into short
distance parts, which involve the weak decay of a heavy quark or its
weak annihilation with an antiquark in the meson, and long-distance
parts, which can be expressed as NRQCD matrix elements.  The main
difference between heavy-quarkonium annihilation and heavy-light meson
decay is in the
relative importance of the various matrix elements.  Since the typical
momentum of a heavy quark in a  heavy-light meson is of order
$\Lambda_{QCD}$ and is independent of $M$,
the relative importance of matrix elements is determined strictly by the
dimension of the operator.

Operator-product-expansion methods have also been used to treat
exclusive decays of heavy quarkonium into light hadrons at leading order
in $v$ \cite{dm}.  The NRQCD formalism might prove to be useful in
extending such analyses to include relativistic corrections. In
exclusive processes, a factorization theorem holds, not only for the
decay rate, but also for the decay amplitude. Thus, just as in the case
of electromagnetic annihilation,  the relevant NRQCD matrix elements for
exclusive decays are vacuum-to-quarkonium matrix elements of
color-singlet operators of the form $\chi^\dagger {\cal K}_n \psi$. 

Operator-product-expansion methods have also been used in a completely
different context in  heavy quarkonium physics \cite{itep,peskin}. These
methods have been used to treat the interactions of heavy quarkonium
with light hadrons whose momenta are small compared to the scale $Mv$ of
quarkonium structure. Voloshin \cite{voloshin} has used this approach to
calculate nonperturbative corrections to quarkonium annihilation rates
that are proportional to the gluon condensate. In our factorization
formula, the gluon-condensate contribution would appear in the
long-distance matrix element.  In some of the cases considered by
Voloshin, the corresponding short-distance part involves the
annihilation of the $Q \overline{Q}$ pair in a color-octet state. His approach
can
therefore be used as a framework for estimating the matrix elements of
color-octet operators.

The general factorization formula (\ref{prod-master}) for the
production cross section of any specific quarkonium state $H$
takes into account the short-distance
production of color-singlet $Q \overline{Q}$ pairs and color-octet $Q
\overline{Q}$ pairs
in all angular-momentum states.
In this respect, our approach has some elements in common with the
``color-evaporation model'' for quarkonium production \cite{frizsch}.
In this model, the total inclusive cross section, summed over all
quarkonium states $H$, is obtained by integrating the perturbative cross
section for inclusive $Q \overline{Q}$ production from the quark threshold $2
M$ up
to the physical threshold for the production of a pair of heavy-light
mesons. No constraints are imposed on the color and angular momentum
states of the $Q \overline{Q}$ pair.  Under the hypothesis of ``semilocal
duality'', the nonperturbative QCD effects that are responsible for the
formation of a color-singlet bound state containing the $Q \overline{Q}$ pair
are
assumed to be negligible after one sums over all quarkonium states $H$.
In the factorization approach, the nonperturbative effects are not
neglected, but are factored into long-distance matrix elements ${\langle 0 |}
{\cal O}^H_n {| 0 \rangle}$. In the color-evaporation model, the production
cross section for a specific quarkonium state $H$ is obtained by
multiplying the total quarkonium cross section by a purely
phenomenological fraction $f_H$. The relative production rates of
different quarkonium states are, therefore, not predicted.  In the
factorization approach, the relative production rates can be calculated
by using perturbative QCD, once the values of the dominant matrix
elements ${\langle 0 |} {\cal O}^H_n {| 0 \rangle}$ have been  determined.

\subsection{Present Status of Calculations}

The possible applications of the factorization formalism
for heavy-quarkonium annihilation and production are almost limitless,
since heavy quarkonia play a role in so many high energy processes.
In order to highlight some of these applications,
we discuss below the present status of calculations of
annihilation and production rates.

In the case of S-wave decays, NLO perturbative corrections have been
calculated for all the annihilation rates.  In many cases, the NLO
corrections are uncomfortably large.  In order to develop a better
understanding of the origin of these large corrections, it would be
desirable to have calculations at NNLO, at least for the simplest
processes $\psi \to e^+ e^-$ and $\eta_c \to \gamma \gamma$.
Relativistic corrections to the S-wave annihilation rates have been
studied by Keung and Muzinich \cite{keung}.  From their results, one can
extract the coefficients of all the matrix elements of relative order
$v^2$ at leading order in $\alpha_s$. A phenomenological analysis of the
decay rates of the lowest-lying S-wave states of charmonium, including
the next-to-leading order corrections in $\alpha_s(M)$ and the
corrections of relative order $v^2$, is in progress \cite{bblz}. 

In the case of P-wave decays, complete NLO perturbative
corrections are available only for the electromagnetic decays
$\chi_{c0} \to \gamma \gamma$ and $\chi_{c2} \to \gamma \gamma$.
For the decays of P-wave states into light hadrons,
complete results are known only to order $\alpha_s^2$ \cite{bbl}.
The coefficients of $|{\overline {R_P'}}|^2$ have been calculated to order
$\alpha_s^3$
\cite{barbe}, but they  contain logarithmic infrared
divergences that should be factored into matrix elements of the
operators ${\cal O}_8({}^1S_0)$ and ${\cal O}_8({}^3S_1)$.   There
are constants under the logarithms that should also be factored into the
matrix elements.  Unfortunately, these constants cannot be determined
readily from the existing calculations.  The relativistic corrections
to P-wave annihilation rates have not yet been analyzed.

In the case of D-wave decays, the only complete LO calculations
are those for the electromagnetic decays of the $^3D_1$ state into
$e^+ e^-$ and the $^1D_2$ state into $\gamma \gamma$ \cite{itep}.
For the decay of the $^1D_2$ state into light hadrons,
the coefficient of the matrix element corresponding to
$|R_D''(0)|^2$ has been calculated at LO \cite{itep}.
For the decays of the $^3D_J$ states into light hadrons,
only the logarithmic infrared divergence in the coefficient of
$|R_D''(0)|^2$ has been extracted \cite{bm}.  This divergence should
be factored into other matrix elements that contribute to the
annihilation rate in the nonrelativistic limit.  These matrix elements
can be identified by using the methods of Section~\ref{sec:vcounting},
and their coefficients can be calculated by using the
methods illustrated in Appendix A.

The status of calculations of the production of heavy quarkonium has
been reviewed recently in Ref. \cite{schuler}, although many aspects of
that review are superceded by the developments described in the present
paper. In the case of S-waves, most production processes have been
computed only to LO. The only processes for which complete NLO
calculations are available are $\psi$ and $\eta_c$ production in
$B$-meson decay \cite{be} and inclusive $\eta_c$ production in hadron
collisions \cite{kuhn,schuler}. It would be desirable to have
calculations of complete NLO corrections for more production processes,
in order to develop a better understanding of the size and behavior of
the perturbative corrections. It is also important to calculate the
relativistic corrections, which are expected to be typically on the
order of 30\% for charmonium. Relativistic corrections have been
calculated for the photoproduction of the $\psi$ \cite{jkgw} within
the model of Keung and Muzinich \cite{keung}.
The factorization formalism can be used to express those results in
terms of well-defined NRQCD matrix elements.

For the production of quarkonia at large transverse momentum $p_T$,
the contributions that are leading in $1/p_T$ sometimes come from
beyond leading order in the perturbation expansion, and they can
be computed without complete calculations of the NLO or NNLO corrections.
These contributions come from fragmentation and
can be expressed in terms of process-independent
fragmentation functions $D_{i \to H}(z,\mu)$ for a parton $i$ with
invariant mass $\mu$  to produce a jet containing the quarkonium $H$
with light-cone momentum fraction $z$.  The fragmentation contribution
to a production cross section sometimes appears in a LO calculation, but it
often appears first at NLO and sometimes even at NNLO.  The fragmentation
functions for producing S-wave quarkonia from the fragmentation of gluons
\cite{gfrag} and heavy quarks \cite{Qfrag}
have been calculated at LO in $\alpha_s$.

For P-wave quarkonia, there are many production processes for which
complete calculations are not even available at LO. For most processes,
the coefficient of $|{\overline {R_P'}}|^2$ has been calculated \cite{schuler}.
Complete LO calculations, including the coefficient of ${\langle 0 |} {\cal
O}_8^H {| 0 \rangle}$, are available only for the production of P-wave
charmonium in $B$-meson decays \cite{bbly}, $\Upsilon$ decays
\cite{trottier}, gluon fragmentation \cite{gfragP}, and charm
fragmentation \cite{QfragP}. Relativistic corrections to P-wave
production processes have not been studied.

In the case of production of $D$-wave quarkonia,
the only perturbative calculations that are available are those for the
coefficients of $|R_D''(0)|^2$ in the decay rates for $Z^0 \to H \gamma$,
where $H$ represents any of the $^1D_2$ or $^3D_J$ states \cite{bgr}.

In the factorization approach, nonperturbative long-distance effects are
organized systematically into well-defined NRQCD matrix elements. This
allows one to go beyond potential model estimates or phenomenological
determinations of the long-distance factors. Instead, they can be
calculated from first principles using lattice simulations of NRQCD.
Such calculations are still in their infancy. At present, the only
vacuum-to-quarkonium matrix elements that have been calculated are 
${\langle 0 |} \chi^\dagger \psi {| \eta_c \rangle}$, ${\langle 0 |}
\chi^\dagger \mbox{\boldmath $\sigma$} \psi {| \psi \rangle}$, and
${\langle 0 |} \chi^\dagger \tensor{\bf D} \chi {| h_c \rangle}$, and
their analogs for the bottomonium system \cite{tl,bks}. The only
4-fermion matrix elements that have been calculated thus far are 
${\langle \eta_c |} {\cal O}_1({}^1S_0) {| \eta_c \rangle}$, ${\langle
h_c |} {\cal O}_1({}^1P_1) {| h_c \rangle}$, and ${\langle h_c |} {\cal
O}_8({}^1S_0) {| h_c \rangle}$ and their analogs for the bottomonium
system \cite{bks}.  Thus far, all matrix elements have been calculated
only up to corrections of relative order $v^2$ and in the absence of
dynamical light quarks. Production matrix elements are much more
difficult to calculate through lattice simulations, unless they can be
related to annihilation matrix elements through the vacuum-saturation
approximation. 

\subsection{Concluding Remarks}

Heavy-quark mesons have long been the best understood of hadrons. Until
recently, our understanding has been based almost exclusively on
phenomenological quark potential models that are motivated by QCD.
Now, lattice QCD simulations are providing
systematic analyses that are based directly upon the QCD lagrangian
\cite{elkhadra-davies}.
Heavy-quark systems are particularly well-suited to lattice simulations,
and, consequently, they are now of central importance to our exploration
of nonperturbative QCD. This new role for quarkonium studies, as a
rigorous testing ground for nonperturbative QCD, demands a much higher
degree of rigor than was necessary in older phenomenological analyses.
Approximations are necessary in tackling most hard problems, but it is
essential in a fundamental analysis that there be systematic procedures
for improving the approximations. In this paper, we have developed a
formalism for studying annihilation decays of heavy-quark mesons that
meets this standard. With our formalism, we can improve upon the
nonrelativistic quark potential model by including relativistic
corrections in a systematic way. We can also go beyond the quark model
to include the dynamical effects of gluons.  Thus, we can, for the first
time, begin to confront the full richness of nonperturbative QCD in
analyses that are systematic and rigorous.

\acknowledgements

One of us (G.T.B.) would like to thank D. Sinclair for numerous useful
discussions.  Another one of us (E.B.) would like to express his
appreciation to the Fermilab Theory Group for its hospitality while this
paper was being written. This work was supported in part by the U.S.
Department of Energy, Division of High Energy Physics, under Contract
W-31-109-ENG-38 and under Grant DE-FG02-91-ER40684, and by the National
Science Foundation.

\vfill\eject

%%%%%%%%%%%%%%%%%%%% appendices %%%%%%%%%%%%%%%%%%%%%%%%%%%%%%%%%%%%%%%%%%

\appendix
\section{Coefficients of 4-fermion operators}
\label{app:coeffs}

The coefficients in the lagrangian for nonrelativistic QCD can be
determined by matching scattering amplitudes in NRQCD with those
in full QCD \cite{lmnmh}.
In this Appendix, we use these techniques to determine the coefficients
for some of the 4-fermion operators that contribute to
quarkonium annihilation rates.
In Section~\ref{app:alpha}, we illustrate the method by
calculating the coefficients of dimension-6 operators
to order $\alpha_s$.  In Section~\ref{app:alphasq},
we apply the method to calculate the
imaginary parts of the coefficients of dimension-6 and dimension-8
operators to order $\alpha_s^2$.  In Section~\ref{app:higher},
we demonstrate how the imaginary parts of some of the
coefficients  can be extracted at next-to-leading order from existing
calculations of the decay rates of bound states.
We also record coefficients that can
be extracted from existing calculations in the literature.
Finally, in Section~\ref{app:electro}, we give the corresponding
coefficients for electromagnetic annihilation rates.

\subsection{Coefficients at Order $\alpha_s$}
\label{app:alpha}

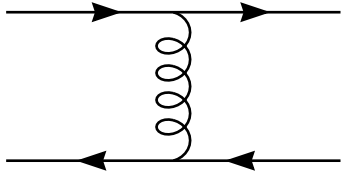
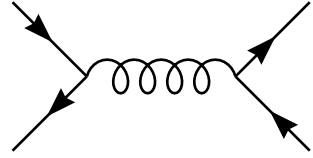
\begin{figure}
\caption{Feynman diagrams for $Q \overline{Q}$ scattering at leading order in
$\alpha_s$.}
\label{fig:qq-leading}
\end{figure}

We wish to determine the coefficients of the dimension-6 and dimension-8
4-fermion operators at order $\alpha_s$ by using the matching condition
(\ref{matching}).  We consider $Q \overline{Q}$ scattering amplitudes, with the
momenta of the heavy quarks and antiquarks small compared to the heavy
quark mass $M$. In full QCD, there are two Feynman diagrams for $Q
\overline{Q}$ scattering at tree level.  The gluon exchange diagram in
Fig.~\ref{fig:qq-leading}(a) is also present in NRQCD.  The annihilation
diagram in Fig.~\ref{fig:qq-leading}(b) is not present in NRQCD, so its
effects must be reproduced by adding 4-fermion terms to the effective
lagrangian. We calculate the annihilation contribution to the amplitude
for $Q \overline{Q}$ scattering in the center of momentum frame.  We take the
incoming $Q$ and $\overline{Q}$ to have momenta ${\bf p}$ and $-{\bf p}$, while
the
outgoing $Q$ and $\overline{Q}$ have momenta ${\bf p}'$ and $-{\bf p}'$. By
conservation of energy, we have $|{\bf p}'| = |{\bf p}| \equiv p$.

The scattering amplitude (T-matrix element) in full QCD corresponding
to the diagram in Fig.~\ref{fig:qq-leading}(b) is
\begin{equation}
{\cal M}_{\rm 6(b)} \;=\; {\pi \alpha_s \over E^2}
	\; \bar u({\bf p}') \gamma^\mu T^a v(- {\bf p}')
	\; \bar v(- {\bf p}) \gamma_\mu T^a u({\bf p}),
\label{Mtree}
\end{equation}
where $E = \sqrt{M^2 + p^2}$.  We have suppressed the color indices on
the 4-component Dirac spinors.
Following Ref. \cite{lmnmh}, we express the 4-component Dirac spinors
in the Dirac representation in terms of
2-component Pauli spinors via the substitutions
\begin{mathletters}
\label{spinor}
\begin{eqnarray}
u({\bf p}) &=& \sqrt{E+M \over 2E}
\left( \begin{array}{c} \xi \\
	{{\bf p} \cdot \mbox{\boldmath $\sigma$} \over E+M} \xi \end{array} \right) ,
\label{uspinor}
\\
v(-{\bf p}) &=& \sqrt{E+M \over 2E}
\left( \begin{array}{c} {(-{\bf p}) \cdot \mbox{\boldmath $\sigma$} \over E+M}
\eta
		\\ \eta  \end{array} \right) ,
\label{vspinor}
\end{eqnarray}
\end{mathletters}
where $\xi$ and $\eta$ are 2-component spinors with suppressed color
indices.  The Dirac spinors
$u({\bf p}')$ and $v(-{\bf p}')$ have similar expressions in terms
of Pauli spinors $\xi'$ and $\eta'$.   The spinors (\ref{uspinor}) and
(\ref{vspinor}) represent fermion states with the standard
nonrelativistic normalization.  Expanding to
second order in the velocity $v = p/E$, we find that the
annihilation contribution to the scattering amplitude
(\ref{Mtree}) from full QCD reduces to
\begin{equation}
{\cal M}_{\rm 6(b)} \;=\; - {\pi \alpha_s \over M^2} \; \Bigg( (1 - v^2)
	{\xi'}^\dagger \mbox{\boldmath $\sigma$} T^a \eta' \cdot \eta^\dagger
\mbox{\boldmath $\sigma$} T^a \xi
	\;-\; {1 \over 2} \; (v^i v^j + v'^i v'^j)
		{\xi'}^\dagger \sigma^i T^a \eta' \;
		\eta^\dagger \sigma^j T^a \xi \Bigg),
\label{MtreeNR}
\end{equation}
where ${\bf v} = {\bf p}/E$ and ${\bf v}' = {\bf p}'/E$.
It is convenient to suppress the spinors and write the above matrix
element as a direct product of color matrices multiplied by a direct
product of spin matrices:
\begin{equation}
{\cal M}_{\rm 6(b)} \;=\; - {\pi \alpha_s \over M^2} \; (T^a \otimes T^a) \;
\left[ (1 - v^2) \sigma^i \otimes \sigma^i
	\;-\; {1 \over 2} (v^i v^j + v'^i v'^j)
		\sigma^i \otimes \sigma^j \right].
\label{Mmatch}
\end{equation}

One can read off the dimension-6 term in the scattering amplitude
in terms of the parameters of NRQCD
by substituting $\xi$, ${\xi'}^\dagger$, $\eta'$, and $\eta^\dagger$
for $\psi$, $\psi^\dagger$, $\chi$, and $\chi^\dagger$
in the effective lagrangian (\ref{Lcontact6}):
\begin{eqnarray}
{\cal M}_{d=6} &=&
{1 \over M^2} \; (1 \otimes 1) \; \left[ f_1({}^1S_0) \; 1 \otimes 1
	\;+\; f_1({}^3S_1) \; \sigma^i \otimes \sigma^i \right] \nonumber \\
&& \;+\; {1 \over M^2} \; (T^a \otimes T^a) \left[ f_8({}^1S_0) \; 1 \otimes 1
	\;+\; f_8({}^3S_1) \; \sigma^i \otimes \sigma^i \right].
\label{M6}
\end{eqnarray}
Comparing (\ref{Mmatch}) and (\ref{M6}), we find that only one of the
four terms in (\ref{Lcontact6}) has a nonvanishing coefficient at order
$\alpha_s$:
\begin{equation}
f_8({}^3S_1) \;=\; - \pi \alpha_s(M) .
\label{f8S}
\end{equation}
The coefficients of the remaining three terms in (\ref{Lcontact6})
are of order $\alpha_s^2$.

To determine the dimension-8 coefficients, we need the scattering
amplitudes from the term (\ref{Lcontact8}) in the effective
lagrangian:
\begin{eqnarray}
{\cal M}_{d=8} &=& {1 \over M^2} \; (1 \otimes 1) \Bigg[
\; f_1({}^1P_1) \; {\bf v}' \cdot {\bf v} \; 1 \otimes 1
\;+\; {f_1({}^3P_{1}) + f_1({}^3P_{2}) \over 2} \;
	{\bf v}' \cdot {\bf v} \; \sigma^i \otimes \sigma^i
\nonumber \\
&& \;+\; {f_1({}^3P_{0}) - f_1({}^3P_{2}) \over 3} \;
	{\bf v}' \cdot \mbox{\boldmath $\sigma$} \otimes {\bf v} \cdot \mbox{\boldmath
$\sigma$}
\;+\; {f_1({}^3P_{2}) - f_1({}^3P_{1}) \over 2} \;
	{\bf v} \cdot \mbox{\boldmath $\sigma$} \otimes {\bf v}' \cdot \mbox{\boldmath
$\sigma$}
\nonumber \\
&& \;+\; g_1({}^1S_0) \; v^2 \; 1 \otimes 1
\;+\; {3 g_1({}^3S_1) - g_1({}^3S_1,{}^3D_{1}) \over 3} \; v^2 \;
	\sigma^i \otimes \sigma^i
\nonumber \\
&& \;+\; {g_1({}^3S_1,{}^3D_{1})\over 2}
	(v^i v^j + v'^i v'^j) \sigma^i \otimes \sigma^j \Bigg] \;+\; \ldots .
\label{M8}
\end{eqnarray}
There are similar terms with color structure $T^a \otimes T^a$ and
coefficients $f_8$ and $g_8$.   Comparing with (\ref{Mmatch}), we find
that
\begin{mathletters}
\begin{eqnarray}
g_8({}^3S_1) &=& {4 \pi \over 3} \alpha_s(M) ,
\label{g8S}
\\
g_8({}^3S_1,{}^3D_{1}) &=& \pi \alpha_s(M) .
\label{g8SD}
\end{eqnarray}
\end{mathletters}
The color-singlet coefficients $g_1$ and the remaining color-octet
coefficients vanish at this order in $\alpha_s(M)$.

\subsection{Imaginary Parts at Order $\alpha_s^2$}
\label{app:alphasq}

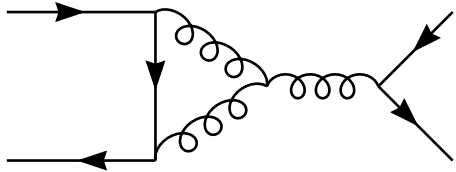
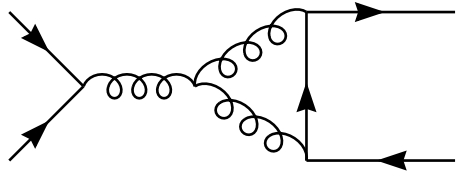
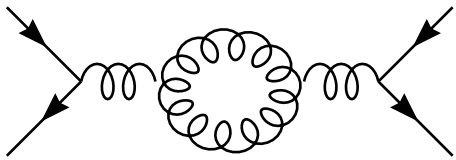
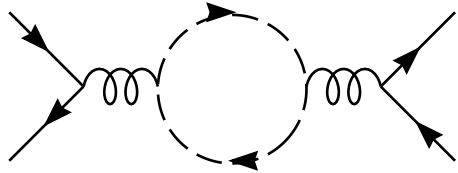
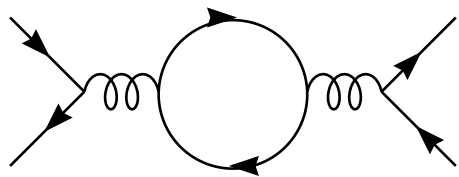
\begin{figure}
\caption{Feynman diagrams that contribute to the imaginary part of the
amplitude for $Q \overline{Q}$ scattering at order $\alpha_s^2$.}
\label{fig:qq-alpha}
\end{figure}

We now turn to the calculation of the imaginary parts of the
coefficients at order $\alpha_s^2$.  They can be determined by matching
the imaginary parts of $Q \overline{Q}$ scattering amplitudes in full
QCD and NRQCD in accordance with (\ref{matching}). In full QCD, the
annihilation contributions to the imaginary parts at order $\alpha_s^2$
come from the one-loop diagrams in Fig.~\ref{fig:qq-alpha}. We will
determine the imaginary parts of the coefficients of the dimension-6
operators in (\ref{Lcontact6}). We will also determine the imaginary
parts of the coefficients of the dimension-8 operators that contribute
to the annihilation of P-wave states at leading order in $v$ 
and S-wave states through 
relative order $v^2$. The dimension-8 terms in the lagrangian are given
in (\ref{Lcontact8}). To determine the coefficients, we consider $Q
\overline{Q}$ scattering in the center of momentum frame, with the $Q$
and $\overline{Q}$ momenta small compared to the heavy-quark mass $M$.
We calculate the imaginary parts of the diagrams in
Fig.~\ref{fig:qq-alpha} in Feynman gauge. After making the substitutions
(\ref{spinor}) for the Dirac spinors, we expand in powers of the
velocity $v$. 

Below, we list the results for the imaginary parts of
each of the diagrams, suppressing the spinors as in (\ref{Mmatch}).
The diagrams in Fig.~\ref{fig:qq-alpha}(a) and \ref{fig:qq-alpha}(b)
yield, on expansion through second order in the velocity $v$,
\begin{mathletters}
\label{A2ab}
\begin{eqnarray}
{\rm Im \,} {\cal M}_{\rm 7(a)} &=&
{\pi \alpha_s^2 \over 2 M^2} \; (T^a T^b \otimes T^b T^a)
\Bigg[ \left( 1 - {4 \over 3} v^2
	+ {1 \over 3} {\bf v} \cdot {\bf v}' \right) \; 1 \otimes 1
\nonumber \\
&& \;+\; \left( {1 \over 3} - {1 \over 5} v^2
		+  {2 \over 5} {\bf v} \cdot {\bf v}' \right)
	\; \sigma^i \otimes \sigma^i
\nonumber \\
&& \;+\; \left( {2 \over 5} v^i v'^j + {11 \over 15} v'^i v^j
		- {11 \over 30} (v^i v^j + v'^i v'^j) \right)
	\; \sigma^i \otimes \sigma^j \Bigg],
\label{A2a}
\\
{\rm Im \,} {\cal M}_{\rm 7(b)} &=&
{\pi \alpha_s^2 \over 2 M^2} \; (T^a T^b \otimes T^a T^b)
\Bigg[ \left( 1 - {4 \over 3} v^2
	- {1 \over 3} {\bf v} \cdot {\bf v}' \right) \; 1 \otimes 1
\nonumber \\
&& \;-\;\left( {1 \over 3} - {1 \over 5} v^2
		- {2 \over 5} {\bf v} \cdot {\bf v}' \right)
	\; \sigma^i \otimes \sigma^i
\nonumber \\
&& \;+\; \left( {2 \over 5} v^i v'^j + {11 \over 15} v'^i v^j
		+ {11 \over 30} (v^i v^j + v'^i v'^j) \right)
	\; \sigma^i \otimes \sigma^j \Bigg] .
\label{A2b}
\end{eqnarray}
\end{mathletters}
The color matrices can be simplified as follows:
\begin{mathletters}
\label{TTTT}
\begin{eqnarray}
T^a T^b \otimes T^b T^a &=& {C_F \over 2 N_c} \; 1 \otimes 1
	\;+\; {N_c^2-2 \over 2 N_c} \; T^a \otimes T^a ,
\label{Tabba}
\\
T^a T^b \otimes T^a T^b &=& {C_F \over 2 N_c} \; 1 \otimes 1
	\;-\; {1\over N_c} \; T^a \otimes T^a ,
\label{Tabab}
\end{eqnarray}
\end{mathletters}
where $C_F = (N_c^2-1)/(2 N_c)$ is the Casimir for the fundamental
representation.
The diagrams in Fig.~\ref{fig:qq-alpha}(c) and ~\ref{fig:qq-alpha}(d),
which involve the triple-gluon vertex,  yield
\begin{mathletters}
\label{A2cd}
\begin{equation}
{\rm Im \,} {\cal M}_{\rm 7(c)} \;=\;
{N_c \pi \alpha_s^2 \over 6 M^2} \; (T^a \otimes T^a) \;
\left[ \left(1 - {11 \over 10} v^2 \right)
	\; \sigma^i \otimes \sigma^i
\;-\; \left({1 \over 5} v^i v^j + {1 \over 2} v'^i v'^j \right)
	\; \sigma^i \otimes \sigma^j \right],
\label{A2c}
\end{equation}
\begin{equation}
{\rm Im \,} {\cal M}_{\rm 7(d)} \;=\;
{N_c \pi \alpha_s^2 \over 6 M^2} \; (T^a \otimes T^a) \;
\left[ \left(1 - {11 \over 10} v^2 \right)
	\; \sigma^i \otimes \sigma^i
\;-\; \left({1 \over 2} v^i v^j + {1 \over 5} v'^i v'^j \right)
	\; \sigma^i \otimes \sigma^j \right].
\label{A2d}
\end{equation}
\end{mathletters}
 From the gluon-loop diagram in Fig.~\ref{fig:qq-alpha}(e)
combined with the associated ghost loop diagram in
Fig.~\ref{fig:qq-alpha}(f), we obtain
\begin{equation}
{\rm Im \,} \left( {\cal M}_{\rm 7(e)} + {\cal M}_{\rm 7(f)} \right) \;=\;
- {5 N_c \pi \alpha_s^2 \over 12 M^2} \; (T^a \otimes T^a) \;
\left[ (1 - v^2) \sigma^i \otimes \sigma^i
	\;-\; {1 \over 2} (v^i v^j + v'^i v'^j)
	\sigma^i \otimes \sigma^j \right].
\label{A2e}
\end{equation}
The quark loop diagram in Fig.~\ref{fig:qq-alpha}(g) gives
\begin{equation}
{\rm Im \,} {\cal M}_{\rm 7(g)} \;=\;
{n_f \pi \alpha_s^2 \over 6 M^2} \; (T^a \otimes T^a) \;
\left[ (1 - v^2) \sigma^i \otimes \sigma^i
	\;-\; {1 \over 2} (v^i v^j + v'^i v'^j)
	\sigma^i \otimes \sigma^j \right].
\label{A2g}
\end{equation}
Adding the amplitudes (\ref{A2ab}), and (\ref{A2cd})--(\ref{A2g}) and
comparing with (\ref{M6}), we can read off the imaginary parts of the
coefficients of the dimension-6 operators:
\begin{mathletters}
\begin{eqnarray}
{\rm Im \,} f_1({}^1S_0) &=& {\pi C_F \over 2 N_c} \; \alpha_s^2(M) ,
\\
{\rm Im \,} f_8({}^1S_0) &=& {\pi (N_c^2 - 4) \over 4 N_c} \; \alpha_s^2(M) ,
\\
{\rm Im \,} f_8({}^3S_1) &=& {\pi n_f \over 6} \; \alpha_s^2(M) .
\label{Imf6}
\end{eqnarray}
\end{mathletters}
The imaginary part of the
coefficient $f_1({}^3S_1)$ vanishes at order $\alpha_s^2$.
Comparing with (\ref{M8}), we see that the coefficients of the
color-singlet dimension-8 operators are
\begin{mathletters}
\begin{eqnarray}
{\rm Im \,} f_1({}^3P_{0}) &=& {3 \pi C_F \over 2 N_c} \; \alpha_s^2(M) ,
\\
{\rm Im \,} f_1({}^3P_{2}) &=&  {2 \pi C_F \over 5 N_c} \; \alpha_s^2(M) ,
\\
{\rm Im \,} g_1({}^1S_0) &=& - {2\pi C_F \over 3 N_c} \; \alpha_s^2(M) .
\label{Imf8}
\end{eqnarray}
\end{mathletters}
The imaginary parts of the
coefficients $f_1({}^1P_1)$, $f_1({}^3P_{1})$,
$g_1({}^3S_1)$ and $g_1({}^3S_1,{}^3D_{1})$
vanish at order $\alpha_s^2$.

\subsection{Imaginary Parts at Higher Order in $\alpha_s$}
\label{app:higher}

According to the matching condition (\ref{matching}),
the coefficients of the 4-fermion operators can be computed
at next-to-leading order in $\alpha_s$ by calculating
scattering amplitudes at next-to-leading order in full QCD and
equating them to the scattering amplitudes in NRQCD, calculated to
next-to-leading order in $\alpha_s$.  For some of the 4-fermion operators,
the imaginary parts of the coefficients can
be extracted at next-to-leading order in $\alpha_s$ from calculations
of heavy-quarkonium annihilation rates that already exist in the literature.

As an illustration of our factorization approach, we discuss in detail
the calculation of ${\rm Im \,} f_1({}^1S_0)$ at next-to-leading order in
$\alpha_s$. In order to determine ${\rm Im \,} f_1({}^1S_0)$, we consider
the matrix element ${\cal M}$ for the forward scattering of a $Q \overline{Q}$
pair
above threshold in a color-singlet spin-singlet state with relative
velocity $2 v$. The imaginary part of ${\cal M}$ can be expressed as a
sum over cuts through the Feynman diagrams for forward scattering.  The
annihilation contribution to ${\rm Im \,} {\cal M}$ is the sum over cuts
through gluon and light quark lines only.  It has been calculated in
full QCD at next-to-leading order in $\alpha_s$ and in the limit $v \to
0$ by Barbieri {\it et al.} \cite{barba}:
\begin{equation}
{\rm Im \,} {\cal M} \; = \;
{\pi C_F \alpha_s^2(2M) \over M^2} \Bigg\{ 1
	\;+\; \left[ \left( {\pi^2 \over 2 v} + {\pi^2 \over 4} - 5 \right)
        C_F
	\;+\; \left( {199 \over 18} -  {13 \pi^2 \over 24} \right) C_A
	\;-\; {8 \over 9} n_f \right]
		{\alpha_s \over \pi} \Bigg\} \;,
\label{MetaNLO}
\end{equation}
where $C_A = N_c$ is the Casimir for the adjoint representation,
$\alpha_s(M)$ is the QCD coupling constant in the modified minimal
subtraction ($\overline{{\rm MS}}$) renormalization scheme for QCD with
$n_f$ flavors of light quarks, and $M$ is the perturbative pole mass of
the heavy quark. We have corrected apparent errors in Ref. \cite{barba}
of 2/3 in the overall coefficient and 1/2 in the coefficient of the
$\pi^2/v$ term. The next-to-leading order correction contains a Coulomb
singularity proportional to $1/v$, which gives an infrared divergence in
the limit $v \to 0$. 

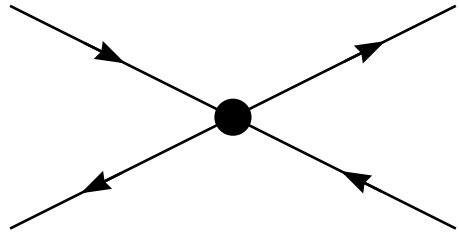
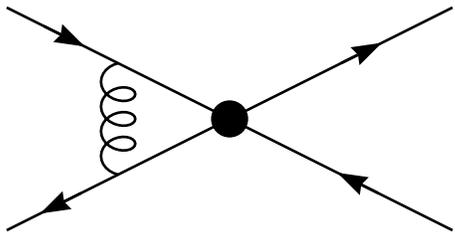
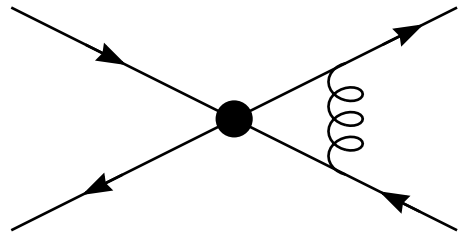
\begin{figure}
\caption{Feynman diagrams in NRQCD for the scattering of a
$Q \overline{Q}$ pair in a color-singlet ${}^1S_0$ state
through the operator $\psi^\dagger \chi \chi^\dagger \psi$.}
\label{fig:sing-swave}
\end{figure}

In order to determine ${\rm Im \,} f_1({}^1S_0)$, we must calculate the
corresponding contribution to ${\rm Im \,} {\cal M}$ in NRQCD at
next-to-leading order in $\alpha_s$. The relevant Feynman diagrams are
shown in Fig.~\ref{fig:sing-swave}. They contain a 4-fermion vertex that
corresponds to the term $\psi^\dagger \chi \chi^\dagger \psi$ in the effective
lagrangian.  The annihilation contribution is the sum over all cuts that
pass through that vertex.  The Cutkosky rules specify that a cut passing
through the 4-fermion vertex is computed by taking the imaginary part of
the coefficient $f_1(^1S_0)$ and complex-conjugating the part of the
diagram to the right of the cut.   The incoming and outgoing states
consist of a $Q \overline{Q}$ pair in a color-singlet spin-singlet state with
relative velocity ${\bf v} \to 0$. Explicitly, the in state is
\begin{equation}
{| Q \overline{Q} \rangle} = \sum_{ij} {1 \over \sqrt{N_c}} \delta^{ij}
	\sum_{m m'} {1 \over \sqrt{2}} (i \sigma_2)_{mm'}
	{|  Q({\bf p},m,i) \overline{Q}(-{\bf p},m',j)  \rangle} \;,
\label{QQbar}
\end{equation}
where ${\bf p} = M {\bf v}$ is the momentum of the quark in the center of
momentum
frame.  The quark states ${| Q({\bf p},m,i) \rangle}$ with momentum ${\bf p}$,
spin
quantum number $m=\pm1/2$, and color $i$ have the standard nonrelativistic
normalization:
\begin{equation}
\langle Q({\bf p}',m',j) | Q({\bf p},m,i) \rangle \;=\;
\delta^{ij} \delta^{mm'} (2 \pi)^3 \delta^3({\bf p}' - {\bf p}) \;.
\end{equation}
For the leading-order diagram in Fig.~\ref{fig:sing-swave}(a),
the cut through the 4-fermion vertex gives simply the imaginary part of the
coefficient $f_1({}^1S_0)/M^2$:
\begin{equation}
{\rm Im \,} {\cal M}_{\rm 8(a)} \;=\; {2 N_c \; {\rm Im \,} f_1({}^1S_0)
\over M^2} \;.
\label{A3a}
\end{equation}
It is convenient to calculate the next-to-leading order diagrams in
Figs.~\ref{fig:sing-swave}(b) and \ref{fig:sing-swave}(c)
by using Coulomb gauge, since then only Coulomb exchange
contributes in the limit $v \to 0$.  For the diagram in
Fig.~\ref{fig:sing-swave}(b),
we obtain
\begin{eqnarray}
\Bigg( {\rm Im \,} {\cal M} \Bigg)_{\rm 8(b)}
&=& {2 N_c \; {\rm Im \,} f_1({}^1S_0) \over M^2} \;
\left( -i 4 \pi C_F \alpha_s \right)
\int{d^4q \over (2 \pi)^4} \; {1 \over {\bf q}^2}
\nonumber \\
&&	{1 \over E + q_0 - ({\bf p} +{\bf q})^2/2M + i \epsilon} \;
	{1 \over E - q_0 - ({\bf p} +{\bf q})^2/2M + i \epsilon} \;,
\label{A3bint4}
\end{eqnarray}
where $E = p^2/2M$.
After using contour integration to integrate over the energy $q_0$ of
the exchanged gluon, we find that the contribution reduces to an
integral over the gluon's 3-momentum:
\begin{equation}
\Bigg( {\rm Im \,} {\cal M} \Bigg)_{\rm 8(b)}
\;=\; {2 N_c \; {\rm Im \,} f_1({}^1S_0) \over M^2} \; 4 \pi C_F \alpha_s M
\int{d^3q \over (2 \pi)^3} {1 \over {\bf q}^2}
	{1 \over {\bf q}^2 + 2 {\bf p} \cdot {\bf q} - i \epsilon} \;.
\label{A3bint}
\end{equation}
The integral is infrared divergent, and can be regularized by using
dimensional regularization.  The integral over ${\bf q}$ is analytically
continued to $D = 3 - 2 \epsilon_{\rm IR}$ spatial dimensions.
Evaluating the regularized integral in (\ref{A3bint}), we obtain
\begin{equation}
\bigg( {\rm Im \,} {\cal M} \bigg)_{\rm 8(b)} \;=\;
{2 N_c \; {\rm Im \,} f_1({}^1S_0) \over M^2} \; {\pi C_F \alpha_s \over 4 v}
\Bigg[ 1 \;-\; {i \over \pi} \left(
	{1 \over\epsilon_{\rm IR}} + \log(4 \pi) - \gamma
		- 2 \log{2Mv \over \mu_{\rm IR}}
\right) \Bigg] \;,
\label{A3b}
\end{equation}
where $\gamma$ is Euler's constant and $\mu_{\rm IR}$ is the arbitrary
regularization scale introduced with dimensional regularization. The
logarithmic infrared divergence appears as a pole in $\epsilon_{\rm IR}$. The
subscripts ${\rm IR}$ on $\epsilon$ and $\mu$ serve as a reminder that they
are associated with infrared divergences. Note that (\ref{A3b}) is
complex valued.   The imaginary part of (\ref{A3b}) arises because it is
possible for the incoming quark and antiquark to scatter on-shell before
annihilating at the 4-fermion vertex.  After summing over all diagrams,
one must, of course, obtain a real result for ${\rm Im \,} {\cal M}$.
The diagram in Fig.~\ref{fig:sing-swave}(c) is evaluated in the same way
as Fig.~\ref{fig:sing-swave}(b), except that the Cutkosky cutting rules
require the complex-conjugation of the part of the diagram that involves
the Coulomb-gluon exchange. The result is
\begin{equation}
\Bigg( {\rm Im \,} {\cal M} \Bigg)_{\rm 8(c)} \;=\;
{2 N_c \; {\rm Im \,} f_1({}^1S_0) \over M^2} \; {\pi C_F \alpha_s \over 4 v}
\Bigg[ 1 \;+\; {i \over \pi} \left(
	{1 \over \epsilon_{\rm IR}} + \log(4 \pi) - \gamma
		- 2 \log{2Mv \over \mu_{\rm IR}}
\right) \Bigg] \;.
\label{A3c}
\end{equation}
Note that the imaginary part of (\ref{A3c}) cancels that of (\ref{A3b}).
Adding (\ref{A3a}), (\ref{A3b}), and (\ref{A3c}), we obtain the complete
result for ${\rm Im \,} {\cal M}$ through next-to-leading order in
$\alpha_s$:
\begin{equation}
{\rm Im \,} {\cal M} \;=\; {2 N_c \; {\rm Im \,} f_1({}^1S_0) \over M^2} \;
\Bigg[ 1 \;+\; {\pi^2 \over 2 v} C_F {\alpha_s \over \pi} \Bigg] \;.
\label{MetaNRQCD}
\end{equation}

Comparing (\ref{MetaNLO}) and (\ref{MetaNRQCD}), we can read off the
imaginary part of $f_1({}^1S_0)$ through next-to-leading order in
$\alpha_s$:
\begin{equation}
{\rm Im \,} f_1({}^1S_0) \; = \;
{\pi C_F \over 2 N_c} \alpha_s^2(2M)
\Bigg\{ 1 \;+\; \left[ \left( {\pi^2 \over 4} - 5 \right) C_F
	\;+\; \left( {199 \over 18} -  {13 \pi^2 \over 24} \right) C_A
	\;-\; {8 \over 9} n_f \right] {\alpha_s \over \pi} \Bigg\} \;.
\label{fetaNLO}
\end{equation}
Note that the factorization approach reproduces the standard
prescription of simply dropping the $1/v$ terms in the perturbatively
calculated annihilation rate \cite{hb}. The factorization approach puts
this prescription on a rigorous footing, and makes it clear how to
extend the calculation systematically to higher orders in $\alpha_s$ and
in $v$.

In (\ref{fetaNLO}), $\alpha_s(M)$ is the $\overline{{\rm MS}}$
coupling constant with renormalization scale $M$.
If we make a different choice for the renormalization scale $\mu$
of $\alpha_s(\mu)$,
then we must differentiate between the $\overline{{\rm MS}}$ coupling constant
$\alpha_s^{(n_f+1)}(\mu)$ for full QCD with
$n_f$ flavors of light quarks and a heavy quark
and the corresponding coupling constant $\alpha_s^{(n_f)}(\mu)$
for only $n_f$ flavors of light quarks, which is the
appropriate running coupling constant below the heavy-quark threshold.
These coupling constants satisfy the matching condition \cite{rs}
$\alpha_s^{(n_f)}(M) = \alpha_s^{(n_f+1)}(M) + O(\alpha_s^3)$.
If we wish to use a different renormalization scale $\mu \ne M$
for $\alpha_s$ in (\ref{fetaNLO}),
then we must make one of the following substitutions:
\begin{mathletters}
\begin{eqnarray}
\alpha_s(M) &=& \alpha_s^{(n_f)}(\mu)
\left[ 1 \;+\; \beta_0 \log{\mu \over M} {\alpha_s \over \pi}
	\;+\; O(\alpha_s^2) \right],
\label{alfl} \\
\alpha_s(M) &=& \alpha_s^{(n_f+1)}(\mu)
\left[ 1 \;+\; \left( \beta_0 - {1 \over 3} \right)
		\log{\mu \over M} {\alpha_s \over \pi}
	\;+\; O(\alpha_s^2) \right],
\label{alflh}
\end{eqnarray}
\end{mathletters}
where $\beta_0 = (33-2n_f)/6$ is the first coefficient in the beta
function for QCD with $n_f$ flavors of light quarks:
$\mu (d/d\mu) \alpha_s(\mu) = - \beta_0 \alpha_s^2/\pi + \ldots$.

The coefficient of the operator ${\cal O}_1({}^1S_0)$ in the NRQCD
lagrangian is $f_1({}^1S_0)/M^2$, and the perturbation series for
$f_1({}^1S_0)$ depends on the definition of the heavy quark mass $M$. The
order-$\alpha_s^3$ correction in (\ref{fetaNLO}) corresponds to the
choice $M = M_{\rm pole}$, where $M_{\rm pole}$ is the perturbative pole
mass, i.e., the location of the pole in the heavy-quark propagator in
perturbation theory.  An alternative choice is the running mass $M(\mu)$
in the $\overline{{\rm MS}}$ renormalization scheme.  Its relation to the pole
mass
through order $\alpha_s$ is \cite{gbgk}
\begin{equation}
M_{\rm pole} \;=\; M(\mu)
\left[ 1 \;+\; \left(1 + {3 \over 2} \log{\mu \over M} \right)
	\; C_F \; {\alpha_s \over \pi} \;+\; O(\alpha_s^2) \right].
\label{Mpole}
\end{equation}
Throughout this paper, we will adopt the choice $M = M_{\rm pole}$
for the heavy quark mass in the coefficient $f_n/M^{d_n-4}$
of a 4-fermion operator with naive scaling dimension $d_n$.

We can obtain the imaginary part of the coefficient
$f_1(^3S_1)$ through next-to-leading order in $\alpha_s$
from a calculation by MacKenzie and Lepage of the
annihilation decay rate of the $J/\psi$ or $\Upsilon$ \cite{ml}.
Their published result is given explicitly only for $N_c = 3$,
but one can insert the appropriate color factors in the various
classes of diagrams and obtain the result
\begin{eqnarray}
&& {\rm Im \,} f_1({}^3S_1)
\nonumber \\
&& \quad \;=\;
{(\pi^2 - 9) (N_c^2 - 4) C_F \over 54 N_c} \alpha_s^3(M)
\Bigg[ 1 \;+\;
	\left( - 9.46(2) C_F \;+\; 4.13(17) C_A \;-\; 1.161(2) n_f \right)
	{\alpha_s \over \pi} \Bigg]
\nonumber \\
&& \quad \quad \quad \quad \quad
\;+\; \pi Q^2 \left( \sum_i Q_i^2 \right) \alpha^2
\Bigg[ 1 - {13 \over 4} C_F {\alpha_s \over \pi} \Bigg] ,
\label{fpsiNLO}
\end{eqnarray}
where $Q$ is the electric charge of the heavy quark ($Q = +2/3$ for the
charmed quark and $Q = -1/3$ for the bottom quark) and the $Q_i$, $i =
1,\ldots, n_f$, are the electric charges of the light quarks. The
perturbative correction in the first term on the right side of
(\ref{fpsiNLO}) was calculated by Mackenzie and Lepage \cite{ml}.  The
term proportional to $\alpha^2$ is due to annihilation of the $Q \overline{Q}$
pair
into a virtual photon, which then decays into light hadrons.  The
order-$\alpha_s$ correction can be calculated as the sum of two terms:
$- 4 C_F \alpha_s/\pi$, which is the order-$\alpha_s$ correction to the
rate for $\psi \to e^+ e^-$, and $3 C_F \alpha_s/(4 \pi)$, which is the
order-$\alpha_s$ correction to the rate for $\gamma^* \to q \bar q$. For
completeness, we also give the coefficient analogous to (\ref{fpsiNLO})
for the decay of the $\psi$ into a photon plus light hadrons:
\begin{eqnarray}
&& {\rm Im \,} f_{\gamma 1}({}^3S_1)
\nonumber \\
&& \quad \;=\;
{2(\pi^2 - 9) C_F Q^2 \alpha \over 3 N_c} \alpha_s^2(M)
\Bigg[ 1 +
	\left( - 9.46(2) C_F + 2.75(11) C_A - 0.774(1) n_f \right)
	{\alpha_s \over \pi} \Bigg].
\label{fpsigamNLO}
\end{eqnarray}

Calculations of the annihilation rates of P-wave states were carried out
through order $\alpha_s^3$ by Barbieri and collaborators
\cite{barbe,barbc,barbd}. They calculated only the coefficients of
$|{\overline {R_P'}}|^2$ in (\ref{HPwave}). These coefficients contain
logarithmic
infrared divergences that should be factored into the color-octet matrix
elements, along with associated constants that can be determined from
calculations in NRQCD. In Ref. \cite{barbe}, the logarithmic infrared
divergences in ${\rm Im \,} f_1({}^3P_{0})$ and ${\rm Im \,}
f_1({}^3P_{2})$ were cut off by taking the heavy quark and antiquark off
their mass-shells and below threshold, in which case the infrared
divergence manifests itself as a logarithm of the binding energy. In
order to extract NRQCD coefficients, it might be necessary to repeat the
next-to-leading order calculations in Ref.~\cite{barbe} using on-shell
scattering amplitudes and dimensional regularization of the infrared
divergences in order to maintain gauge invariance.

\subsection{Coefficients of Electromagnetic Operators}
\label{app:electro}

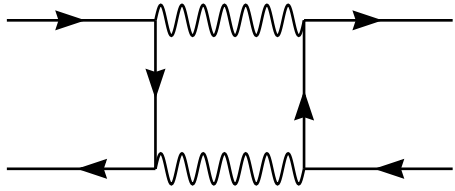
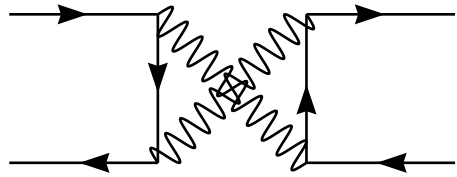
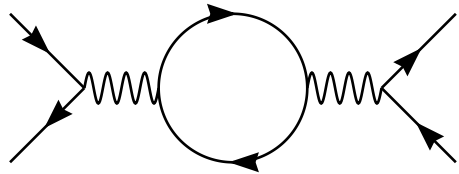
\begin{figure}
\caption{Feynman diagrams that contribute to the imaginary part of the
amplitude for electromagnetic $Q \overline{Q}$ scattering at order
$\alpha^2$.}
\label{fig:qq-imag-alpha}
\end{figure}

The calculation in Section~\ref{app:alphasq} can be easily modified to
give the imaginary parts of the coefficients of the electromagnetic
4-fermion operators at order $\alpha^2$ and at leading order in
$\alpha_s$. The Feynman diagrams in Figs.~\ref{fig:qq-imag-alpha}(a) and
\ref{fig:qq-imag-alpha}(b) yield imaginary parts that correspond to
annihilation into two photons. These imaginary parts can be
obtained from (\ref{A2ab}) by replacing the color matrices
$T^a$ by the unit color matrix and by substituting
$\alpha_s \to Q^2 \alpha$, where $Q$ is the electric charge of the
heavy quark: $Q = +2/3$ for the charmed quark, and $Q = -1/3$
for the bottom quark.  The sum of the 2 diagrams yields
\begin{eqnarray}
{\rm Im \,} \left({\cal M}_{\rm 9(a)}  + {\cal M}_{\rm 9(b)} \right) &=&
{\pi Q^4 \alpha^2 \over M^2} \; (1 \otimes 1)
\Bigg[ \left( 1 - {4 \over 3} v^2 \right) \; 1 \otimes 1
\;+\; {2 \over 5} {\bf v} \cdot {\bf v}' \; \sigma^i \otimes \sigma^i
\nonumber \\
&& \;+\; \left( {2 \over 5} v^i {v'}^j \;+\; {11 \over 15} {v'}^i v^j \right)
	\sigma^i \otimes \sigma^j \Bigg].
\label{A4a}
\end{eqnarray}
Comparing to the NRQCD scattering amplitudes analogous to (\ref{M6}),
we find that the only nonzero coefficient for the dimension-6
operators is
\begin{equation}
{\rm Im \,} f_{\gamma \gamma}({}^1S_0) \; = \;  \pi Q^4 \alpha^2 .
\label{fEMeta}
\end{equation}
Comparing to the NRQCD scattering amplitudes analogous to (\ref{M8}),
we can read off the nonzero coefficients of the dimension-8 operators:
\begin{mathletters}
\begin{eqnarray}
{\rm Im \,} f_{\gamma \gamma}({}^3P_{0}) &=& 3 \pi Q^4 \alpha^2 ,
\label{gEMchi0}
\\
{\rm Im \,} f_{\gamma \gamma}({}^3P_{2}) &=& {4 \pi Q^4 \alpha^2 \over 5} ,
\label{gEMchi2}
\\
{\rm Im \,} g_{\gamma \gamma}({}^1S_0) &=&  - {4 \pi Q^4 \alpha^2 \over 3} .
\label{gEMeta}
\end{eqnarray}
\end{mathletters}

The diagram in Fig.~\ref{fig:qq-imag-alpha}(c) yields an imaginary part
that corresponds to the annihilation into lepton pairs. The imaginary
part can be obtained from (\ref{A2g}) by replacing $T^a$ by the unit
color matrix, and by substituting $(n_f/2) \alpha_s \to - Q \alpha$. The
resulting matrix element is
\begin{equation}
{\rm Im \,} {\cal M}_{\rm 9(c)} \;=\;
{\pi Q^2 \alpha^2 \over 3 M^2} \; (1 \otimes 1) \;
\left[ (1 - v^2) \sigma^i \otimes \sigma^i
	\;-\; {1 \over 2} (v^i v^j + v'^i v'^j)
	\sigma^i \otimes \sigma^j \right].
\label{A4c}
\end{equation}
Comparing to the NRQCD scattering amplitudes analogous to (\ref{M6}),
we find that the only nonzero coefficient of the dimension-6
operators is
\begin{equation}
{\rm Im \,} f_{\rm ee}({}^3S_1) \; = \; {\pi Q^2 \alpha^2 \over 3} .
\label{fEMpsi}
\end{equation}
Comparing to the NRQCD scattering amplitudes analogous to (\ref{M8}),
we can read off the nonzero coefficients of the dimension-8 operators:
\begin{mathletters}
\begin{eqnarray}
{\rm Im \,} g_{\rm ee}({}^3S_1) &=& - {4 \pi Q^2 \alpha^2 \over 9} \;,
\label{gEMSD}
\\
{\rm Im \,} g_{\rm ee}({}^3S_1,{}^3D_{1}) &=& - {\pi Q^2 \alpha^2 \over 3} \;.
\label{gEMpsi}
\end{eqnarray}
\end{mathletters}

Several of the electromagnetic coefficients can be determined through
next-to-leading order in $\alpha_s$ from calculations that are available
in the literature. The annihilation rates for $\eta_c$, $\chi_{c0}$, and
$\chi_{c2}$ into two photons have been calculated through next-to-leading
order in $\alpha_s$ by Barbieri {\it et al.} \cite{barba,barbe}.  The
corresponding coefficients are
\begin{mathletters}
\begin{eqnarray}
{\rm Im \,} f_{\gamma \gamma}({}^1S_0)
&=&  \pi Q^4 \alpha^2 \Bigg[ 1 \;+\;
	\left( {\pi^2 \over 4} - 5 \right) C_F {\alpha_s \over \pi} \Bigg] \;,
\label{fEMetaNLO}
\\
{\rm Im \,} f_{\gamma \gamma}({}^3P_{0})
&=& 3 \pi Q^4 \alpha^2 \Bigg[ 1 \;+\; \left( {\pi^2 \over 4}
	- {7 \over 3} \right) C_F {\alpha_s \over \pi} \Bigg] \;,
\label{gEMchi0NLO}
\\
{\rm Im \,} f_{\gamma \gamma}({}^3P_{2})
&=& {4 \pi Q^4 \alpha^2 \over 5}
	\Bigg[ 1 \;-\; 4 C_F {\alpha_s \over \pi} \Bigg] \;.
\label{gEMchi2NLO}
\end{eqnarray}
\end{mathletters}
The rate for $\psi \to e^+ e^-$ is known through
next-to-leading order in $\alpha_s$ \cite{wc}:
\begin{equation}
{\rm Im \,} f_{\rm ee}({}^3S_1) \; = \;
{\pi Q^2 \alpha^2 \over 3}
\Bigg[ 1 \;-\; 4 C_F {\alpha_s \over \pi} \Bigg] \;.
\label{fEMpsiNLO}
\end{equation}
Finally, Mackenzie and Lepage \cite{ml} have calculated the rate for
$\psi \to \gamma \gamma \gamma$ to next-to-leading order in $\alpha_s$.
The corresponding coefficient is
\begin{equation}
{\rm Im \,} f_{3 \gamma}({}^3S_1) \; = \;
{4 (\pi^2 - 9) Q^6 \alpha^3 \over 9}
\Bigg[ 1 \;-\; 9.46(2) C_F {\alpha_s \over \pi} \Bigg] \;.
\label{fpsi3gNLO}
\end{equation}

\vfill \eject

\section{Evolution of 4-fermion operators}
\label{app:evol}

As we mentioned in Section~\ref{sec:fact-scale}, loop corrections to the
4-fermion operators in the NRQCD lagrangian are, in general, ultraviolet
divergent, and, therefore, must be regularized. One can remove power
divergences, either by employing a mass-independent regularization
scheme, such as dimensional regularization, or by making explicit
subtractions.  Once this has been done, the 4-fermion operators satisfy
simple evolution equations of the form (\ref{evolop}). The evolution
equation for an operator ${\cal O}_n$ with naive scaling dimension $d_n$
involves only operators ${\cal O}_k$ with dimensions $d_k \ge d_n$. The
coefficients $\gamma_{nk}$ in the evolution equation can be computed as
power series in $\alpha_s$.  For $d_k = d_n$, the coefficients
$\gamma_{nk}$ are at most of order $\alpha_s^2$, because the logarithmic
ultraviolet divergences at order $\alpha_s$ come only from corrections
of relative order $v^2$, which correspond to operators ${\cal O}_k$ of
dimension $d_n+2$ or larger. In this Appendix, we compute at order
$\alpha_s$ the coefficients of the dimension-8 operators that appear in
the evolution of the dimension-6 4-fermion operators.

\subsection{Heavy Quark Self-energy}
\label{app:renorm}

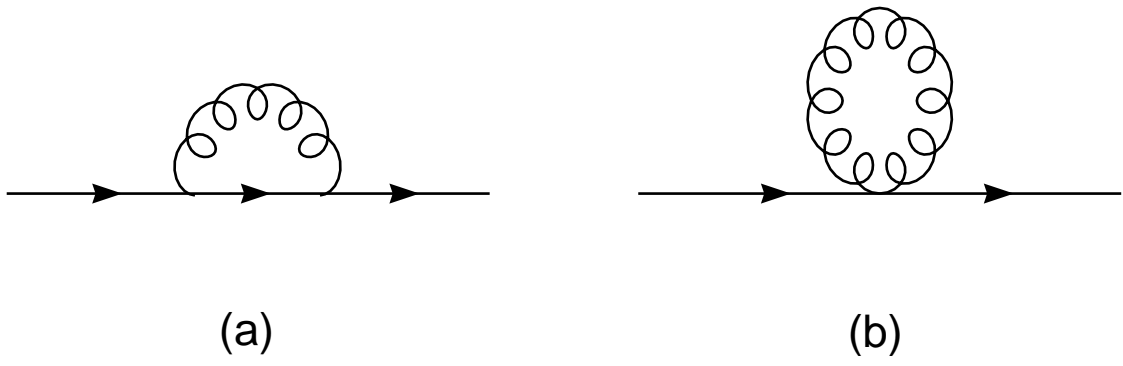
\begin{figure}
\caption{Feynman diagrams in NRQCD for the self-energy of a heavy quark
at order $\alpha_s$.}
\label{fig:self-energy}
\end{figure}

In order to
illustrate the methods that are used to calculate the coefficients in the
evolution equations,  we first calculate the self-energy of the heavy quark
in NRQCD through order $\alpha_s$. From this calculation, we determine the
relation between the perturbative pole mass $M_{\rm pole}$
and the mass parameter $M$ in the NRQCD lagrangian,
and we extract the residue $Z(p)$ of the pole
in the heavy-quark propagator through order $\alpha_s v^2$.
The residue is given by
\begin{equation}
Z(p)^{-1} \; = \; 1 \;-\;
	{\partial \Sigma \over \partial E}(E = p^2/2M_{\rm pole},p) \; ,
\label{Zdef}
\end{equation}
where $\Sigma(E,p)$ is the self-energy correction. To determine $Z(p)$
to order $\alpha_s$ and to order $v^2$, we must calculate the
self-energy correction $\Sigma$ that arises from the one-loop diagrams in
Fig.~\ref{fig:self-energy}. We calculate these diagrams in Coulomb
gauge, because it facilitates the extraction of the dependence on $v$.
The seagull diagram in Fig.~\ref{fig:self-energy}(b) gives only power
ultraviolet divergences, which are subtracted as part of the
regularization scheme. Interactions from ${\cal L}_{\rm bilinear}$ also
need not be included, because the terms of order $\alpha_s v^2$ that
they produce are all proportional to power ultraviolet divergences.

The contribution to the self-energy from the diagram in
Fig.~\ref{fig:self-energy}(a) is
\begin{equation}
\Sigma(E,p) \;=\; i 4 \pi C_F \alpha_s \; \int{d^4q \over (2 \pi)^4} \;
{1 \over E - q_0 - ({\bf p} - {\bf q})^2/2M + i \epsilon}
\Bigg({1 \over {\bf q}^2} \;+\; {p^2 - ({\bf p} \cdot {\hat {\bf q}})^2
	\over M^2 (q_0^2 - {\bf q}^2 + i \epsilon)} \Bigg) .
\label{A5aint4}
\end{equation}
The integral of the term containing $1/{\bf q}^2$, which comes from Coulomb
exchange, gives rise to an ill-defined power divergence, which can be
dropped. After using contour integration to integrate over the energy
$q_0$ of the gluon, we find that the contribution reduces to
\begin{equation}
\Sigma(E,p) \;=\; {2 \pi C_F \alpha_s \over M^2} \;
\int{d^3q \over (2 \pi)^3} \; {1 \over q} \;
	{p^2 - ({\bf p} \cdot {\hat {\bf q}})^2
		\over E - q - ({\bf p} - {\bf q})^2/2M + i \epsilon} \;.
\label{A5aint3}
\end{equation}
In order to identify the power divergences in (\ref{A5aint3}), we expand
the denominator in a Taylor series in $1/M$:
\begin{equation}
\Sigma(E,p) \;=\; - {2 \pi C_F \alpha_s \over M^2} \;
\int{d^3q \over (2 \pi)^3} \; {p^2 - ({\bf p} \cdot {\hat {\bf q}})^2 \over
q^2}
	\left( 1 + {E - p^2/2M + (2{\bf p} \cdot {\bf q} - q^2)/2M
		\over q} + \ldots \right).
\label{A5aint}
\end{equation}
Setting $E=p^2/2M$, we find that every remaining term in the integrand
in (\ref{A5aint}) yields a power divergence, which is subtracted in our
regularization scheme.  Thus, after regularization, the self-energy
vanishes on the energy shell, and there is no correction to the
energy-momentum relation $E= p^2/2M$.  In full QCD, the energy-momentum
relation defined by the pole in the perturbative heavy-quark propagator
is $E^2 = p^2 +M_{\rm pole}^2$. Matching the coefficients of $p^2$ in
these energy-momentum relations, we obtain
\begin{equation}
M \;=\; M_{\rm pole} \left( 1 \;+\; O(\alpha_s^2) \right) .
\label{MMpole}
\end{equation}
Thus, through order $\alpha_s$, the mass parameter $M$ in the lagrangian
for NRQCD can be identified with the perturbative pole mass.

We proceed to compute the residue of the pole in the heavy-quark
propagator, which is given by (\ref{Zdef}). After we subtract the power
divergences, the only term remaining in (\ref{A5aint}) that contributes
at order $\alpha_s v^2$ is the term proportional to $E-p^2/2M$.
Consequently, the expression for the residue is
\begin{equation}
Z(p) \;=\; 1 \;-\; {2 \pi C_F \alpha_s \over M^2} \;
\int{d^3q \over (2 \pi)^3}  {p^2 - ({\bf p} \cdot {\hat {\bf q}})^2 \over q^3}
\;.
\label{Zint}
\end{equation}
Imposing a momentum cutoff $|{\bf q}| < \Lambda$ on the magnitude of the
gluon momentum and keeping only the logarithmic ultraviolet divergence
at order $\alpha_s$, we find that the residue $Z(p)$ is
\begin{equation}
Z(p) \;\approx\; 1 \;-\;
	{2 C_F \alpha_s \log \Lambda \over 3 \pi} v^2 \; ,
\label{ZQ}
\end{equation}
where $v^2 = p^2/M^2$.

\subsection{One-loop Ultraviolet Divergences}
\label{app:one-loop}

The coefficients in the evolution equation for the operator ${\cal
O}_8({}^1S_0) = \psi^\dagger T^a \chi \chi^\dagger T^a \psi$ can be determined
at
order $\alpha_s$ by computing one-loop corrections to scattering
amplitudes in NRQCD that involve this operator.
We consider the amplitude for the scattering
of a $Q \overline{Q}$ pair with momenta ${\bf p}$ and $-{\bf p}$ into a $Q
\overline{Q}$ pair with
momenta ${\bf p}'$ and $-{\bf p}'$.  We use the compact notation with
suppressed
Pauli spinors that was introduced
in Eq. (\ref{Mmatch}).  The matrix element corresponding to the leading-order
diagram in Fig.~\ref{fig:sing-swave}(a) is then written
\begin{equation}
{\cal M} \;=\; \left( T^a\otimes T^a \right) \; ( 1 \otimes 1 )\;,
\label{A60}
\end{equation}
where the first factor gives the color structure and the second factor
gives the spin structure. The one-loop correction to the matrix element
is given by the sum of the contributions of the 10 diagrams in
Figs.~\ref{fig:octet-swave}(a)--\ref{fig:octet-swave}(j). We wish to
calculate the terms in these contributions that are proportional to
$\log \Lambda$, where $\Lambda$ is an ultraviolet cutoff.  For
higher-order calculations, it might be wise to impose the cutoff by using
dimensional regularization, in order to maintain gauge invariance, but
for our purposes it is sufficient to impose a cutoff on the magnitude
of the gluon 3-momentum:  $|{\bf q}| > \Lambda$.

The 4 diagrams in
Fig.~\ref{fig:octet-swave}(a)--\ref{fig:octet-swave}(d)
are self-energy corrections to the
external quark lines.  Each diagram contributes
$\sqrt{Z} - 1$ times the leading order amplitude in (\ref{A60}).
Using the expression (\ref{ZQ}) for the renormalization constant $Z$,
we find that the sum of the contributions of the 4 diagrams is
\begin{equation}
{\cal M}_{\rm 11(a-d)} \;\approx\;
- {4 C_F \alpha_s \log \Lambda \over 3 \pi} \; v^2 \;
	\left( T^a \otimes T^a \right) \; ( 1 \otimes 1 ) \;.
\label{A6a}
\end{equation}
The diagram in Fig.~\ref{fig:octet-swave}(e)
represents the exchange of a transverse gluon between the incoming
quark and antiquark.  (The exchange of a Coulomb gluon does not lead to
an ultraviolet divergence.) This diagram yields the contribution
\begin{eqnarray}
{\cal M}_{\rm 11(e)} &=& i {4 \pi \alpha_s \over M^2} \;
\left( T^a \otimes T^b T^a T^b \right) \; (1 \otimes 1) \;
\int{d^4q \over (2 \pi)^4} \;
	{p^2 - ({\bf p} \cdot {\hat {\bf q}})^2 \over q_0^2 - {\bf q}^2 + i \epsilon}
\;
\nonumber \\
&&	{1 \over E + q_0 - ({\bf p} +{\bf q})^2/2M + i \epsilon} \;
	{1 \over E - q_0 - ({\bf p} +{\bf q})^2/2M + i \epsilon} \; ,
\label{A6eint4}
\end{eqnarray}
where $E = p^2/2M$.   We integrate over the energy $q_0$ of the
exchanged gluon and identify the power divergences by expanding the
denominators in a Taylor series in $1/M$. Keeping only the term that
gives a logarithmic ultraviolet divergence, we find that the
contribution reduces to
\begin{equation}
{\cal M}_{\rm 11(e)} \;=\; - {2 \pi \alpha_s \over M^2} \;
\left( T^a \otimes T^b T^a T^b \right) \; ( 1 \otimes 1 ) \;
\int{d^3q \over (2 \pi)^3} {p^2 - ({\bf p} \cdot {\hat {\bf q}})^2 \over q^3}
\; .
\label{A6eint}
\end{equation}
The integral is the same as in (\ref{Zint}).
The diagram in Fig.~\ref{fig:octet-swave}(f) gives an
identical contribution:
\begin{equation}
{\cal M}_{\rm 11(e)} \;\approx\; {\cal M}_{\rm 11(f)} \;\approx\;
- {2 \alpha_s \log \Lambda \over 3 \pi} \; v^2 \;
	\left( T^a \otimes T^b T^a T^b \right) \; (1 \otimes 1) \;.
\label{A6e}
\end{equation}

The diagrams in Figs.~\ref{fig:octet-swave}(g)--\ref{fig:octet-swave}(j)
involve the exchange of a transverse gluon between initial and final
quark or antiquark lines.  (The exchange of a Coulomb gluon leads to a
vanishing contribution.) These diagrams are evaluated in the same way as
those in Fig.~\ref{fig:octet-swave}(e). The results are
\begin{mathletters}
\label{A6gi}
\begin{equation}
{\cal M}_{\rm 11(g)} \;\approx\; {\cal M}_{\rm 11(h)} \;\approx\;
{2 \alpha_s \log \Lambda \over 3 \pi} \; {\bf v} \cdot {\bf v}' \;
	\left( T^a T^b \otimes T^a  T^b \right) \; (1 \otimes 1) \;,
\label{A6g}
\end{equation}
\begin{equation}
{\cal M}_{\rm 11(i)} \;\approx\; {\cal M}_{\rm 11(j)} \;\approx\;
{2 \alpha_s \log \Lambda \over 3 \pi} \; {\bf v} \cdot {\bf v}'  \;
	\left( T^a T^b \otimes T^b T^a \right) \; (1 \otimes 1) \;.
\label{A6i}
\end{equation}
\end{mathletters}
The color factors in (\ref{A6gi}) can be simplified by using
the identities in (\ref{TTTT}).
Adding up the results for the diagrams in (\ref{A6a})
and (\ref{A6e})--(\ref{A6gi}),
we find that the sum of the logarithmically divergent terms of order
$\alpha_s v^2$ is
\begin{eqnarray}
{\cal M}_8({}^1S_0) &\approx& {2 \alpha_s \log \Lambda \over 3 \pi N_c}
\Bigg( 2 C_F \; {\bf v} \cdot {\bf v}' \; (1 \otimes 1)
\nonumber \\
&& \;+\;
\left[ (N_c^2 - 4) \; {\bf v} \cdot {\bf v}' \;-\; (N_c^2 - 2) \; v^2 \right]
	\;( T^a \otimes T^a) \Bigg)
\; (1 \otimes 1) \;.
\label{Mdiv1S08}
\end{eqnarray}

The logarithmically divergent part of the diagrams for scattering through
the color-singlet operator ${\cal O}_1({}^1S_0) = \psi^\dagger \chi
\chi^\dagger \psi$
can be obtained from the expressions
(\ref{A6a}) and (\ref{A6e})--(\ref{A6i}) simply by replacing the color matrix
$T^a$ by the unit matrix $1$.  Adding up these contributions, we obtain
\begin{equation}
{\cal M}_1({}^1S_0) \;\approx\; {8 \alpha_s \log \Lambda \over 3 \pi} \;
\left( {\bf v} \cdot {\bf v}' \;  (T^a \otimes T^a)
	\;-\; C_F v^2 \; (1 \otimes 1) \right)
\; (1 \otimes 1) \;.
\label{Mdiv1S0}
\end{equation}
The ultraviolet divergent parts of
the matrix elements ${\cal M}_8({}^3S_1)$ and ${\cal M}_1({}^3S_1)$, which
correspond to scattering through the spin-triplet operators
${\cal O}_8({}^3S_1)$ and ${\cal O}_1({}^3S_1)$, can be obtained by
replacing the spin factor $1 \otimes 1$ by $\sigma^i \otimes \sigma^i$
in (\ref{Mdiv1S08}) and (\ref{Mdiv1S0}), respectively.

\subsection{Evolution Equations}
\label{app:evoleq}

The logarithmically divergent contributions to the scattering amplitudes
in Section~\ref{app:one-loop} can be expressed to leading order in $v$
as the matrix elements of dimension-8 operators.  Differentiating the
operator equation corresponding to (\ref{Mdiv1S0}) with respect to
$\Lambda$, we obtain the evolution equation for the operator ${\cal
O}_1({}^1S_0)$: 
\begin{equation}
\Lambda {d \ \over d \Lambda} {\cal O}_1({}^1S_0)
\;=\; {8 \alpha_s \over 3 \pi M^2} \; {\cal O}_8({}^1P_1)
\;-\; {8 C_F \alpha_s \over 3 \pi M^2} \; {\cal P}_1({}^1S_0) .
\label{evol1S0}
\end{equation}
By differentiating the operator equation corresponding to
(\ref{Mdiv1S08}) with respect to $\Lambda$, we obtain the evolution
equation for the operator ${\cal O}_8({}^1S_0)$:
\begin{equation}
\Lambda {d \ \over d \Lambda} {\cal O}_8({}^1S_0)
\;=\; {4 C_F \alpha_s \over 3 \pi N_c M^2} \; {\cal O}_1({}^1P_1)
\;+\; {2 (N_c^2-4) \alpha_s \over 3 \pi N_c M^2} \; {\cal O}_8({}^1P_1)
\;-\; {2(N_c^2 -2) \alpha_s \over 3 \pi N_c M^2} \; {\cal P}_8({}^1S_0)
\; .
\label{evol1S08}
\end{equation}
The evolution equations for the corresponding spin-triplet
operators can be obtained from (\ref{evol1S08}) and (\ref{evol1S0})
simply by inserting $\sigma^i$
between $\psi^\dagger$ and $\chi$ and also between $\chi^\dagger$ and $\psi$.
It is convenient to express the resulting operators in terms of
the combinations that appear in (\ref{Lcontact8}) by using the identity
\begin{equation}
D^i \sigma^j \otimes D^i \sigma^j
\;=\; {1 \over 3} {\bf D} \cdot \mbox{\boldmath $\sigma$} \otimes {\bf D} \cdot
\mbox{\boldmath $\sigma$}
\;+\; {1 \over 2} ({\bf D} \times \mbox{\boldmath $\sigma$})^i \otimes ({\bf D}
\times \mbox{\boldmath $\sigma$})^i
\;+\; D^{(i} \sigma^{j)} \otimes D^{(i} \sigma^{j)} \;.
\label{irrep}
\end{equation}
The resulting evolution equations are
\begin{mathletters}
\begin{eqnarray}
&& \Lambda {d \ \over d \Lambda} {\cal O}_1({}^3S_1)
\;=\; {8 \alpha_s \over 3 \pi M^2} \;
\left( {\cal O}_8({}^3P_{0}) \;+\; {\cal O}_8({}^3P_{1})
\;+\; {\cal O}_8({}^3P_{2}) \right)
\;-\; {8 C_F \alpha_s \over 3 \pi M^2} \; {\cal P}_1({}^3S_1) ,
\label{evol3S1}
\\
&& \Lambda {d \ \over d \Lambda} {\cal O}_8({}^3S_1)
\;=\; {4 C_F \alpha_s \over 3 \pi N_c M^2} \;
\left( {\cal O}_1({}^3P_{0})
\;+\; {\cal O}_1({}^3P_{1}) \;+\; {\cal O}_1({}^3P_{2}) \right)
\nonumber \\
&& \quad \;+\; {2 (N_c^2-4) \alpha_s \over 3 \pi N_c M^2} \;
\left( {\cal O}_8({}^3P_{0})
\;+\; {\cal O}_8({}^3P_{1}) \;+\; {\cal O}_8({}^3P_{2}) \right)
\;-\; {2(N_c^2 -2) \alpha_s \over 3 \pi N_c M^2} \; {\cal P}_8({}^3S_1) .
\label{evol3S18}
\end{eqnarray}
\end{mathletters}

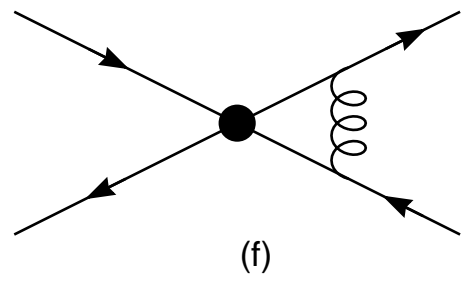
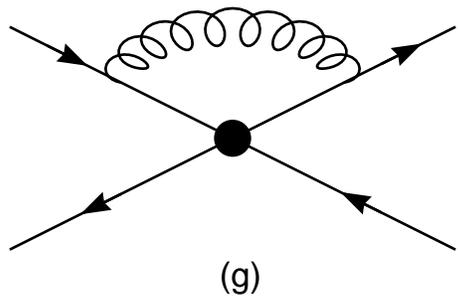
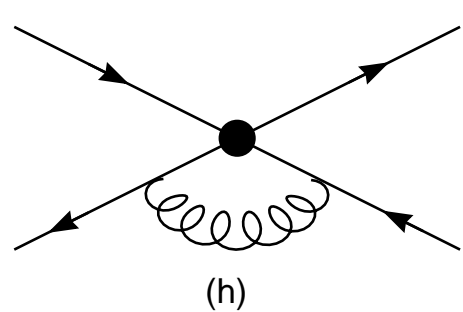
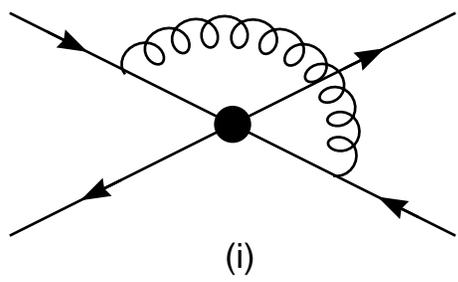
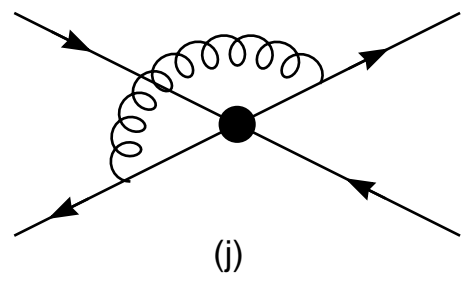
\begin{figure}
\caption{Feynman diagrams in NRQCD that contribute to the evolution of
an S-wave 4-fermion operator, such as $\psi^\dagger T^a \chi \chi^\dagger T^a
\psi$
or $\psi^\dagger \chi \chi^\dagger \psi$.}
\label{fig:octet-swave}
\end{figure}

The evolution equations for electromagnetic operators can be calculated
in the same way, except that there are no contributions from diagrams
such as those in
Figs.~\ref{fig:octet-swave}(g)-\ref{fig:octet-swave}(j), which involve
exchange of gluons between initial and final quark lines.  The evolution
equations for the dimension-6 electromagnetic operators can be obtained
from (\ref{evol1S0}) and (\ref{evol3S1}) by dropping the color-octet
terms on the right sides and inserting vacuum projections:
\begin{mathletters}
\begin{eqnarray}
\Lambda {d \ \over d \Lambda}
	\left( \psi^\dagger \chi {| 0 \rangle} {\langle 0 |} \chi^\dagger \psi \right)
&=& \;-\; {4 C_F \alpha_s \over 3 \pi M^2} \;
\left[ \psi^\dagger \chi {| 0 \rangle} {\langle 0 |} \chi^\dagger
(-\mbox{$\frac{i}{2}$} \tensor{\bf D})^2 \psi
	+ {\rm h.c.} \right],
\label{evolEM1S0}
\\
\Lambda {d \ \over d \Lambda}
\left[ \psi^\dagger \mbox{\boldmath $\sigma$} \chi {| 0 \rangle} \cdot {\langle
0 |} \chi^\dagger \mbox{\boldmath $\sigma$} \psi \right)
&=& \;-\; {4 C_F \alpha_s \over 3 \pi M^2} \;
\left[ \psi^\dagger \mbox{\boldmath $\sigma$} \chi {| 0 \rangle} \cdot
	{\langle 0 |} \chi^\dagger \mbox{\boldmath $\sigma$} (-\mbox{$\frac{i}{2}$}
\tensor{\bf D})^2 \psi + {\rm h.c.} \right].
\label{evolEM3S1}
\end{eqnarray}
\end{mathletters}

%%%%%%%%%%%%%%%%%%%%%%%%%%%%%%  REFERENCES  %%%%%%%%%%%%%%%%%%%%%%%%%%%%%%

%%%%%%%%%%%%%%%%%%%%%%%%%%%%  FIGURE CAPTIONS  %%%%%%%%%%%%%%%%%%%%%%%%%%%%%%

\begin{figure}

\caption{Example of a diagram that  contributes to the quarkonium
annihilation rate at order $\alpha_s^3$.  The three cuts of the diagram
participate in a KLN cancellation.}
\label{fig:kln}
\end{figure}

\begin{figure}
\caption{Schematic representation of the topological factorization
of the rate for quarkonium annihilation.  The short distance part
is represented by the circle labelled {\bf H}.  The quarkonium
wavefunctions are represented by the shaded ovals.  The wavefunctions
can be connected by light partons, such as the two gluons that are
shown explicitly.  Soft gluon interactions between the light partons
are represented by the circle labelled {\bf S}.}
\label{fig:top}
\end{figure}

\begin{figure}
\caption{Example of a Feynman diagram for quarkonium annihilation at order
$\alpha_s^2$. The shaded ovals represent the quarkonium wavefunctions.}
\label{fig:leading}
\end{figure}

\begin{figure}
\caption{Examples of real-gluon emission in quarkonium decay at order
$\alpha_s^3$. The shaded ovals represent the quarkonium wavefunctions.}
\label{fig:real}
\end{figure}

\begin{figure}
\caption{Examples of virtual-gluon emission in quarkonium decay at order
$\alpha_s^3$.  The shaded ovals represent the quarkonium wavefunctions.}
\label{fig:virtual}
\end{figure}

%%%%%%%%%%%%%%%%%%%%%%%%%%%%%%  TABLE  %%%%%%%%%%%%%%%%%%%%%%%%%%%%%%

 \begin{table}
 \begin{center}
 \begin{tabular}{lcl}
 Operator  & Estimate      & Description \\ \hline
 $\alpha_s$&  $v$          & effective quark-gluon coupling constant \\
 $\psi$    &  $(Mv)^{3/2}$ & heavy-quark (annihilation) field \\
 $\chi$    &  $(Mv)^{3/2}$ & heavy-antiquark (creation) field \\
 $D_t$ (acting on $\psi$ or $\chi$)
           &  $M v^2$      & gauge-covariant time derivative \\
 ${\bf D}$ (acting on $\psi$ or $\chi$)
	   &  $M v$        & gauge-covariant spatial derivative \\
 $g{\bf E}$    &  $M^2v^3$     & chromoelectric field \\
 $g{\bf B}$    &  $M^2v^4$     & chromomagnetic field \\
 $g\phi$ (in Coulomb gauge)  &  $M v^2$      & scalar potential \\
 $g{\bf A}$  (in Coulomb gauge)  &  $M v^3$      & vector potential \\
 \end{tabular}
 \caption{
Estimates of the magnitudes of NRQCD operators for matrix elements
between heavy-quarkonium states in terms of the heavy-quark mass $M$ and
the typical heavy-quark velocity $v$.  The estimates shown apply to
matrix elements in a quarkonium state ${| H \rangle}$ whose position is
localized to a region of size $1/Mv$ or less.  If the states are
normalized to $\langle H|H \rangle=1$, then the product of the
magnitudes of the operators gives the magnitude of the matrix element.
(In order to obtain estimates for matrix elements between momentum
eigenstates that are normalized to $\langle H| H\rangle=V$, where $V$ is
the volume of space, one should multiply the estimates for localized
states of unit norm by $(Mv)^{-3}$.)
}
\label{tab}
\end{center}
\end{table}

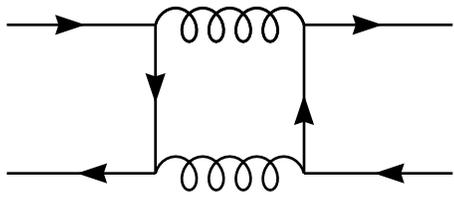
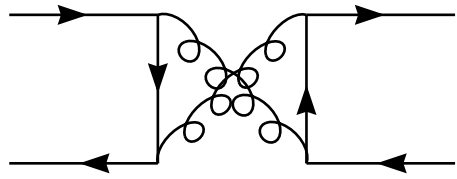
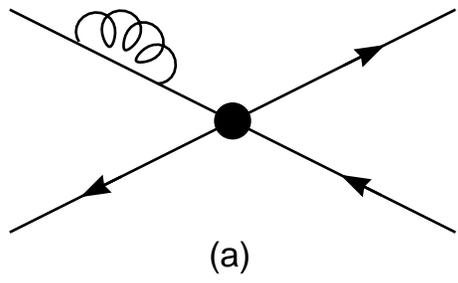
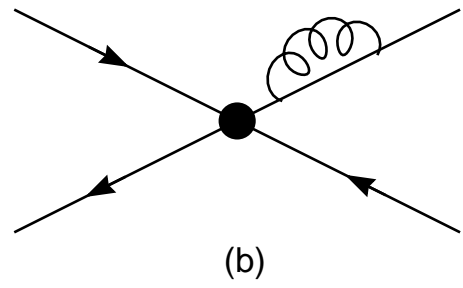
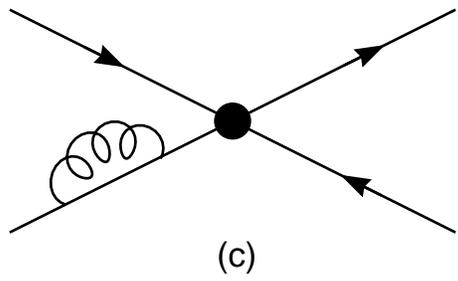
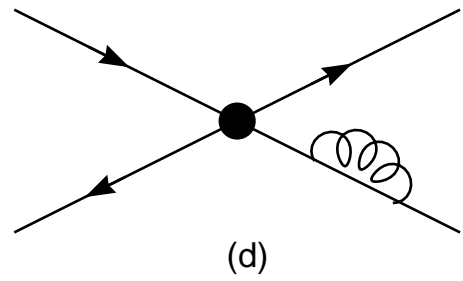
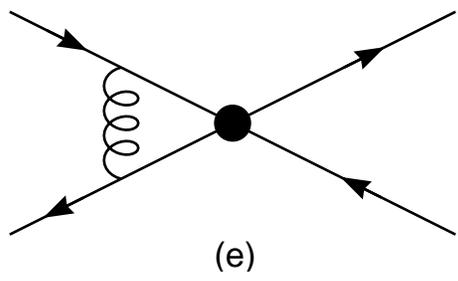
\end{document}